\documentclass[aps,prd,nofootinbib,twocolumn,superscriptaddress,preprintnumbers,balancelastpage,longbibliography]{revtex4-2}

\usepackage{style}

\newcommand{\dropfig}[2]{\IfFileExists{#1}{\includegraphics[width=\linewidth]{#1}}{\fbox{\parbox[c][0.4\linewidth][c]{\dimexpr\linewidth-2\fboxsep-2\fboxrule\relax}{\centering\itshape #2\\[6pt]\normalfont\small (figure to be added; mock analysis in progress)}}}}

\begin{document}
\preprint{Imperial--TP--2026--AM--03}
\title{Correlated signals of ultralight scalar dark matter in pulsar timing}

\author{Joshua~W.~Foster\,\orcidlink{0000-0002-7399-2608}}
\email{jwfoster@wisc.edu}
\affiliation{Department of Physics, University of Wisconsin-Madison, Madison, WI 53706, USA}

\author{Kimberly~K.~Boddy\,\orcidlink{0000-0003-1928-4667}}
\affiliation{Texas Center for Cosmology and Astroparticle Physics, Weinberg Institute,
Department of Physics, The University of Texas at Austin, Austin, TX 78712, USA}

\author{Jeff~A.~Dror\,\orcidlink{0000-0003-0110-6184}}
\affiliation{Institute for Fundamental Theory, Physics Department, University of Florida, Gainesville, FL 32611, USA}

\author{Vincent~S.~H.~Lee\,\orcidlink{0000-0002-3481-3590}}
\affiliation{Department of Physics, University of California Berkeley, Berkeley, CA 94720, USA}
\affiliation{Department of Physics, University of California, San Diego, La Jolla, CA 92093-0319, USA}

\author{Andrea Mitridate\,\orcidlink{0000-0003-2898-5844}}
\affiliation{Abdus Salam Centre for Theoretical Physics, Imperial College, London, SW7 2AZ, UK}

\author{Tristan~L.~Smith\,\orcidlink{0000-0003-2685-5405}}
\affiliation{Department of Physics and Astronomy, Swarthmore College, Swarthmore, PA 19081, USA}

\author{Kelsie~Taylor\,\orcidlink{0009-0001-7510-2306}}
\affiliation{Texas Center for Cosmology and Astroparticle Physics, Weinberg Institute, Department of Physics, The University of Texas at Austin, Austin, TX 78712, USA}

\author{Tanner Trickle\,\orcidlink{0000-0003-1371-4988}}
\affiliation{Department of Physics, Grainger College of Engineering, University of Illinois Urbana-Champaign, Urbana, IL 61801, USA}

\date{\today}
\begin{abstract}
Pulsar timing arrays (PTAs) are sensitive to ultralight dark matter (ULDM) in the $10^{-24}$--$10^{-20}\,\mathrm{eV}$ mass range, with existing datasets already probing otherwise open parameter space and future PTAs promising substantial improvements in reach.
Thus far, however, PTA searches for ULDM have typically been formulated using limiting descriptions. Analyses are performed in either the fully correlated limit, in which the local ULDM amplitude is shared across the array, or the fully uncorrelated limit, in which each pulsar has an independent local amplitude.
Because the transition between these regimes occurs within the PTA-sensitive mass range, projected sensitivities and data-derived constraints can depend on which limiting description is assumed.
For the first time, we develop a self-consistent analysis that treats the ULDM field as a Gaussian random field with finite spatial correlations, allowing the amplitude prior used in PTA signal models to interpolate continuously between the fully correlated and fully uncorrelated limits.
We apply the framework to both linearly and quadratically coupled scalar ULDM, the latter including the universal gravitational signal sourced by the oscillating ULDM pressure.
Pulsar-distance uncertainties are incorporated through an augmented latent-field prior, and the resulting distance-marginalized latent-amplitude distribution is represented with a normalizing-flow surrogate.
We validate the method on mock PTA datasets, including blinded signal injection tests.
\end{abstract}
\maketitle

\section{Introduction}
Ultralight dark matter (ULDM) generates novel experimental signatures through its coherent oscillations over length scales comparable to its de Broglie wavelength (e.g., see Ref.~\cite{Antypas:2022asj}). These oscillations produce metric fluctuations that can have observable gravitational effects in the local region of our Galaxy~\cite{Khmelnitsky:2013lxt}. Furthermore, if a scalar ULDM field is coupled to the Standard Model (SM), it can induce an oscillatory variation of fundamental constants~\cite{Damour:1994zq,Damour:2002nv,Cho:2007cy,Damour:2010rp}.
Such variations can be tested in laboratory experiments~\cite{Olive:2002tz,Flambaum:2004tm,Damour:2010rm,VanTilburg:2015oza,Hees:2016gop,Berge:2017ovy,Hees:2018fpg,Wcislo:2018ojh,Kennedy:2020bac,Davoudiasl:2024xnq} and can impact cosmological~\cite{Barrow:2005qf,Stadnik:2015kia,Sibiryakov:2020eir,Bouley:2022eer,Baryakhtar:2024rky,Baryakhtar:2025uxs,Ghosh:2025pbn} and astrophysical~\cite{Chiba:2006xx,Graham:2015ifn,Kaplan:2022lmz,NANOGrav:2023hvm,Smarra:2024kvv,Gan:2025icr} observations.
In particular, these variations affect the arrival times of pulses from millisecond pulsars monitored by pulsar timing array (PTA) experiments~\cite{Graham:2015ifn,Kaplan:2022lmz,NANOGrav:2023hvm,Smarra:2024kvv,Gan:2025icr}.
These PTA signals arise from fluctuations in the pulsar spin, deriving from changes in its moment of inertia and in the transition frequencies of the atomic clocks that serve as a reference to determine pulse arrival times.

Most PTA searches for ULDM couplings to the SM (as well as searches for ULDM metric fluctuations~\cite{Porayko:2018sfa,NANOGrav:2023hvm,EPTA:2023xxk,EPTA:2023xiy, Hu:2026yop}) focus on the ``fast'' mode, which imprints coherent oscillations in timing residuals that are approximately monochromatic, at frequency $\omega \approx m_\phi$ or $\omega \approx 2m_\phi$ for PTA signals that scale linearly or quadratically with the ULDM field, respectively, where $m_\phi$ is the ULDM mass.
There is also a ULDM slow mode, arising from the interference pattern of the ULDM field~\cite{Kim:2023kyy,Gan:2025icr}, but we do not consider it here.

\begin{figure}[ht]
    \centering
    \includegraphics[width=\linewidth]{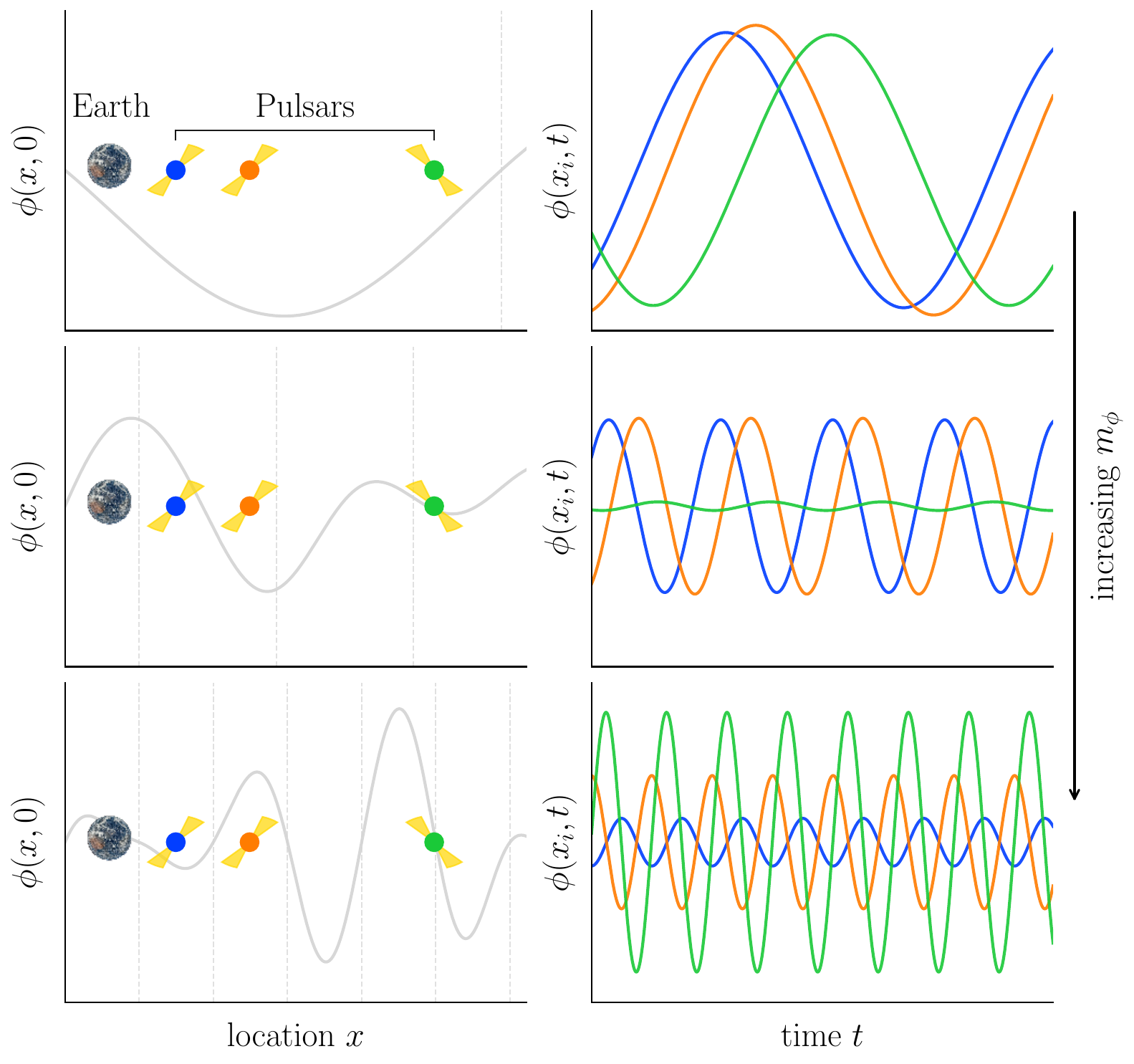}
    \caption{Illustration of the ULDM scalar field $\phi$ for three ULDM masses $m_\phi$ (rows, increasing downward).
    We consider only one spatial dimension for clarity.
    \textbf{Left panels:} Snapshots of the ULDM field at a given time $\phi(x, t{=}0)$ (gray solid lines), overlaid with the locations of Earth (globe icon) and three representative millisecond pulsars (colored icons) at various distances. Dashed vertical lines schematically indicate coherence regions of characteristic length \(\ell = 1/(m_\phi v_0)\) as in Eq.~\eqref{eq:ell}.
    The field amplitude is approximately constant within each patch but exhibits stochastic variations between patches.
    \textbf{Right panels:} Time series $\phi(x_i, t)$ of the ULDM field sampled at each pulsar (colors match the left panels), oscillating at angular frequency $\omega \approx m_\phi$ with an amplitude set by the coherence patch in which the object resides.
    At low masses (top panels), the coherence length exceeds the relevant Earth-pulsar and pulsar-pulsar separations, and the local amplitudes are highly correlated.
    At high masses (bottom panels), the coherence length is shorter than any Earth-pulsar or pulsar-pulsar separation, so all objects reside in distinct patches with uncorrelated amplitudes.
    At intermediate masses (middle panels), some but not all objects may reside in the same coherence patch.
    Our work addresses the full range of possible spatial correlations.}
    \label{fig:ULDM_cartoon}
\end{figure}

Figure~\ref{fig:ULDM_cartoon} illustrates how the ULDM field differs based on the locations of the pulsars and Earth versus the ULDM coherence regions of characteristic length $\ell = 1/(m_\phi v_0)$, where $v_0$ is the local velocity dispersion of the ULDM field. There are two useful limiting cases.  In the fully correlated limit, the coherence length is large compared to the relevant Earth-pulsar and pulsar-pulsar separations, so the local ULDM amplitudes are shared across the array.  In the fully uncorrelated limit, the coherence length is short compared to these separations, so the local amplitudes are effectively independent.

Most current PTA searches extract limits by analyzing these two extreme cases over the full range of ULDM-mass sensitivity, $10^{-24} \, \text{eV} \lesssim m_\phi \lesssim 10^{-20} \, \text{eV}$, but the regime of validity for the fully correlated (uncorrelated) limit applies only to the smallest (largest) masses in this range.%
\footnote{The PPTA collaboration considered only the uncorrelated limit for analyzing metric perturbations~\cite{Porayko:2018sfa}.
The EPTA collaboration considered three cases for analyzing metric perturbations~\cite{EPTA:2023xiy, Hu:2026yop} and universal ULDM couplings~\cite{Smarra:2024kvv, Hu:2026yop}: ``uncorrelated,'' ``correlated,'' and ``pulsar-correlated''~\cite{EPTA:2023xiy,Smarra:2024kvv, Hu:2026yop}. The ``correlated'' analysis combines two degenerate parameters into one, while the ``pulsar-correlated'' analysis leaves them as two parameters and is referred to as the correlated analysis by the NANOGrav collaboration~\cite{NANOGrav:2023hvm}.}

A proper treatment of the ULDM-induced correlations is needed for an analysis that applies across the full PTA-sensitive ULDM mass range, including the intermediate regime between the fully correlated and fully uncorrelated limits, illustrated in the middle panels of Fig.~\ref{fig:ULDM_cartoon}. References~\cite{Boddy:2025oxn,Dror:2025nvg} established the theoretical framework for understanding ULDM spatial correlations relevant for PTA searches, but an end-to-end numerical implementation suitable for Bayesian data analysis has not yet been developed.

In this paper, we develop such an implementation for the ULDM fast mode and apply it to both linearly and quadratically coupled scalar ULDM. For a linear coupling, the timing residual is linear in the field evaluated at the Earth and at the pulsars and oscillates at the Compton frequency $m_\phi$. For a quadratic coupling, the residual is quadratic in the field and oscillates at $2m_\phi$; this class includes both quadratic couplings of the field to the SM and the universal gravitational signal sourced by the oscillating ULDM pressure~\cite{Khmelnitsky:2013lxt}. Because the two cases share the same underlying Gaussian field, they are described by a common finite-correlation prior and differ only in the deterministic map from the field to the timing residual. More generally, the formalism we develop is applicable to a broader range of signals, including vector dark matter, whose treatment we defer to upcoming work.

Our analysis generalizes the limiting amplitude-phase approaches used in previous PTA searches for ULDM. These analyses typically impose one of two limiting descriptions: a fully correlated model, in which the local ULDM amplitude is shared across the array, or a fully uncorrelated model, in which each pulsar has an independent local amplitude. Rather than adopting either limit, we derive the joint prior over local ULDM amplitudes from the finite-distance ULDM covariance. Pulsar-distance uncertainties are incorporated into this prior, and the resulting distance-marginalized amplitude distribution is represented with a normalizing-flow surrogate. 

We test the robustness of our framework on mock PTA datasets, including null analyses, unblinded signal injections, and blinded injections with unknown input masses. 
Together, these tests demonstrate that the method provides a practical and robust approach to the inference challenges introduced by finite ULDM spatial correlations. Our implementation builds on \texttt{enterprise}~\cite{enterprise}, using it as the foundation for the PTA likelihood evaluation while adding the finite-correlation ULDM signal model, augmented latent-field prior, and normalizing-flow amplitude prior developed in this work. 

This paper is organized as follows.
We describe the impact of linear and quadratic ULDM couplings, including the gravitational signal, on pulsar timing in Sec.~\ref{sec:linear_signals}, and we discuss the correlations of the ULDM field in Sec.~\ref{sec:uldm_signal_correlations}.
We present our deterministic analysis in Sec.~\ref{sec:deterministic_analysis}.
We validate the analysis on mock datasets and present projected sensitivities in Sec.~\ref{sec:mock_data_analysis}. We conclude in Sec.~\ref{sec:conclusion}.

\section{ULDM signals in pulsar timing}
\label{sec:linear_signals}

An oscillating ULDM field can affect pulsar timing whenever it induces a time-dependent change in the pulse arrival times. For linearly coupled signals, these effects are especially simple: the timing residual is proportional to the ULDM field evaluated at the Earth and at the pulsar at the appropriate retarded time. Quadratically coupled signals, including the gravitational signal sourced by the oscillating ULDM pressure, are nearly as simple: the residual is then a quadratic function of the same field, with a fast mode oscillating at twice the Compton frequency. In either case, the simple response nevertheless has a nontrivial statistical structure, because a PTA does not observe a single field value, but rather many samples of the same stochastic field at widely separated spacetime points.

We therefore begin by describing how an oscillating ULDM field appears in pulsar timing data and then identify the field statistics needed to model the signal consistently across a PTA.  In Secs.~\ref{subsec:linear_timing_response} and~\ref{subsec:quadratic_couplings}, we describe the Earth-term and pulsar-term structure of the timing residual for the linear and quadratic scalar couplings.  In Sec.~\ref{subsec:signal_statistics_limiting_parametrizations}, we discuss the statistical assumptions implicit in the fully correlated and fully uncorrelated signal models used in previous PTA analyses, and we show how their limitations motivate a description based on the correlation structure of the underlying Gaussian ULDM field.  In Sec.~\ref{subsec:signal_parametrization}, we collect these ingredients into the signal parametrization used in our analyses, where we also restrict the quadratic signal to its fast mode.  Finally, in Sec.~\ref{subsec:gravitational_couplings}, we treat the universal gravitational signal as a special case of the quadratic fast mode.

\subsection{Linear couplings}
\label{subsec:linear_timing_response}
We first make the Earth-term and pulsar-term structure of the timing response explicit.
A linearly coupled signal is proportional to the ULDM field sampled at the spacetime points relevant to pulse emission and detection.
In a PTA, this gives an Earth term, associated with the field at the terrestrial reference clock, and a pulsar term, associated with the field at the pulsar at the retarded time of pulse emission.
The relative importance of these two terms is fixed by the physical coupling under consideration.

Specializing to scalar ULDM coupled linearly to the SM, we parametrize the couplings relevant to pulsar timing as
\begin{align}
    \mathcal{L} \supset\frac{\phi}{\Lambda}& \Bigg[ \frac{ d_\gamma}{4e^2}F_{\mu\nu}F^{\mu\nu}+\frac{d_g\beta_3}{2g_3}G_{\mu\nu}^{A}G_{A}^{\mu\nu}\nonumber\\
    - & \sum_{f=e,\mu} d_{f} \, m_f\, \bar f f -\sum_{q=u,d}(d_q+\gamma_q d_g)\, m_q \, \bar qq\Bigg] \, ,
    \label{eq:uldm_L}
\end{align}
where $\Lambda \equiv M_\text{Pl} / \sqrt{4\pi}$, $M_\text{Pl} = 1.22 \times 10^{19} \, \text{GeV}$ is the Planck mass, $F^{\mu \nu}$ is the electromagnetic field strength tensor, $G_A^{\mu \nu}$ is the QCD field strength tensor, $f$ are the Dirac fermion lepton fields, $q$ are the Dirac fermion light quark fields, $e$ is the electromagnetic coupling, $\beta_3$ is the QCD beta function, $g_3$ is the QCD gauge coupling, $m_f$ are the lepton masses, $m_q$ are the light quark masses, $\gamma_q$ are the light quark anomalous dimensions, and the various $d_i$ are the dimensionless coupling coefficients. These interactions induce oscillatory variations of SM
parameters~\cite{Damour:2010rp,Damour:2010rm},
\begin{equation}
\begin{gathered}
    \frac{\delta\alpha}{\alpha}
    =
    d_\gamma\frac{\phi}{\Lambda}, \qquad 
    \frac{\delta m_{e,\mu}}{m_{e,\mu}}
    =
    d_{e,\mu}\frac{\phi}{\Lambda},
    \\
    \frac{\delta m_{p,n}}{m_{p,n}}
    \approx
    \left(d_g+C_n d_{\hat m}\right)
    \frac{\phi}{\Lambda},
    \label{eq:fluctuations}
\end{gathered}
\end{equation}
where $\alpha$ is the fine-structure constant, $C_n=0.048$~\cite{Damour:2010rp} and $d_{\hat m}=(m_u d_u+m_d d_d)/(m_u+m_d)$.

These variations affect pulsar timing through two effects~\cite{Kaplan:2022lmz}.  First,
changes in the masses of pulsar constituents alter the pulsar moment of
inertia and therefore its spin frequency.  This produces a pulsar term,
evaluated at the retarded time of pulse emission.  Second, changes in
atomic energy levels alter the terrestrial clock standard used to record
pulse arrival times.  This produces an Earth term, common to all pulsars.
For weak couplings, both effects are linear in the ULDM field, and the
timing residual for pulsar $I$ can be written as
\begin{align}
    r_I(t)
    =
    \frac{\mathbf d_\mathrm{SM}}{m_\phi\Lambda}
    \cdot
    \left[
        \mathbf y^E\,
        \phi(\mathbf x_E,t)
        +
        \mathbf y^P\,
        \phi(\mathbf x_I,t-x_I)
    \right],
    \label{eq:general_residual}
\end{align}
where $\mathbf x_E$ and $\mathbf x_I$ are the locations of the Earth and
the $I^{\rm th}$ pulsar, respectively, and $x_I\equiv |\mathbf x_I|$.
Here
\begin{equation}
    \mathbf d_\mathrm{SM}
    =
    \{d_\gamma,d_e,d_\mu,d_g,d_{\hat m}\},
\end{equation}
while
\begin{equation}
    \mathbf y^i
    =
    \{y_\gamma^i,y_e^i,y_\mu^i,y_g^i,y_{\hat m}^i\},
    \qquad
    i\in\{E,P\},
\end{equation}
parametrizes the response of the Earth and pulsar terms to the different
scalar couplings.

The values of $\mathbf y^i$ depend on the physical effect induced by ULDM.  For the scalar effects considered here, the pulsar-spin response due to the pulsar moment of inertia fluctuation is
approximately~\cite{Kaplan:2022lmz,NANOGrav:2023hvm}
\begin{equation}
\begin{gathered}
    y_g^P
    \approx -5,
    \qquad\qquad
    y_{\hat m}^P
    \approx -0.24,\\
    y_\mu^P
    \approx 2\times 10^{-3},
    \qquad
    y_e^P
    \approx 1.7\times 10^{-5}.
    \label{eq:y_P}
\end{gathered}
\end{equation}
with $y_\gamma^P=0$.
The clock response arises because pulse arrival times are referenced
against an atomic clock standard, primarily based on cesium
atoms~\cite{McCarthy2009}.  Therefore, any effect that shifts the relevant atomic
energy levels appears as a fluctuation in the measured arrival times,
common to all pulsars.  For the scalar effects considered here, the clock
response is approximately~\cite{Kaplan:2022lmz,NANOGrav:2023hvm}:
\begin{equation}
\begin{gathered}
    y_g^E
    \approx 1,
    \qquad\qquad
    y_{\hat m}^E
    \approx 0.16,
	\\
    y_e^E
    \approx 2,
    \qquad\qquad
    y_\gamma^E
    \approx 4.8 .
    \label{eq:y_E}
\end{gathered}
\end{equation}
with $y_\mu^E=0$.
Substituting these response coefficients into
Eq.~\eqref{eq:general_residual} specifies the timing signal for any choice
of the underlying scalar couplings.

\subsection{Quadratic couplings}
\label{subsec:quadratic_couplings}

The ULDM field may instead couple to the SM quadratically, which is the leading interaction whenever a symmetry such as $\phi\to-\phi$ forbids the linear term~\cite{Stadnik:2015kia,Hees:2018fpg}. Following Ref.~\cite{Gan:2025icr}, the quadratic interactions relevant to pulsar timing parallel Eq.~\eqref{eq:uldm_L} with $\phi/\Lambda$ replaced by $\phi^2/(2\Lambda^2)$:
\begin{equation}
\begin{split}
    \mathcal{L} &\supset\frac{\phi^2}{2\Lambda^2}\Bigg[ \frac{ d_\gamma^{(2)}}{4e^2}F_{\mu\nu}F^{\mu\nu}+\frac{d_g^{(2)}\beta_3}{2g_3}G_{\mu\nu}^{A}G_{A}^{\mu\nu} \\
    &- \sum_{f=e,\mu} d_{f}^{(2)}  m_f\, \bar f f -\sum_{q=u,d}(d_q^{(2)}+\gamma_q d_g^{(2)})\, m_q \, \bar qq\Bigg] \, ,
    \label{eq:uldm_L_quadratic}
\end{split}
\end{equation}
where $\Lambda=M_\text{Pl}/\sqrt{4\pi}$, as in the linear case, and the $d_i^{(2)}$ are the dimensionless quadratic coupling coefficients. These induce fractional variations of the SM parameters that are quadratic in the field,
\begin{equation}
\begin{gathered}
    \frac{\delta\alpha}{\alpha} = d_\gamma^{(2)}\frac{\phi^2}{2\Lambda^2}, \qquad 
    \frac{\delta m_{e,\mu}}{m_{e,\mu}} = d_{e,\mu}^{(2)}\frac{\phi^2}{2\Lambda^2},\\
    \frac{\delta m_{p,n}}{m_{p,n}} \approx \left(d_g^{(2)}+C_n d_{\hat m}^{(2)}\right)\frac{\phi^2}{2\Lambda^2},
    \label{eq:fluctuation_quadratic}
\end{gathered}
\end{equation}
in direct analogy with Eqs.~\eqref{eq:fluctuations}.

Because these variations act on the pulsar spin and the terrestrial clock through the same mechanisms as in the linear case, the response vectors $\mathbf y^P$ and $\mathbf y^E$ of Eqs.~\eqref{eq:y_P} and~\eqref{eq:y_E} are unchanged, but the variation is now quadratic in the field. The timing residual can then be written as
\begin{equation}
    r_I(t) = \frac{\mathbf d^{(2)}_{\rm SM}}{2m_\phi\Lambda^2}\cdot
    \left[ \mathbf y^E\,\phi(\mathbf x_E,t)^2
    + \mathbf y^P\,\phi(\mathbf x_I,t-x_I)^2 \right],
    \label{eq:quadratic_general_residual}
\end{equation}
with $\mathbf d^{(2)}_{\rm SM}=\{d_\gamma^{(2)},d_e^{(2)},d_\mu^{(2)},d_g^{(2)},d_{\hat m}^{(2)}\}$. Unlike the linearly coupled field, the squared field is not monochromatic: it contains a fast component at $2m_\phi$, itself monochromatic whenever the field is, together with a slowly varying part. A monochromatic field therefore produces a monochromatic fast-mode signal at $2m_\phi$, just as the linear coupling produces one at $m_\phi$. We restrict to this fast mode when parameterizing the signal in Sec.~\ref{subsec:signal_parametrization}.

The slow mode is the low-frequency component at $\omega\sim\omega_\mathbf{p}-\omega_\mathbf{q}\sim m_\phi v_0^2$, the beat between two field modes of momenta $\mathbf{p},\mathbf{q}$ and nearly equal frequencies $\omega_\mathbf{p},\omega_\mathbf{q}$. Its pulsar-timing phenomenology was studied in Ref.~\cite{Gan:2025icr}, which argued that it is approximately Gaussian, so we do not treat it in detail here. We show in App.~\ref{app:slowmode} that it is in fact exactly Gaussian in the continuum limit, strengthening that result and allowing it to be analyzed as an ordinary stationary Gaussian process.

A quadratic coupling also makes the field's dispersion relation density dependent, shifting the in-medium mass by $\Delta m^2\simeq g\,\rho/\Lambda^2$ inside a body of density $\rho$, with $g$ the relevant effective coupling. For sufficiently large couplings, this screens the field within the Earth, Sun, or pulsar and suppresses the signal~\cite{Gan:2025icr}; we restrict to the unscreened regime throughout.

\subsection{Signal statistics and limiting behaviors}
\label{subsec:signal_statistics_limiting_parametrizations}

Equation~\eqref{eq:general_residual} shows that, once the scalar response
coefficients have been specified, the remaining signal uncertainty is the
ULDM field realization itself.  The standard parametrization of this
realization starts from a superposition of nonrelativistic waves with
random phases and Rayleigh-distributed amplitudes.  Following
Ref.~\cite{Foster:2017hbq}, we write
\begin{align}
    \phi(\mathbf{x}, t)
    \approx
    \frac{\sqrt{2\rho_\phi}}{m_\phi}
    \sum_{\mathbf v}
    \alpha_{\mathbf v}
    \cos\!\left[
        \omega_{\mathbf v} t
        -
        m_\phi \mathbf v\cdot\mathbf x
        +
        \theta_{\mathbf v}
    \right],
    \label{eq:ULDM_field}
\end{align}
where $\rho_\phi\simeq 0.4\,\mathrm{GeV}/\mathrm{cm}^3$ is the local ULDM
mass density, which we approximate as spatially-constant over the PTA. The phases $\theta_{\mathbf v}$ are independent random
variables uniformly distributed on $[0,2\pi]$, and the amplitudes
$\alpha_{\mathbf v}$ are independent Rayleigh-distributed random variables
with
\begin{equation}
    \left\langle
        \alpha_{\mathbf v}\alpha_{\mathbf v'}
    \right\rangle
    =
    f(\mathbf v)\,\Delta\mathbf v\,
    \delta_{\mathbf v\mathbf v'} .
\end{equation}
The spacing $\Delta\mathbf v$ discretizes the velocity sum weighted by the
velocity distribution $f(\mathbf v)$.

The Rayleigh-amplitude and random-phase parametrization in
Eq.~\eqref{eq:ULDM_field} can equivalently be written in terms of
Gaussian quadratures,
\begin{equation}
\begin{split}
    \phi(\mathbf{x}, t)
    \approx
    \frac{\sqrt{2\rho_\phi}}{m_\phi}
    \sum_{\mathbf v}\bigg[&
    C_{\mathbf v}
    \cos\!\left(
        \omega_{\mathbf v} t
        -
        m_\phi \mathbf v\cdot\mathbf x
    \right)
    \\
    &+
    S_{\mathbf v}
    \sin\!\left(
        \omega_{\mathbf v} t
        -
        m_\phi \mathbf v\cdot\mathbf x
    \right)
    \bigg],
    \label{eq:ULDM_field_Gaussian}
\end{split}
\end{equation}
where $C_{\mathbf v}$ and $S_{\mathbf v}$ are independent Gaussian random variables with variances fixed by $f(\mathbf v)\Delta\mathbf v$. We review this equivalence in App.~\ref{app:rayleigh_gaussian_equivalence}.  

The important consequence is that the ULDM field is Gaussian, since it is a linear
  combination of Gaussian mode variables. Its statistics are therefore fixed by the
  spatio-temporal two-point function $\langle \phi(\mathbf x,t)\,\phi(\mathbf
  x',t')\rangle$. We derive this correlation structure in Sec.~\ref{sec:uldm_signal_correlations}. Here we first identify the spatial coherence scale that controls its qualitative behavior. For cold dark matter following an isotropic Maxwell-Boltzmann velocity distribution,
\begin{equation}
    f(\mathbf v)
    =
    \frac{1}{(2\pi v_0^2)^{3/2}}
    \exp\!\left[-\frac{\mathbf v^2}{2v_0^2}\right],
    \label{eq:Maxwell-Boltzmann}
\end{equation}
with $v_0 \approx 155\,\mathrm{km/s}$ in the Standard Halo Model~\cite{Freese:2012xd}, the characteristic coherence length is
\begin{equation}
    \ell
    \equiv
    \frac{1}{m_\phi v_0}
    \approx
    1.2\,\mathrm{kpc}
    \left(\frac{10^{-23}\,\mathrm{eV}}{m_\phi}\right)
    \left(\frac{155\,\mathrm{km/s}}{v_0}\right).
    \label{eq:ell}
\end{equation}
This scale determines whether the Earth and pulsars sample nearly the same field realization or effectively independent ones.

Previous PTA searches have often adopted one of two corresponding limits~\cite{Graham:2015ifn,Kaplan:2022lmz,NANOGrav:2023hvm,Smarra:2024kvv,Gan:2025icr}. In the amplitude-phase formulation used in these analyses, the rapidly varying retarded phases are treated as independent nuisance parameters, while the limiting assumptions enter through the prior on the local ULDM amplitudes. When all relevant Earth-pulsar and pulsar-pulsar separations are small compared to $\ell$, the array samples a fully correlated field realization and the sites share a single Rayleigh-distributed amplitude. When those separations are large compared to $\ell$, the field values decohere and each pulsar receives an independent Rayleigh-distributed amplitude.

These amplitude priors are useful limiting cases, but they are not separate signal models. They arise from the same underlying Gaussian random field in Eq.~\eqref{eq:ULDM_field}. Across the PTA-sensitive mass range, $\ell$ can be comparable to Earth-pulsar and pulsar-pulsar separations, so the sampled field values need not be either perfectly correlated or completely uncorrelated. In this intermediate regime, the correlations among the Earth and pulsar terms must instead be determined from the finite-distance two-point function of the ULDM field.

In Sec.~\ref{sec:uldm_signal_correlations}, we derive this correlation
structure explicitly for the ULDM field sampled across the PTA. We also show that the monochromatic description used in the analysis is an excellent approximation on the relevant timescales.  This allows the finite-distance field correlations to be translated into correlations among the local amplitudes and phases used in the Bayesian analysis.

\subsection{Signal parametrization}
\label{subsec:signal_parametrization}

For the Bayesian analysis developed below, it is convenient to express the timing residual in a compact form that separates the overall signal strength and coupling dependence from the underlying field realization. Both the linear and quadratic responses admit such a form: the residual is the sum of an Earth term and a pulsar term, each oscillating at the relevant frequency with an amplitude and phase fixed by the local field, modulated by an overall amplitude and a pair of coupling-dependent coefficients. We introduce this parametrization for the linear signal first and then for the quadratic and gravitational signals, which reuse the same field variables and differ only in the response map.

\subsubsection{Linear signals}

Starting from Eq.~\eqref{eq:general_residual}, the response to a fixed choice of scalar couplings is controlled by the two scalar combinations
\begin{align}
    y_E
    &\equiv
    \mathbf d_\mathrm{SM}\cdot\mathbf y^E,
    &
    y_P
    &\equiv
    \mathbf d_\mathrm{SM}\cdot\mathbf y^P .
    \label{eq:effective_response_coefficients}
\end{align}
The timing residual can therefore be written as
\begin{align}
    r_I(t)
    =
    \frac{1}{m_\phi\Lambda}
    \left[
        y_E\,\phi(\mathbf x_E,t)
        +
        y_P\,\phi(\mathbf x_I,t-x_I)
    \right].
    \label{eq:scalar_residual_yEyP}
\end{align}

To make the signal amplitude explicit, we define the dimensionless retarded-time field
\begin{align}
    \hat\phi_I(t)
    \equiv
    \frac{m_\phi}{\sqrt{2\rho_\phi}}
    \phi(\mathbf x_I,t-x_I),
    \label{eq:dimensionless-phi}
\end{align}
with the convention that $I=0$ denotes the Earth.  Thus $\hat\phi_0(t)$ is the Earth-term field, while $\hat\phi_I(t)$ for $I=1,\ldots,N_p$, where $N_p$ is the number of pulsars, is the field sampled at the pulsars.

In the monochromatic approximation used for the deterministic signal model, the dimensionless field sampled at each site can be written in amplitude--phase form as
\begin{equation}
    \hat\phi_J(t)
    =
    A_J\cos(m_\phi t-\alpha_J),
    \qquad
    J=0,\ldots,N_p .
    \label{eq:linear_field_amplitude_phase}
\end{equation}
The latent\footnote{Throughout, ``latent'' variables are quantities that are not measured directly but are introduced into the model, inferred as part of the fit, and then marginalized over.} amplitudes and phases $(A_J,\alpha_J)$ are fixed by the local realization of the ULDM field, with their joint distribution determined by the finite-distance field correlations derived in Sec.~\ref{sec:uldm_signal_correlations}.
In terms of these dimensionless fields,
\begin{align}
    r_I(t)
    =
    A_E\,\hat\phi_0(t)
    +
    A_P\,\hat\phi_I(t),
    \label{eq:free_ratio_residual}
\end{align}
where
\begin{align}
    A_E
    =
    \frac{y_E\sqrt{2\rho_\phi}}
    {m_\phi^2\Lambda},
    \qquad
    A_P
    =
    \frac{y_P\sqrt{2\rho_\phi}}
    {m_\phi^2\Lambda}.
    \label{eq:free_ratio_amplitudes}
\end{align}

Equivalently, using Eq.~\eqref{eq:linear_field_amplitude_phase}, the free coupling-ratio residual is
\begin{equation}
    r_I(t)
    =
    A_E A_0\cos(m_\phi t-\alpha_0)
    +
    A_P A_I\cos(m_\phi t-\alpha_I).
    \label{eq:linear_free_residual_amplitude_phase}
\end{equation}
Here $A_E$ and $A_P$ are timing-residual amplitude scales, while $A_0$ and $A_I$ are dimensionless ULDM field amplitudes at the Earth and at pulsar $I$.
The parametrization in Eq.~\eqref{eq:free_ratio_residual} lends itself to an agnostic analysis, in which the Earth-term and pulsar-term signal strengths are allowed to vary independently.
Since this approach does not impose a fixed relation between the Earth and pulsar responses, we refer to it as the free coupling-ratio parametrization.

We also use an alternative parametrization in which the relative Earth-term and pulsar-term responses are fixed by a specified scalar-coupling scenario.
In this case, the overall signal strength can be factored out by defining
\begin{equation}
    y_{\rm max}
    =
    \max\{|y_E|,|y_P|\},
    \label{eq:ymax_definition}
\end{equation}
and the dimensionless relative weights
\begin{align}
    w_E
    =
    \frac{y_E}{y_{\rm max}},
    \qquad
    w_P
    =
    \frac{y_P}{y_{\rm max}} .
    \label{eq:relative_couplings}
\end{align}
The corresponding timing-residual amplitude is
\begin{align}
    A_\phi
    =
    \frac{y_{\rm max}\sqrt{2\rho_\phi}}
    {m_\phi^2\Lambda}
    \sim
    5\,\mathrm{ns}
    \left(
        \frac{y_{\rm max}}{10^{-7}}
    \right)
    \left(
        \frac{10^{-22}\,\mathrm{eV}}{m_\phi}
    \right)^2 .
    \label{eq:amplitude_parametrics}
\end{align}
With these definitions, Eq.~\eqref{eq:free_ratio_residual} becomes
\begin{align}
    r_I(t)
    =
    A_\phi
    \left[
        w_E\hat\phi_0(t)
        +
        w_P\hat\phi_I(t)
    \right].
    \label{eq:useful_residual}
\end{align}
In amplitude--phase form, Eq.~\eqref{eq:useful_residual} becomes
\begin{equation}
\begin{split}
    r_I(t) = A_\phi\big[ &w_E A_0\cos(m_\phi t-\alpha_0)\\
    &+ w_P A_I\cos(m_\phi t-\alpha_I)\big].
\end{split}
    \label{eq:linear_fixed_residual_amplitude_phase}
\end{equation}
We refer to this as the fixed coupling-ratio parametrization: $w_E$ and $w_P$ are fixed numbers for a given coupling scenario in an analysis, and $A_\phi$ is a varying parameter that controls the overall signal strength.

\subsubsection{Quadratic signals}

The quadratically coupled signals of Sec.~\ref{subsec:quadratic_couplings} admit the same compact parametrization once we restrict to the fast mode. We write $[X]_{\rm fast}$ for the part of a quantity $X$ quadratic in the field that oscillates at $2m_\phi$, as distinct from the slowly varying remainder, namely the constant and difference-frequency parts that source the slow mode. Using the monochromatic form in Eq.~\eqref{eq:linear_field_amplitude_phase}, the fast part of the squared field is
\begin{equation}
    [\hat\phi_J^2(t)]_{\rm fast}
    =
    \tfrac12 A_J^2\cos(2m_\phi t - 2\alpha_J).
    \label{eq:phihat_squared}
\end{equation}
Absorbing the factor of $\tfrac12$ into the amplitude, the residual takes the same Earth- and pulsar-term form as Eq.~\eqref{eq:useful_residual}, now oscillating at $2m_\phi$ with phases $2\alpha_0$ and $2\alpha_I$,
\begin{equation}
\begin{split}
    r_I(t) = A_\phi\big[ &w_E A_0^2\cos(2m_\phi t-2\alpha_0)\\
    &+ w_P A_I^2\cos(2m_\phi t-2\alpha_I)\big],
\end{split}
    \label{eq:quadratic_residual_common}
\end{equation}
and analogously, in the free coupling-ratio parametrization,
\begin{equation}
\begin{split}
    r_I(t) ={}& A_E A_0^2\cos(2m_\phi t-2\alpha_0)\\
    &+ A_P A_I^2\cos(2m_\phi t-2\alpha_I).
\end{split}
    \label{eq:quadratic_free_residual}
\end{equation}
We retain the symbols $A_\phi$, $A_E$, and $A_P$ used in the linear case, now with the quadratic amplitude scale. For the SM quadratic coupling, with $y_{E,P}=\mathbf d_{\rm SM}^{(2)}\cdot\mathbf y^{E,P}$, the free coupling-ratio amplitudes are
\begin{equation}
    A_E = \frac{y_E\,\rho_\phi}{2 m_\phi^3\Lambda^2},
    \qquad
    A_P = \frac{y_P\,\rho_\phi}{2 m_\phi^3\Lambda^2},
    \label{eq:quad_free_amplitudes}
\end{equation}
mirroring Eq.~\eqref{eq:free_ratio_amplitudes}, while the fixed coupling-ratio amplitude is
\begin{equation}
    A_\phi = \frac{y_{\rm max}\,\rho_\phi}{2 m_\phi^3\Lambda^2},
    \label{eq:quad_amplitude}
\end{equation}
with relative weights $w_{E,P}=y_{E,P}/y_{\rm max}$ defined as in Eqs.~\eqref{eq:ymax_definition} and~\eqref{eq:relative_couplings}. The universal gravitational signal is the special case $w_E=-w_P=1$, treated in Sec.~\ref{subsec:gravitational_couplings}. In all cases the residual depends on the field only through the latent amplitudes and phases $(A_J,\alpha_J)$, now through $A_J^2$ at frequency $2m_\phi$, so the field statistics of Sec.~\ref{sec:uldm_signal_correlations} and the latent-amplitude prior of Sec.~\ref{sec:deterministic_analysis} carry over unchanged, with only the deterministic response map modified.

\subsection{Fast mode gravitational fluctuations as a quadratic observable}
\label{subsec:gravitational_couplings}

Independently of any coupling to the SM, every ULDM produces a gravitational signal: the oscillating pressure sources an oscillating gravitational potential $\Psi$. Its fast part $[\Psi]_{\rm fast}$ is local in the field, proportional to $\phi^2$ at each point, because the Poisson inversion of the sum-frequency source is cancelled by its kinetic structure. In the subcoherence limit,
\begin{equation}
    [\Psi(\mathbf x,t)]_{\rm fast} = \pi G\,[\phi(\mathbf x,t)^2]_{\rm fast},
    \label{eq:grav_potential}
\end{equation}
as we derive in App.~\ref{app:gravitational}, with $G$ Newton's constant; we write $\Psi_{\rm fast}\equiv[\Psi]_{\rm fast}$ hereafter. The potential perturbs pulse arrival times through the metric, and for the fast mode, the gauge-invariant Einstein delay (the gravitational redshift) dominates, while the Doppler and Shapiro contributions are suppressed by the ULDM velocity; the hierarchy is reversed for the slow mode~\cite{Kim:2023kyy}. Integrating the redshift over time and differencing the Earth and pulsar contributions gives
\begin{equation}
    r_I(t) = \frac{\pi G}{2m_\phi}\left[\phi(\mathbf x_E,t)^2-\phi(\mathbf x_I,t-x_I)^2\right]_{\rm fast},
    \label{eq:grav_residual}
\end{equation}
which is the universal-coupling special case of the quadratic residual of Eq.~\eqref{eq:quadratic_general_residual}, restricted to the fast mode as in Sec.~\ref{subsec:signal_parametrization}. In the fixed coupling-ratio parametrization, it reads
\begin{equation}
    r_I(t) = A_\phi\left[A_0^2\cos(2m_\phi t-2\alpha_0) - A_I^2\cos(2m_\phi t-2\alpha_I)\right],
    \label{eq:grav_residual_param}
\end{equation}
that is, $w_E=-w_P=1$, with amplitude
\begin{equation}
    A_\phi = \frac{\pi G\rho_\phi}{2m_\phi^3}
    \sim 20\,\mathrm{ns}
    \left(\frac{\rho_\phi}{0.4\,\mathrm{GeV/cm^3}}\right)
    \left(\frac{10^{-23}\,\mathrm{eV}}{m_\phi}\right)^3,
    \label{eq:grav_amplitude}
\end{equation}
up to an $\mathcal{O}(1)$ factor set by the redshift projection~\cite{Khmelnitsky:2013lxt,Porayko:2018sfa}.

Equation~\eqref{eq:grav_residual} is the fast mode, for which the potential is local in $\phi^2$. The corresponding slow mode, at $\omega\sim m_\phi v_0^2$, instead has a potential set by the ordinary Poisson equation, nonlocal in the field; it was studied in Ref.~\cite{Kim:2023kyy} and, as a special case of the quadratic slow mode of Sec.~\ref{subsec:quadratic_couplings}, is also exactly Gaussian in the continuum limit (App.~\ref{app:slowmode}).

\section{ULDM correlations}
\label{sec:uldm_signal_correlations}

Having established that ULDM signals depend on the field sampled at the Earth and at the pulsars at the appropriate retarded times, we now determine the correlations among these field values. These correlations are fixed by the spatio-temporal two-point function of the underlying ULDM field, and they control how the local amplitude prior interpolates between the fully correlated and fully uncorrelated limits.

  We derive this correlation structure in three steps. We first compute the retarded-time field covariance in the time domain in Sec.~\ref{subsec:time_domain_covariance}. We then express the same correlation structure in the frequency domain in Sec.~\ref{subsec:fourier_domain_covariance}. Finally, in Sec.~\ref{subsec:latent_covariance}, we translate the field covariance into the covariance of the local sine and cosine quadratures, which determines the amplitude and phase correlations used in the Bayesian analysis.

\subsection{Covariance in the time domain}
\label{subsec:time_domain_covariance}

The two-point function for a Gaussian ULDM field $\phi$, using Eq.~\eqref{eq:ULDM_field} in the continuum limit, is
\begin{align}
    \langle\phi(\mathbf{x}, t)\phi(\mathbf{x}', t') \rangle
    &= \frac{\rho_\phi}{m_\phi^2}\int d^3 \mathbf{v}\, f(\mathbf{v})\, \nonumber\\
    &\qquad\times \cos\!\left[\omega_{\mathbf{v}}\Delta t - m_\phi\mathbf{v} \cdot \Delta \mathbf{x} \right] \,,
    \label{eq:nonrelativistic_correlation}    
\end{align}
where $f(\mathbf{v})$ is the unit-normalized three-velocity distribution, $\omega_\mathbf{v} \approx m_\phi(1 +v^2/2)$ is the ULDM energy in the nonrelativistic limit, and $\langle \cdots\rangle$ indicates averages over the random variables $\alpha_\mathbf{v}$ and $\theta_\mathbf{v}$. Making the substitutions $\mathbf{x} \to \mathbf{x}_I$, $\mathbf{x}' \to \mathbf{x}_J$, $t \to t - x_I$, and $t' \to t' - x_J$, and using Eq.~\eqref{eq:nonrelativistic_correlation}, we obtain the cross-correlation for the dimensionless retarded-time field at the $I^\mathrm{th}$ and $J^\mathrm{th}$ sites, the rescaling of Eq.~\eqref{eq:dimensionless-phi} having reduced the $\rho_\phi/m_\phi^2$ prefactor to $1/2$:
\begin{equation}
    \langle \hat{\phi}_I(t) \hat{\phi}_J(t') \rangle
    = \frac{1}{2} \int d^3 \mathbf{v}\, f(\mathbf{v})
    \cos\!\left[\omega_{\mathbf{v}}\Delta  t_{IJ}- m_\phi\mathbf{v} \cdot \Delta \mathbf{x}_{IJ}\right],
    \label{eq:full_velocity_integral_covariance}
\end{equation}
where \(\Delta \mathbf{x}_{IJ}\equiv \mathbf{x}_I-\mathbf{x}_J\),
\(\Delta t\equiv t-t'\), and
\begin{equation}
    \Delta t_{IJ}
    \equiv
    \Delta t - x_I + x_J
\end{equation}
is the retarded-time separation.

The monochromatic approximation, which corresponds to taking $\omega_\mathbf{v} \to m_\phi$ in Eq.~\eqref{eq:full_velocity_integral_covariance}, is accurate when the dispersive contribution to the temporal phase satisfies
$(\omega_{\mathbf v}-m_\phi)\Delta t_{IJ} \simeq m_\phi v^2\Delta t_{IJ}/2 \ll 1$.
For $v \sim 10^{-3}$ and typical inter-pulsar distances of $1\, \mathrm{kpc}$, this quantity can become $\mathcal{O}(1)$ or greater for $m_\phi \gtrsim 10^{-21}\,\mathrm{eV}$.
Physically, when the light-travel time between pulsars exceeds the coherence time $\tau = 1/(m_\phi v^2)$, retaining the full velocity dependence in the phase can drive correlations between retarded-time fields toward zero.
However, note that the spatial phase $m_\phi \mathbf{v}\cdot\Delta\mathbf{x}_{IJ}$ is also velocity dependent and is parametrically larger (by $1/v$) than the dispersive contribution to the temporal phase.
As a result, for sufficiently large velocities that allow $(\omega_{\mathbf v}-m_\phi)\Delta t_{IJ}$ to be non-negligible, the larger
$m_\phi \mathbf{v}\cdot\Delta\mathbf{x}_{IJ}$ already causes the integrand in Eq.~\eqref{eq:full_velocity_integral_covariance} to be rapidly oscillatory, driving the covariance toward zero.
Therefore, while replacing $\omega_{\mathbf v}\to m_\phi$ in the cosine argument can induce a large relative error at sufficiently large $m_\phi$, it produces a negligible absolute error, because the covariance is already exponentially suppressed by spatial dephasing (i.e., finite coherence length).

\begin{figure}[t]
    \centering
    \includegraphics[width=\linewidth]{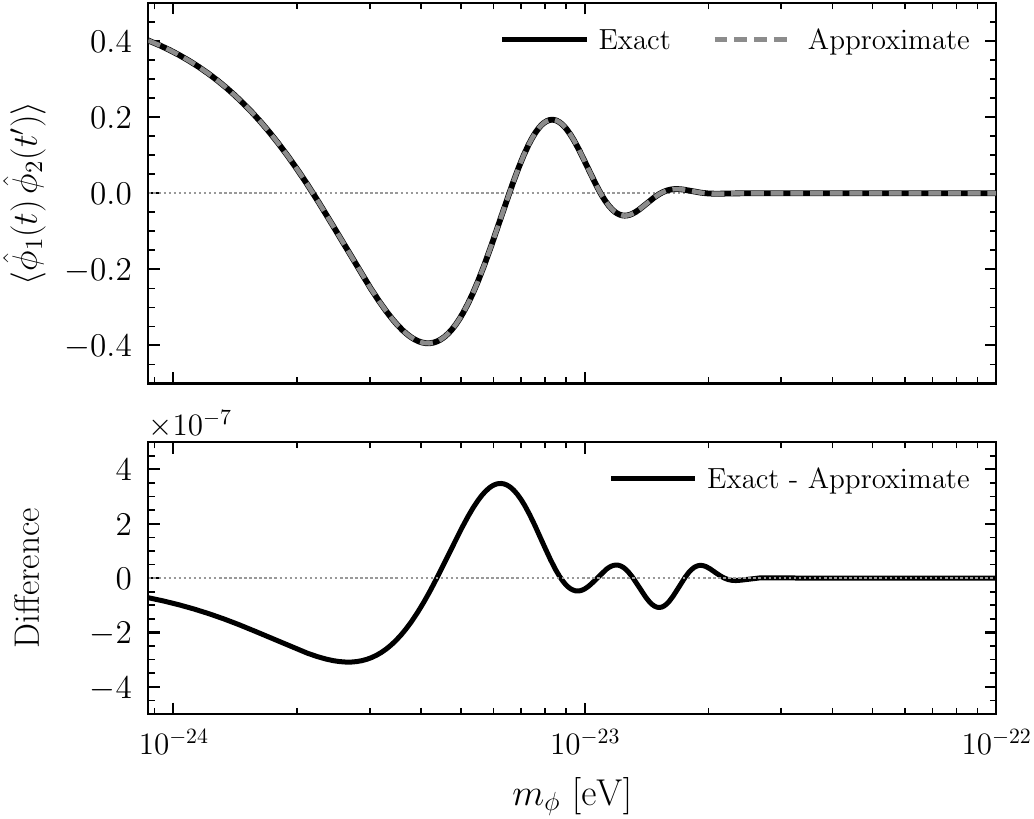}
    \caption{
    Numerical comparison of the full nonrelativistic velocity integral and the monochromatic approximation \(\omega_{\mathbf v}\simeq m_\phi\) for the retarded-time ULDM field covariance. We place two pulsars \(2\,\mathrm{kpc}\) apart, with Earth at the midpoint, and evaluate the covariance between the retarded-time field at the first pulsar at \(t=0\,\mathrm{yr}\) and the retarded-time field at the second pulsar at \(t'=15\,\mathrm{yr}\). \textbf{Top panel:} The dimensionless covariance \(\langle \hat{\phi}_1(t)\hat{\phi}_2(t')\rangle\) as a function of ULDM mass for a Maxwell--Boltzmann velocity distribution with \(v_0=155\,\mathrm{km/s}\). The solid curve shows the full nonrelativistic velocity integral in Eq.~\eqref{eq:full_velocity_integral_covariance}, while the dashed curve shows the monochromatic approximation in Eq.~\eqref{eq:Approximation}. \textbf{Bottom panel:} The signed difference, plotted in the same absolute units as the covariance. Any differences are negligibly small.
    }
    \label{fig:Demonstration}
\end{figure}

Substituting $\omega_\mathbf{v} \to m_\phi$ in Eq.~\eqref{eq:full_velocity_integral_covariance} and assuming the ULDM field in the Galaxy has the isotropic velocity distribution in Eq.~\eqref{eq:Maxwell-Boltzmann}, the two-point correlation function for the dimensionless retarded-time field becomes
\begin{equation}
    \langle \hat{\phi}_I(t) \hat{\phi}_J(t') \rangle
    = \frac{1}{2}\cos\!\left[ m_\phi \Delta t _{IJ}\right] \exp\!\left[-\frac{\Delta\mathbf{x}_{IJ}^2}{2 \ell^2}\right]
    \label{eq:Approximation}
\end{equation}
on all relevant timescales and spatial scales, where $\ell$ is the coherence length from Eq.~\eqref{eq:ell}.
We define the magnitude of the spatial correlation $R_{IJ}$ and an effective distance phase $\psi_I$ as
\begin{align}
    R_{IJ} &\equiv \exp\!\left[-\frac{\Delta\mathbf{x}_{IJ}^2}{2\ell^2}\right]
    \label{eq:corr-amplitude} \\
    \psi_{I}
    &\equiv
    m_\phi x_I
    \label{eq:corr-phase}
    \end{align}
and the two-point function takes the compact form
\begin{equation}
    \langle \hat{\phi}_I(t) \hat{\phi}_J(t') \rangle
    = \frac{1}{2}\cos\!\left[m_\phi\Delta t - \psi_{I} +\psi_{J} \right] R_{IJ} \,,
    \label{eq:compact_correlation}
\end{equation}
where we leave the dependence of $R_{IJ}$ and $\psi_{I}$ on $m_\phi$ and $\mathbf{x}_I$ implicit. We verify the accuracy of this approximation in the sense relevant for our analysis in Fig.~\ref{fig:Demonstration}, which compares Eq.~\eqref{eq:Approximation} with a numerical evaluation of the full nonrelativistic velocity integral in Eq.~\eqref{eq:full_velocity_integral_covariance}. The absolute difference remains small throughout the mass range shown. At larger masses, the relative difference can become large, but only after spatial decoherence has already strongly suppressed the covariance. 

\subsection{Covariance in the frequency domain}
\label{subsec:fourier_domain_covariance}

The correlation function in Eq.~\eqref{eq:compact_correlation} depends only on the time difference $\Delta t = t - t'$ and therefore describes a stationary process, making it natural to work in the frequency domain. We use the Fourier convention
\begin{equation}
    \hat\phi_I(f)
    =
    \int dt\, e^{-2\pi i f t}\hat\phi_I(t).
\end{equation}
In Fourier space, the correlation function takes the general form
\begin{equation}
    \langle \hat{\phi}_I(f) \hat{\phi}_J(f')  \rangle = \delta(f + f') \, \hat P_{IJ}(f) \, .
    \label{eq:FreqDomainCovariance}
\end{equation}
For a strictly monochromatic signal at angular frequency $m_\phi$, the support for \(\hat P_{IJ}(f)\) is confined to $f=\pm m_\phi/(2\pi)$, and an explicit Fourier transform of Eq.~\eqref{eq:compact_correlation} yields
\begin{equation}
    \hat P_{IJ}(f) = \frac{1}{4} \left[ \delta\!\left(f-\frac{m_\phi}{2\pi}\right) \,\hat \Gamma_{IJ}+\delta\!\left(f+\frac{m_\phi}{2\pi}\right)\,\hat \Gamma_{IJ}^{\ast}\right] \,,
\end{equation}
where
\begin{equation}
    \hat\Gamma_{IJ}
    =
    e^{-i(\psi_I-\psi_J)}R_{IJ}
    \label{eq:complex_orf}
\end{equation}
is the complex overlap reduction function (ORF), which is a function of $\{\mathbf{x}_I\}$ through its dependence on $\psi_I$ and $R_{IJ}$.
The complex nature of $\hat\Gamma_{IJ}$ arises from the retarded-time evaluation of the field at the pulsar locations, corresponding to a distance-dependent phase delay. However, the matrix with entries \(\hat\Gamma_{IJ}\) is Hermitian, which
enforces the real-valuedness of the time-domain field.

\subsection{Gaussian amplitude covariance}
\label{subsec:latent_covariance}

Since the ULDM field is approximately monochromatic, we can write the dimensionless retarded-time field as
\begin{equation}
    \hat{\phi}_I(t) = \hat{C}_I \cos(m_\phi t) + \hat{S}_I \sin(m_\phi t) \,,
    \label{eq:monochromatic_field}
\end{equation}
where $\hat{C}_I$ and $\hat{S}_I$ are constant-valued, i.e., time-independent, random variables associated with the $I^\mathrm{th}$ location. We can evaluate these amplitudes as
\begin{equation}
  \begin{aligned}
      \hat{C}_I &= \lim_{T\rightarrow \infty} \frac{1}{T} \int_{-T}^T \hat{\phi}_I(t)
  \cos(m_\phi t)\, dt , \\
      \hat{S}_I &= \lim_{T\rightarrow \infty} \frac{1}{T} \int_{-T}^T \hat{\phi}_I(t)
  \sin(m_\phi t)\, dt .
  \end{aligned}
  \label{eq:projectors}
  \end{equation}
Since they are obtained by linear projections of the Gaussian fields \(\hat\phi_I(t)\), they also obey Gaussian statistics. The covariances of these amplitudes are
\begin{equation}
  \begin{aligned}
      \langle \hat{C}_I \hat{C}_J \rangle &= \langle \hat{S}_I \hat{S}_J \rangle =
  \tfrac{1}{2}\mathrm{Re}(\hat\Gamma_{IJ}), \\
      \langle \hat{C}_I \hat{S}_J \rangle &= - \langle \hat{S}_I \hat{C}_J \rangle =
  \tfrac{1}{2}\mathrm{Im}(\hat\Gamma_{IJ}) .
  \end{aligned}
  \label{eq:FourierCoefficientCovariances}
  \end{equation}
  We provide the full details of this calculation in Appendix~\ref{app:FourierCovariances}.

\section{Deterministic analysis with finite ULDM correlations}
\label{sec:deterministic_analysis}

In this section, we develop the deterministic analysis used in this work, including the finite-correlation latent-field prior needed to describe the ULDM field across the full PTA-sensitive mass range. We first review the background model for the non-ULDM timing residuals in Sec.~\ref{subsec:background_model}. We then construct the PTA likelihood for a fixed latent ULDM field realization in Sec.~\ref{subsec:pta_likelihood_uldm}. In Sec.~\ref{subsec:augmented_latent_prior}, we specify the augmented latent-field prior that incorporates finite ULDM spatial correlations and pulsar-distance uncertainties. We describe the normalizing-flow surrogate for the resulting amplitude prior in Sec.~\ref{subsec:flow_amplitude_prior}. Finally, in Sec.~\ref{subsec:bayesian_analysis_flow_prior}, we summarize how the flow prior is used in the Bayesian analysis.

We note that, for linearly coupled signals, the Gaussian statistics of the ULDM field permit an alternative but mathematically equivalent formulation in which the signal is modeled as a correlated time-domain Gaussian process. In this description, a signal covariance matrix encodes the spatial correlation structure set jointly by the pulsar locations and the ULDM mass, while its overall normalization is controlled by the ULDM coupling strength. Such an analysis is possible, and we describe it in more detail in App.~\ref{app:stochastic_analysis}. In practice, however, we find that it offers limited advantages over the latent-field formulation used in this work. The stochastic formulation requires the same distance-induced nuisance structure, often drives Monte Carlo sampling into numerically ill-conditioned regions of covariance-parameter space, and does not generalize to the nonlinear couplings considered in this work.

\subsection{Background model}
\label{subsec:background_model}

Our background model consists of three components: detector white noise, uncorrelated pulsar red noise, and a Hellings--Downs-correlated red-noise process due to a stochastic gravitational-wave background (SGWB).
Our implementation of each component follows standard PTA analysis conventions, so we review their definitions only briefly.
For more details, see, e.g., Refs.~\cite{Taylor:2021yjx, NANOGrav:2023hvm, NANOGrav:2023icp}.

The white noise is assumed Gaussian and described by a covariance matrix $\mathbf{N}$.
By assumption, the white-noise contributions are uncorrelated between pulsars, endowing $\mathbf{N}$ with a block-diagonal structure.
The white-noise covariance is specified by the standard PTA parameters: EFAC, EQUAD, and ECORR.
For each pulsar, we model the white noise independently for each receiver backend with its own set of $\{\mathrm{EFAC},\mathrm{EQUAD},\mathrm{ECORR}\}$ parameters.
Thus, the white-noise model includes three parameters per backend per pulsar.

The uncorrelated red noise is also taken to be Gaussian and statistically independent between pulsars.
It is, therefore, described by a block-diagonal covariance matrix $\bm\Sigma_{\rm RN}$.
At each pulsar, we model the red-noise power spectral density as a power law with amplitude $A_{\rm RN}$ and spectral index $\gamma_{\rm RN}$.
Unlike the white noise, the red-noise parameters are shared across receiver backends for a given pulsar.
Hence, the number of uncorrelated red-noise parameters in our model is $2N_p$.
We represent this process using a Fourier basis with the first 30 harmonics, $f_k = k/T_{\rm obs}$ for $k\in \{1,\ldots,30\}$, where $T_{\rm obs}$ is the total observing time.

In addition to the uncorrelated processes modeled by $\mathbf{N}$ and $\bm\Sigma_{\rm RN}$, we include a Hellings--Downs correlated red-noise process $\bm\Sigma_{\rm GW}$ which models a possible SGWB. While the Hellings--Downs correlation couples the process across pulsars,
\(\bm\Sigma_{\rm GW}\) is otherwise constructed in analogy to
\(\bm\Sigma_{\rm RN}\).
We model the correlated red noise as a power law, and because it is common across pulsars, it is specified by a single amplitude $A_{\rm GW}$ and spectral index $\gamma_{\rm GW}$.
As with the uncorrelated red noise, we represent this process using 30 Fourier modes with frequencies $f_k = k/T_{\rm obs}$ for $k\in\{1,\ldots,30\}$.

In the interest of compact notational convenience, we denote the total background covariance by
\begin{equation}
\bm{\Sigma}_\mathrm{bkg}(\bm{\theta}_\mathrm{bkg})
=
\mathbf{N}(\bm{\theta}_\mathrm{WN})
+
\bm\Sigma_{\rm RN}(\bm{\theta}_\mathrm{RN})
+
\bm\Sigma_{\rm GW}(\bm{\theta}_\mathrm{GW}) ,
\end{equation}
where $\bm{\theta}_\mathrm{bkg}$ is the total background parameter vector containing
the parameter vectors $\bm{\theta}_\mathrm{WN}$,
$\bm{\theta}_\mathrm{RN}$, and $\bm{\theta}_\mathrm{GW}$ for the white
noise, uncorrelated red noise, and correlated red noise, respectively.
The background model parameters and associated priors used in this analysis are summarized in Tab.~\ref{tab:parameters}.
We denote the joint prior for the background parameter vector $\bm{\theta}_\mathrm{bkg}$ by $\pi_\mathrm{bkg}$. 

Additionally, using standard methods implemented in \texttt{enterprise}, we analytically marginalize over the timing-model parameters for each pulsar. This procedure modifies the likelihood evaluation but does not affect the ULDM signal parametrization or the latent-prior construction \cite{fastshermanmorrison-pulsar}. Because timing-model marginalization is standard in PTA analyses, we do not discuss it further.

\begin{table*}[!ht]{
\renewcommand{\arraystretch}{1.2}
    \ra{1.3}
    \begin{center}
    \tabcolsep=0.25cm
    \begin{tabular}{l l l l}
    \hlinewd{1pt}
     & \textbf{Description} & \textbf{Prior} & \textbf{Comments} \\
    \hlinewd{1pt}
    \multicolumn{4}{c}{\textbf{Uncorrelated Red Noise}} \\
    $A_{\rm RN}$ & red noise power-law amplitude & log-uniform $[-20, -11]$ & one parameter per pulsar \\
    $\gamma_{\rm RN}$ & red noise power-law spectral index & uniform $[0, 7]$ & one parameter per pulsar \\
    \hlinewd{0.5pt}
    \multicolumn{4}{c}{\textbf{Hellings--Downs Correlated Noise}} \\
    $A_{\rm GW}$ & common process strain amplitude & log-uniform $[-18, -11]$ & one parameter for PTA \\
    $\gamma_{\rm GW}$ & common process spectral index & uniform $[0, 7]$ &  one parameter for PTA \\
    \hlinewd{0.5pt}
    \multicolumn{4}{c}{\textbf{ULDM Signal [Fixed Coupling Ratio]}} \\
    $A_\phi$ & overall ULDM signal strength & log-uniform, mass-dependent & one parameter for PTA \\
    $A_I$ & local ULDM field amplitude & normalizing flow & one parameter for Earth and one per pulsar\\
    $\alpha_I$ & local ULDM phase & uniform  $[0, 2\pi)$ & one parameter for Earth and one per pulsar \\
    \hlinewd{0.5pt}
    \multicolumn{4}{c}{\textbf{ULDM Signal [Free Coupling Ratio]}} \\
    $A_E$ & Earth-term signal strength & log-uniform, mass-dependent & one parameter for PTA \\
    $A_P$ & pulsar-term signal strength & log-uniform, mass-dependent & one parameter for PTA \\
    $A_I$ & local ULDM field amplitude & normalizing flow & one parameter for Earth and one per pulsar\\
    $\alpha_I$ & local ULDM phase & uniform  $[0, 2\pi)$ & one parameter for Earth and one per pulsar \\
    \hlinewd{1pt}   
    \end{tabular}
    \end{center}}
\caption{Model parameters, priors, and comments relevant for the analysis. While always log-uniform, the signal amplitude priors are taken to have mass-dependent bounds; see text for details.}
\label{tab:parameters}
\end{table*}

\subsection{PTA likelihood for ULDM signals}
\label{subsec:pta_likelihood_uldm}

We now construct the PTA likelihood, separating the stochastic ULDM field realization from the deterministic map that turns it into timing residuals. The field statistics are fixed by the ULDM covariance, while the map depends on the coupling: it is linear in the field for a linear coupling and quadratic for the quadratic and gravitational signals. Conditioned on a realization, the likelihood has the same form in every case, and only this response map differs.

Let $\mathbf d$ denote the timing residuals of the full PTA. For fixed background parameters $\bm\theta_\mathrm{bkg}$, the non-ULDM contribution is a zero-mean Gaussian process with covariance $\bm\Sigma_\mathrm{bkg}(\bm\theta_\mathrm{bkg})$. The approximately monochromatic ULDM realization is described by an amplitude and phase at each site,
\begin{equation}
    \hat\phi_I(t; A_I,\alpha_I,m_\phi)
    =
    A_I \cos(m_\phi t-\alpha_I),
    \label{eq:amplitude_phase_field}
\end{equation}
where $A_I \ge 0$ and $\alpha_I \in [0,2\pi)$.
We collect the amplitudes and phases at Earth and at the pulsars into the vectors
\begin{equation}
\begin{gathered}
    \bm A
    =
    \begin{bmatrix}
        A_0 & A_1 & \cdots & A_{N_p}
    \end{bmatrix}^{T},
    \\
    \bm\alpha
    =
    \begin{bmatrix}
        \alpha_0 & \alpha_1 & \cdots & \alpha_{N_p}
    \end{bmatrix}^{T}.
\end{gathered}
\label{eq:amplitude_phase_vectors}
\end{equation}
Here $I=0$ denotes Earth and $I=1,\ldots,N_p$ denote the pulsars.

The ULDM contribution to the residuals is the response map $\mathbf f(\bm A,\bm\alpha;\bm\theta_\mathrm{sig})$, with $\bm\theta_\mathrm{sig}$ the signal parameters. For the linear signal, in the fixed coupling-ratio parametrization $\bm\theta_\mathrm{sig}^{\rm fix}=\{A_\phi,m_\phi,w_E,w_P\}$ ($w_E$ and $w_P$ are fixed once a coupling scenario is chosen), the component-wise residual is
\begin{equation}
\begin{split}
    r_I^{\rm fix}(t)
    = A_\phi\big[ &w_E A_0\cos(m_\phi t-\alpha_0)\\
    &+ w_P A_I\cos(m_\phi t-\alpha_I)\big].
\end{split}
    \label{eq:fixed_ratio_residual_amp_phase}
\end{equation}
In the free coupling-ratio parametrization $\bm\theta_\mathrm{sig}^{\rm free}=\{A_E,A_P,m_\phi\}$, the component-wise residual is
\begin{equation}
\begin{split}
    r_I^{\rm free}(t)
    ={}& A_E A_0\cos(m_\phi t-\alpha_0)\\
    &+ A_P A_I\cos(m_\phi t-\alpha_I).
\end{split}
    \label{eq:free_ratio_residual_amp_phase}
\end{equation}
The quadratic and gravitational signals share this structure with the field entering quadratically. In the fixed parametrization the component-wise residual is
\begin{equation}
\begin{split}
    r_I^{\rm fix}(t)
    = A_\phi\big[ &w_E A_0^2\cos(2m_\phi t-2\alpha_0)\\
    &+ w_P A_I^2\cos(2m_\phi t-2\alpha_I)\big].
\end{split}
    \label{eq:quad_fixed_residual_amp_phase}
\end{equation}
In the free parametrization it is
\begin{equation}
\begin{split}
    r_I^{\rm free}(t)
    ={}& A_E A_0^2\cos(2m_\phi t-2\alpha_0)\\
    &+ A_P A_I^2\cos(2m_\phi t-2\alpha_I).
\end{split}
    \label{eq:quad_free_residual_amp_phase}
\end{equation}
The full vector $\mathbf f$ follows by evaluating the appropriate residual at every observing time and pulsar. The latent variables $(\bm A,\bm\alpha)$ are the same in all cases; only $\mathbf f$ changes with the coupling.

The data model is therefore
\begin{equation}
    \mathbf d
    =
    \mathbf n
    +
    \mathbf f
    \!\left(
        \bm A,
        \bm\alpha;
        \bm\theta_\mathrm{sig}
    \right),
    \quad
    \mathbf n
    \sim
    \mathcal N
    \!\left(
        0,
        \bm\Sigma_\mathrm{bkg}(\bm\theta_\mathrm{bkg})
    \right).
\end{equation}
Conditioned on the latent realization, the likelihood is Gaussian in the shifted residual:
\begin{widetext}
\begin{equation}
\begin{split}
    \mathcal L
    (&\mathbf d
    \mid
    \bm\theta_\mathrm{bkg},
    \bm\theta_\mathrm{sig},
    \bm A,
    \bm\alpha)
    =
    \frac{1}
    {|2\pi\bm\Sigma_\mathrm{bkg}|^{1/2}}
    \exp\!\left[
        -\frac12
        \left(
            \mathbf d
            -
            \mathbf f(\bm A,\bm\alpha;\bm\theta_\mathrm{sig})
        \right)^T
        \bm\Sigma_\mathrm{bkg}^{-1}
        \left(
            \mathbf d
            -
            \mathbf f(\bm A,\bm\alpha;\bm\theta_\mathrm{sig})
        \right)
    \right],
\end{split}
\label{eq:conditional_pta_likelihood_amp_phase}
\end{equation}
\end{widetext}
where we suppress the dependence of $\bm\Sigma_\mathrm{bkg}$ on $\bm\theta_\mathrm{bkg}$. It remains to specify the prior over the latent realization $(\bm A,\bm\alpha)$, which we turn to next.

\subsection{Augmented latent-field prior}
\label{subsec:augmented_latent_prior}

The likelihood in Eq.~\eqref{eq:conditional_pta_likelihood_amp_phase} is conditional on the latent ULDM realization, represented by
$(\bm A,\bm\alpha)$.
The prior for these variables is fixed by the Gaussian statistics of the field.
At fixed pulsar distances, the cosine-sine coefficients in Eq.~\eqref{eq:monochromatic_field} are jointly Gaussian, with covariance given by Eq.~\eqref{eq:FourierCoefficientCovariances}.
Transforming to the amplitude-phase variables defined in Eqs.~\eqref{eq:amplitude_phase_field} and \eqref{eq:amplitude_phase_vectors} induces the prior
$\pi(\bm A,\bm\alpha\mid \{x_I\}, m_\phi)$.

The complication is that the pulsar distances are not known exactly.
For a fixed set of distances, the same distances determine both the correlation magnitudes $R_{IJ}$ and the effective distance phases $\psi_I$.
The exact distance-marginalized latent prior is therefore
\begin{equation}
    \pi_{\rm exact}(\bm A,\bm\alpha\mid m_\phi)
    =
    \int
    \prod_{I=1}^{N_p}
    \mathrm{d} x_I\,
    \pi_I(x_I)\,
    \pi(\bm A,\bm\alpha\mid \{x_I\},m_\phi).
    \label{eq:exact_distance_marginalized_prior}
\end{equation}
This expression is straightforward to define, but difficult to use directly because changing the distances changes both the spatial-correlation envelope and the retarded-time phases.

These two distance dependences have very different practical consequences.
The correlation magnitudes $R_{IJ}$ vary on the coherence length scale $\ell$ and determine whether the local ULDM amplitudes at two PTA sites are correlated.
The phases $\psi_I$, by contrast, vary on the much shorter Compton wavelength scale and determine the phase of the oscillating field at the retarded pulsar time. At the lowest mass considered in this work,
\(m_\phi=10^{-24}\,\mathrm{eV}\), the reduced Compton wavelength
\(m_\phi^{-1}\) is \(6.4\,\mathrm{pc}\).
Existing pulsar-distance uncertainties are typically larger than this scale, so the retarded-time phases are effectively unconstrained even when the same distance measurements still contain useful information about the correlation magnitudes.
For larger masses, the phase scale becomes still shorter, while the distance uncertainties can also become comparable to the coherence length, so that the expected correlation magnitudes may vary appreciably across the distance posterior.

If one samples directly in the pulsar distances, this creates a multi-scale inference problem. Changing a distance by \(\Delta x_I\sim 2\pi m_\phi^{-1}\) shifts the pulsar-term phase by \(\Delta\psi_I\sim 2\pi\), while often leaving the correlation envelope nearly unchanged. The likelihood can therefore develop many nearly degenerate modes in each distance direction, separated by Compton-scale intervals.

At the same time, the sampler must still track coherence-length variations in $R_{IJ}$ whenever the correlation envelope changes over the distance posterior. This motivates an augmented latent-field prior that separates the slowly varying correlation magnitudes from the rapidly varying retarded-time phases. This is the same phase-augmentation logic underlying the limiting amplitude-phase models used in previous searches; the difference in this work is that we do not collapse the amplitude prior to either the independent-Rayleigh (fully uncorrelated) or common-Rayleigh (fully correlated) limit.

In the augmented model, we keep the distance dependence of the correlation magnitudes but treat the effective pulsar phases as independent nuisance variables.
A draw from the augmented latent prior is defined by the following procedure:
\begin{enumerate}
    \item Draw pulsar distances $\{x_I\}$ from their distance priors $\pi_I(x_I)$.
    \item Use these distances to compute the correlation magnitudes $R_{IJ}$.
    \item Draw effective pulsar phases $\psi_I$ independently and uniformly on $[0,2\pi)$ for $I\ge 1$, with $\psi_0=0$ at Earth.
    \item Construct the covariance matrix for the cosine-sine coefficients using Eq.~\eqref{eq:FourierCoefficientCovariances}.
    \item Draw the cosine-sine coefficients from this Gaussian covariance and transform them to $(\bm A,\bm\alpha)$.
\end{enumerate}
This procedure preserves the distance-dependent spatial correlations while avoiding the need to resolve Compton-scale phase windings when exploring pulsar distances.

The phase augmentation has an important simplifying consequence.
After marginalizing over the independent nuisance phases $\{\psi_I\}$, the physical phases $\{\alpha_I\}$ are uniformly distributed on $[0,2\pi)$ and are independent of the amplitudes.
As shown in Appendix~\ref{app:amplitude_phase_statistics}, the augmented latent prior factorizes as
\begin{equation}
    \pi(\bm A,\bm\alpha\mid m_\phi)
    =
    \pi(\bm A\mid m_\phi)
    \prod_{I=0}^{N_p}
    \frac{1}{2\pi}.
    \label{eq:latent_prior_factorized}
\end{equation}
All nontrivial information about ULDM spatial coherence is, therefore, contained in the joint amplitude prior $\pi(\bm A\mid m_\phi)$.

The same prior applies to the quadratic and gravitational signals. Their residuals carry the phase through $2\alpha_I$; see Eqs.~\eqref{eq:quad_fixed_residual_amp_phase} and~\eqref{eq:quad_free_residual_amp_phase}. Because each $\alpha_I$ is uniform on $[0,2\pi)$, so is $2\alpha_I$, and relabeling $2\alpha_I\to\alpha_I$ leaves the phase prior unchanged. The same invariance implies that the analysis depends on the coupling weights only through their magnitudes $|w_E|$ and $|w_P|$: a sign flip of $w_E$ or $w_P$ amounts to a constant shift of the corresponding phase $\alpha_0$ or $\alpha_I$ by $\pi/2$, which the uniform prior absorbs. In particular, the equal-weight case $w_E=w_P=1$ and the gravitational case $w_E=-w_P=1$ yield identical latent-marginalized likelihoods. The latent-amplitude prior $\pi(\bm A\mid m_\phi)$ is thus identical for the linear and quadratic signals, and only the deterministic response map differs.

The likelihood marginalized over the latent ULDM realization is formally
\begin{equation}
\begin{split}
    \mathcal L_\mathrm{marg}
    (&\mathbf d
    \mid
    \bm\theta_\mathrm{bkg},
    \bm\theta_\mathrm{sig})
    =
    \int d\bm A\,d\bm\alpha\,
    \mathcal L
    (
        \mathbf d
        \mid
        \bm\theta_\mathrm{bkg},
        \bm\theta_\mathrm{sig},
        \bm A,
        \bm\alpha
    )
    \\
    &\times
    \pi(\bm A,\bm\alpha\mid m_\phi).
\end{split}
\label{eq:latent_marginal_likelihood}
\end{equation}
This expression is useful conceptually, but we do not evaluate the latent integral explicitly. Instead, we sample the latent amplitudes and phases together with the signal and background parameters.
For each run, we fix \(m_\phi\) and choose either the fixed or free coupling-ratio parametrization.  The sampled posterior is proportional to
\begin{equation}
\begin{split}
    p(&\bm\theta_\mathrm{amp},\bm\theta_\mathrm{bkg},\bm A,\bm\alpha
    \mid
    \mathbf d,m_\phi)
    \propto
    \mathcal L
    (
        \mathbf d
        \mid
        \bm\theta_\mathrm{bkg},
        \bm\theta_\mathrm{sig},
        \bm A,
        \bm\alpha
    )
    \\
    &\times
    \pi(\bm A\mid m_\phi)
    \prod_{I=0}^{N_p}
    \frac{1}{2\pi}
    \,
    \pi(\bm\theta_\mathrm{amp})
    \,
    \pi_\mathrm{bkg}(\bm\theta_\mathrm{bkg}) .
\end{split}
\label{eq:sampled_posterior_deterministic}
\end{equation}
Here
\begin{equation}
    \bm\theta_\mathrm{amp}
    =
    \begin{cases}
        \{A_\phi\}, & \text{fixed coupling ratio},\\
        \{A_E,A_P\}, & \text{free coupling ratio},
    \end{cases}
\end{equation}
with \(m_\phi\) fixed in each run.  In the fixed coupling-ratio case, \(w_E\) and \(w_P\) are also fixed by the chosen coupling scenario; therefore, the posterior in Eq.~\eqref{eq:sampled_posterior_deterministic} depends on the signal parameters \(\bm\theta_\mathrm{sig}\) only through the sampled amplitudes \(\bm\theta_\mathrm{amp}\).

Thus, the analysis is deterministic at fixed latent field realization, while the stochastic ULDM field statistics enter through the finite-correlation prior over the latent amplitudes and phases.
The remaining technical task is to represent the correlated amplitude prior $\pi(\bm A\mid m_\phi)$ accurately enough for Bayesian inference.

\subsection{Normalizing flow surrogates for latent field amplitudes}
\label{subsec:flow_amplitude_prior}

\begin{figure*}[!htb]
    \centering
    \includegraphics[width=.49\linewidth]{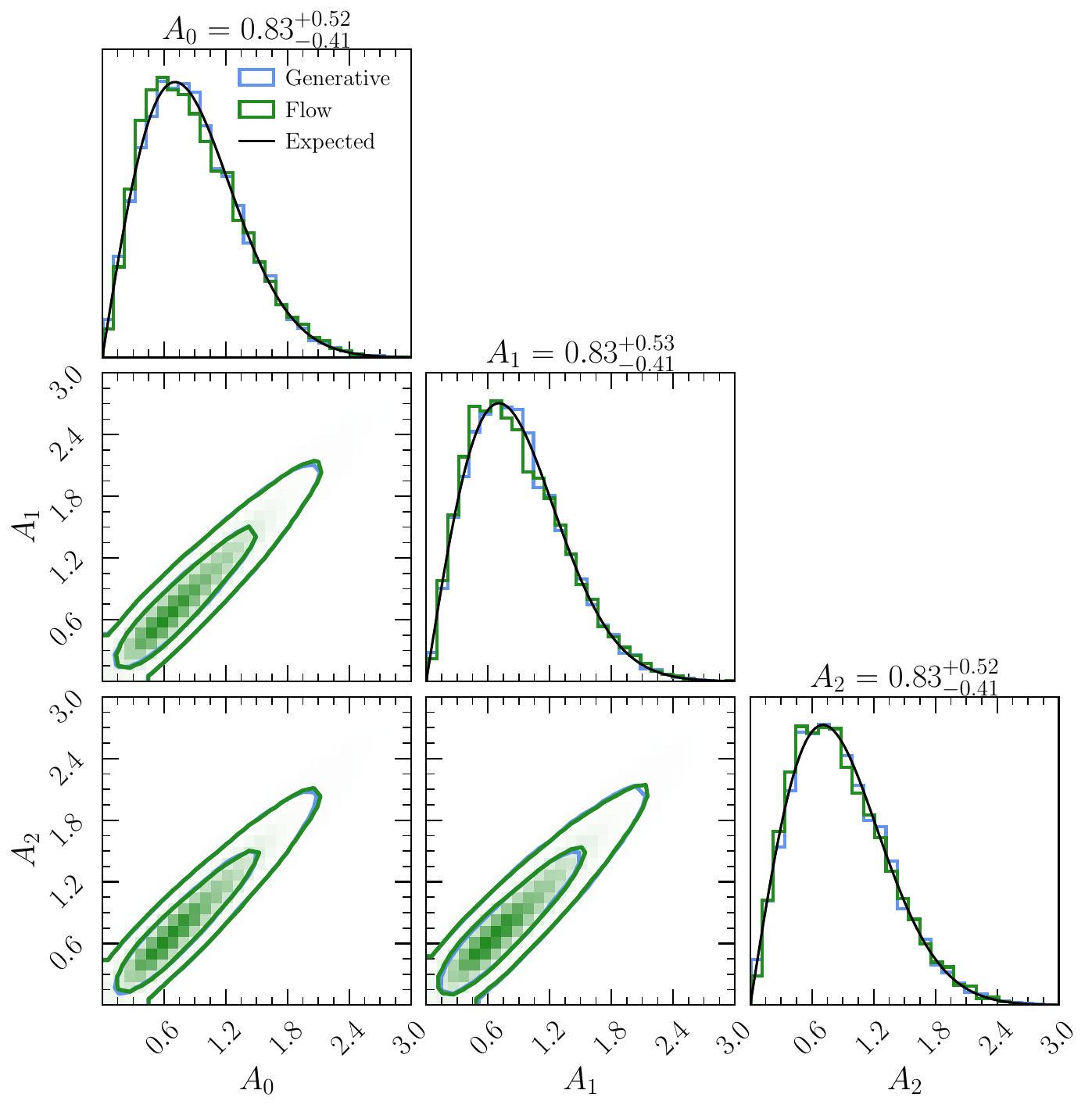}
    \includegraphics[width=.49\linewidth]{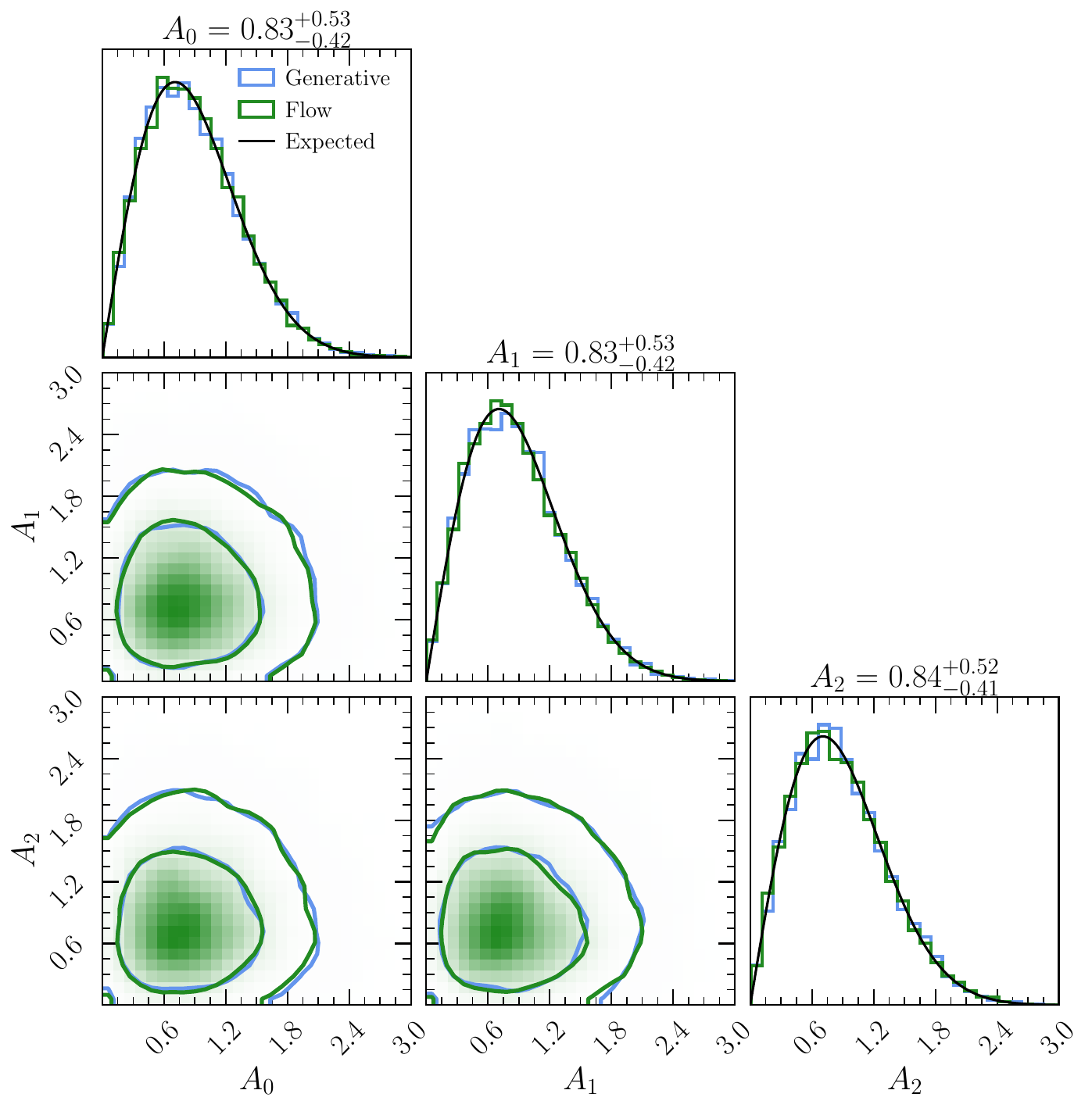}
    \caption{
    Validation of the normalizing-flow surrogate for the amplitude prior.
    We compare samples from the augmented generative model used for training with samples drawn from the trained flow, showing the first three amplitudes in the latent vector $\bm A$.
    \textbf{Left panel}: Results for $m_\phi = 10^{-24}\,\mathrm{eV}$, where the coherence length is large compared to the inter-pulsar separations and the amplitudes are visibly correlated.
    \textbf{Right panel}: Results for $m_\phi = 10^{-20}\,\mathrm{eV}$, where the coherence length is shorter and the amplitude correlations are suppressed.
    In each panel, blue contours show $10,000$ samples from the augmented generative model, green contours show $10,000$ samples from the normalizing flow, and black curves show the Rayleigh distribution expected for the one-dimensional amplitude marginals.
    The agreement demonstrates that the flow captures both the fixed Rayleigh marginals and the mass-dependent amplitude correlations.
    }
    \label{fig:NF_Examples}
\end{figure*}

The augmented latent-field prior reduces the nontrivial part of the latent distribution to the joint amplitude prior $\pi(\bm A\mid m_\phi)$. We approximate this distribution with a normalizing flow trained separately at each ULDM mass. A normalizing flow is a machine-learning model that represents a complicated probability distribution as an invertible, differentiable transformation of a simple base distribution (here a multivariate Gaussian). Because the transformation is invertible and has a tractable Jacobian, a trained flow can both generate samples and return the exact probability density at any point in the phase space; ``training'' the flow means tuning the parameters of this transformation so that its density matches a target distribution, given a set of samples drawn from that target. We refer to the trained flow as a ``surrogate'' because it serves as a fast, reusable stand-in for a distribution that would otherwise be expensive to evaluate. Normalizing flows have previously been applied to PTA analyses to accelerate posterior inference for the SGWB~\cite{Shih:2023jme}; here we instead use a flow
  to represent the latent ULDM amplitude prior $\pi(\bm A\mid m_\phi)$. The training data are generated by repeatedly drawing latent-field realizations from the augmented prior
  described in Sec.~\ref{subsec:augmented_latent_prior} and retaining only the amplitudes $\bm A$. Figure~\ref{fig:NF_Examples} illustrates the performance of the flow surrogate relative to the augmented generative model used to produce the training data. We show results at two representative masses, one in the highly correlated regime and one in the
  nearly uncorrelated regime. In both cases, the flow accurately reproduces the one-dimensional Rayleigh marginals and the leading correlations among the amplitudes.

In principle, one could model $\pi(\bm A\mid m_\phi)$ with a single flow conditioned on $m_\phi$.
In practice, we find that this performs poorly.
Changing $m_\phi$ changes the ULDM coherence length and therefore changes the structure of the correlation-envelope matrix $R_{IJ}$.
The target distribution must then interpolate between qualitatively different regimes, including nearly low-rank behavior in the highly correlated limit and nearly independent Rayleigh amplitudes in the uncorrelated limit.
We therefore train a separate flow for each mass value used in the analysis.

This mass-by-mass strategy is more stable and gives better likelihood fidelity in our tests.
Our flow models are implemented as neural spline flows~\cite{durkan2019neuralsplineflows} using the \texttt{zuko} package~\cite{rozet2022zuko}. A neural spline flow is a particular normalizing-flow architecture in which the invertible transformation is built from monotonic spline functions whose shapes are set by the outputs of a neural network. The splines make the map flexible enough to capture the correlated, non-Gaussian amplitude distribution while remaining easy to invert. We denote the resulting surrogate by $\pi_{\rm flow}(\bm A\mid m_\phi)$ and use it in place of $\pi(\bm A\mid m_\phi)$ in the sampled posterior.
Appendix~\ref{app:nsf} gives additional details on the training set generation, model architecture, and validation tests.

\subsection{Bayesian analysis with the flow prior}
\label{subsec:bayesian_analysis_flow_prior}

The finite-correlation deterministic analysis used in our inference is obtained by replacing the amplitude prior in Eq.~\eqref{eq:sampled_posterior_deterministic} with the trained surrogate \(\pi_{\rm flow}(\bm A\mid m_\phi)\).  For each analysis run, we fix the ULDM mass \(m_\phi\) and choose either the fixed or free coupling-ratio parametrization.  In the fixed coupling-ratio case, we also fix \(w_E\) and \(w_P\) and sample the overall amplitude \(A_\phi\).  In the free coupling-ratio case, we instead sample the two residual-level amplitudes \(A_E\) and \(A_P\).  In both cases, we sample the latent field realization and the background parameters \(\bm\theta_\mathrm{bkg}\) using the likelihood in Eq.~\eqref{eq:conditional_pta_likelihood_amp_phase}.  Marginalizing over the sampled nuisance and latent parameters gives the posterior for the relevant signal-amplitude parameters at fixed \(m_\phi\), from which we infer upper limits or characterize signal detections.

In practice, we do not sample the physical amplitudes $\bm A$ directly. Instead, we sample in the base space of the normalizing flow and use the flow as a forward transformation into amplitude space. Writing the base variables as $\bm z$, we take
\begin{equation}
    \bm A
    =
    T_{\rm flow}(\bm z;m_\phi),
    \label{eq:flow_forward_transform}
\end{equation}
where $T_{\rm flow}$ is the trained flow map at fixed ULDM mass. We find this parametrization more stable than sampling directly in amplitude space and evaluating the flow density $\pi_{\rm flow}(\bm A\mid m_\phi)$. The remaining latent variables are the phases $\bm\alpha$, which are sampled with their periodic domains specified explicitly.

The normalizing flow is not required in principle. One could instead sample the pulsar distances $\{x_I\}$ and effective phases $\{\psi_I\}$ explicitly, construct the conditional Gaussian prior for the cosine-sine coefficients at each point in the chain, and marginalize over these additional nuisance parameters directly. The flow performs this marginalization up front: the distance and phase nuisance variables are integrated out during training, leaving a reusable prior for the latent amplitudes. This reduces the dimensionality of the Bayesian analysis by $2N_p$ nuisance parameters while preserving the distance-dependent amplitude correlations relevant for the ULDM signal.

Finally, this construction connects smoothly to the limiting amplitude priors commonly used in previous ULDM searches.  As shown in Appendix~\ref{app:amplitude_phase_statistics}, the augmented prior reduces to independent Rayleigh amplitudes in the fully uncorrelated limit \(\ell\to 0\), and to a single shared Rayleigh amplitude in the fully correlated limit \(\ell\to\infty\).  The present analysis therefore keeps the familiar augmented amplitude-phase parametrization while extending it to the intermediate regime, where finite ULDM spatial correlations produce a nontrivial joint amplitude prior.

\section{Projected sensitivities and mock analyses}
\label{sec:mock_data_analysis}

In this section, we apply the finite-correlation deterministic analysis developed in Sec.~\ref{sec:deterministic_analysis} to synthetic PTA datasets. These mock analyses serve two purposes. First, they validate the latent-prior construction in controlled settings where the injected ULDM signal, if present, is known. Second, they illustrate the projected sensitivity of PTA data analyses that retain finite ULDM spatial correlations rather than imposing the fully correlated or fully uncorrelated limits.

We first summarize the mock data generation procedure in Sec.~\ref{sec:mock_data_generation}. We describe our Bayesian inference procedure in Sec.~\ref{sec:bayesian_procedure}. The mock-data studies then proceed in three stages.  First, in Sec.~\ref{sec:NullData}, we analyze null datasets to characterize the expected sensitivity of the finite-correlation analysis in the absence of an injected ULDM signal.  Second, in Sec.~\ref{sec:UnblindedTests}, we analyze unblinded signal injections to test signal recovery when the injected mass and coupling structure are known.  These first two studies use the fixed coupling-ratio parametrization, which allows us to examine representative limiting response scenarios with specified Earth-term and  pulsar-term weights.  Finally, in Sec.~\ref{sec:BlindedTests}, we consider blinded injection tests in which the signal masses are not specified in advance. For these blinded tests, we use the free coupling-ratio parametrization, allowing the Earth-term and pulsar-term amplitudes to vary independently, as would be appropriate in a search where the relative response of a possible signal is not known \textit{a priori}. Together, these tests demonstrate both the fidelity of the latent-prior construction and the practical power of the full analysis pipeline across null, unblinded, and blinded analysis settings. Within each study we treat the linearly coupled signal first and then the quadratically coupled and gravitational signals, which share the same pipeline with only the response map squared.

\subsection{Mock data generation}
\label{sec:mock_data_generation}

The mock data are based on pulsars from the NANOGrav 15-year dataset~\cite{NANOGrav:2023hde}. To create a more manageable dataset, we chose the 30 best-timed pulsars. For our purposes, the ability to detect a sinusoidal signal with frequency $f>1/T_{\rm obs}$ in the timing residuals scales with the root-mean-square noise, $\sigma_{\rm rms}$, and the inverse root of the effective number of independent samples, $N=1/(\Delta f\, \Delta t) = T_{\rm obs}/\Delta t$, where $\Delta f = 1/T_{\rm obs}$ is the frequency resolution, $\Delta t$ is the cadence, which is roughly a few weeks for each pulsar, and $T_{\rm obs}$ is the length of time the pulsar has been observed. Given this scaling, and the fact that the cadence is similar for all pulsars, we keep the 30 pulsars with the lowest $\sigma_{\rm rms}/\sqrt{T_{\rm obs}}$ and $T_{\rm obs}\geq 6.5$ years.

We generate the mock data using a modified version of the Python package \texttt{PTA Replicator} to implement the timing residual given in Eq.~\eqref{eq:free_ratio_residual}.\footnote{This package can be found at \url{https://github.com/bencebecsy/pta_replicator.git}. We also use the deterministic timing package \texttt{PINT}, which can be found at \url{https://github.com/nanograv/PINT.git}.} The spatial correlation, given in Eq.~\eqref{eq:Approximation}, requires estimates of the distances between Earth and the pulsars. We use a combination of distance estimates from parallax and dispersion measure, as recorded in Table~\ref{tab:distances}.

The times of arrival and deterministic parameters for each of these pulsars are then used to generate a mock dataset. The intrinsic white- and red-noise parameters correspond to maximum-likelihood values taken from individual pulsar analyses. In addition, each mock dataset includes an SGWB with $A_{\rm GW}= 6.4 \times 10^{-15}$ and $\gamma_{\rm GW} = 3.2$. These values correspond to the posterior means for these parameters when an isotropic SGWB is fit to the NANOGrav 15-year data \cite{NANOGrav:2023gor}.

\subsection{Bayesian sampling and limit-setting}
\label{sec:bayesian_procedure}

In standard PTA analyses, the EFAC, EQUAD, and ECORR parameters are inferred in single-pulsar analyses before performing a full-array analysis. We emulate this procedure for our mock datasets. For each pulsar, we first infer the white-noise parameters independently, and then fix those values in the multi-pulsar ULDM analyses described below. This allows the full-array analyses to focus on the red-noise, correlated-background, and ULDM signal parameters.

For the fixed coupling-ratio analyses, each coupling scenario fixes \(w_E\) and \(w_P\).  We then perform Bayesian inference on a grid of fixed ULDM masses \(m_\phi\), obtaining the marginal posterior for the overall signal amplitude \(A_\phi\) at each mass.  This is the parametrization used for the null analyses and unblinded signal-injection tests.  For the blinded injection tests, we instead use the free coupling-ratio parametrization, sampling \(A_E\) and \(A_P\) at fixed \(m_\phi\).  In either parametrization, the latent-field prior \(\pi_{\rm flow}(\bm A\mid m_\phi)\) is the same; only the deterministic response map from the latent field to timing residuals changes.

This mass-by-mass strategy differs from previous PTA analyses, which have typically sampled jointly in amplitude and mass for a fixed coupling scenario and then extracted amplitude posteriors conditioned on \(m_\phi\). As discussed in Appendix~\ref{app:mass_resolution}, obtaining a converged joint analysis over both amplitude and mass is computationally expensive, especially when the likelihood contains many latent field parameters. Working at fixed mass therefore provides a more controlled and computationally efficient way to obtain limits as a function of \(m_\phi\). The full set of nuisance and signal parameters used in the analysis is summarized in Table~\ref{tab:parameters}. Unless otherwise stated, the bounds reported in this paper correspond to 95\% upper credible limits on the relevant marginalized signal-amplitude posterior at fixed \(m_\phi\). For fixed coupling-ratio analyses, this is the posterior for \(A_\phi\); for free coupling-ratio analyses, we report limits in the two-dimensional \((A_E,A_P)\) amplitude space, or on one of these amplitudes after marginalizing over the other. This follows the limit-setting convention used in previous PTA searches for linearly coupled ULDM~\cite{Kaplan:2022lmz,NANOGrav:2023hvm}.

For the signal-amplitude parameters, we use log-uniform priors with mass-dependent bounds.  Denoting by \(A\) any of the residual-level amplitudes \(A_\phi\), \(A_E\), or \(A_P\), we take
\begin{equation}
\begin{gathered}
    A_{\rm max}(m_\phi)
    =
    \left[
    10^{-5}
    +
    10^{-1}
    \left(
        \frac{m_\phi}{10^{-24}\,\mathrm{eV}}
    \right)^{-3}
    \right]\mathrm{s},
    \\
    A_{\rm min}(m_\phi)
    =
    10^{-5}\,
    A_{\rm max}(m_\phi).
\end{gathered}
\end{equation}
For the free coupling-ratio parametrization, the same mass-dependent prior is applied independently to \(A_E\) and \(A_P\). These bounds were chosen to span the range from undetectable to detectable signals at each mass in the mock datasets considered here, while avoiding an unnecessarily large amplitude range with substantial unused prior volume. The resulting prior is sufficient for the limit-setting procedure used in this work, but it should not be interpreted as a universal prescription: analyses with different datasets, noise levels, observing baselines, or signal parametrizations may require different amplitude-prior bounds.

We perform the Bayesian sampling with \texttt{pocomc}~\cite{karamanis2022pocomc}.
This sampler is well suited to the present analysis because our latent-field formulation introduces one phase latent and one amplitude latent for Earth and for each pulsar, increasing the dimensionality relative to typical PTA analyses. The \texttt{pocomc} algorithm evolves an ensemble of particles using $t$-preconditioned Crank--Nicolson proposals together with a finite-temperature ladder, allowing it to explore broad and correlated posterior structure. Its ensemble-based structure is also naturally parallelizable, making it advantageous compared to the standard \texttt{PTMCMC} workflow for the higher-dimensional analyses considered here.

\subsection{Null data tests}
\label{sec:NullData}

\begin{figure*}[!htb]
    \centering
    \includegraphics[width=\linewidth]{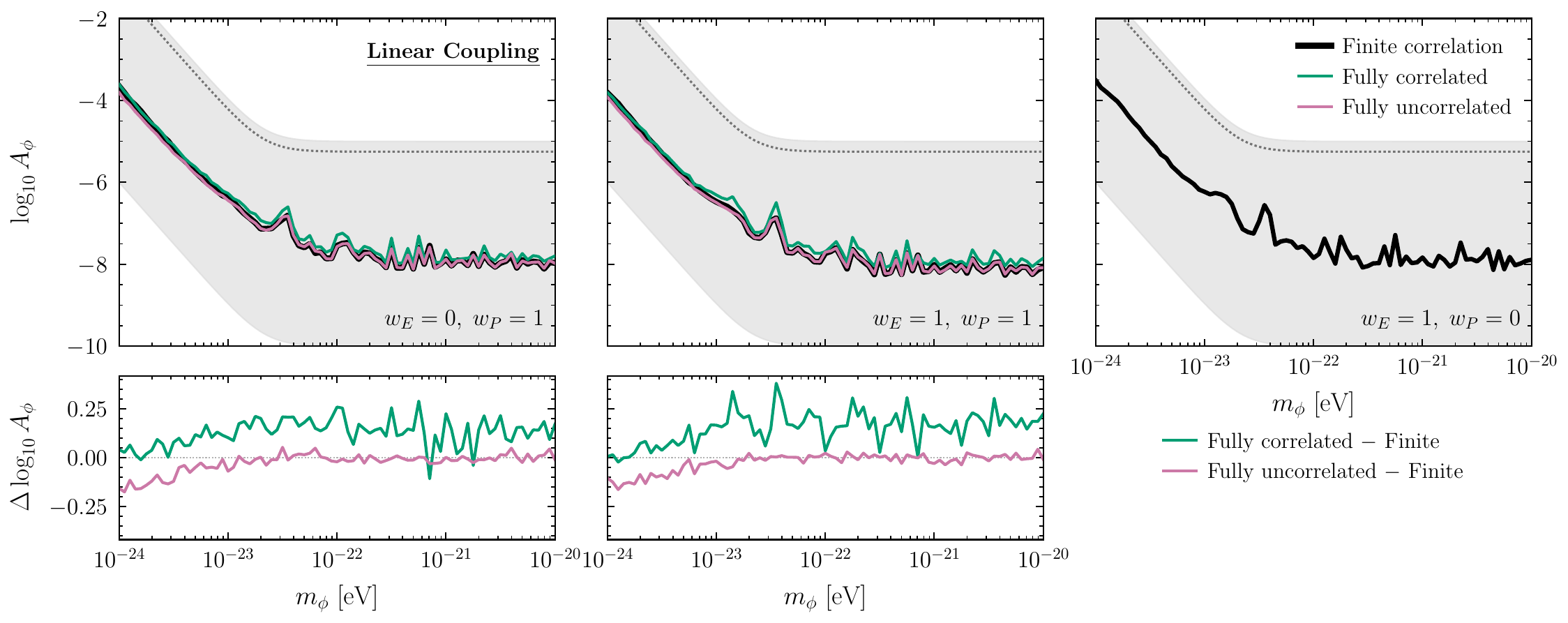}
    \caption{
    Projected sensitivity to linearly coupled ULDM in a synthetic null dataset with no injected ULDM signal.
    The curves show the 95\% upper credible limit on the fixed-ratio signal amplitude \(A_\phi\) as a function of the ULDM mass \(m_\phi\).
    The three columns correspond to representative fixed coupling-ratio scenarios: pulsar-term only, \(w_E=0,\;w_P=1\); equal Earth and pulsar response, \(w_E=1,\;w_P=1\); and Earth-term only, \(w_E=1,\;w_P=0\).
    The gray band shows the support of the mass-dependent prior on \(A_\phi\) used in this analysis, while the dotted gray curve shows the corresponding prior-only 95th percentile.
    The posterior upper limits lie within the prior support but remain distinct from the prior-only percentile, indicating that the projected limits are not dominated by the prior cutoff.
    In the first two columns, we compare the finite-correlation analysis developed in this work with the two limiting amplitude priors commonly used in previous searches: the fully correlated and fully uncorrelated limits.
    These limiting analyses are recovered only in their appropriate asymptotic regimes; away from those regimes, they impose an incorrect prior on the latent ULDM field amplitudes.
    The lower panels show the difference between each limiting analysis and the finite-correlation result.
    At masses above \(10^{-23}\,\mathrm{eV}\), the fully correlated prior gives inferred upper limits that are systematically shifted relative to the finite-correlation result by up to about a factor of \(2\).
    Likewise, at masses below \(10^{-23}\,\mathrm{eV}\), the fully uncorrelated prior gives inferred upper limits that are shifted by up to about a factor of \(2\).
    For the Earth-term-only case, the signal depends only on the ULDM field at Earth, so the pulsar spatial-correlation structure does not enter; we include it for completeness.
    }
    \label{fig:linear_null}
\end{figure*}   

We first apply the analysis to synthetic datasets with no injected ULDM signal. The data are generated following the procedure described in Sec.~\ref{sec:mock_data_generation}, including white noise, pulsar-intrinsic red noise, and a Hellings--Downs-correlated SGWB, but with $A_\phi=0$. These null tests quantify the expected sensitivity of the analysis: at each fixed ULDM mass, we use the marginalized posterior for $A_\phi$ to derive a 95\% upper credible limit. Because these limits are derived from a single background-only realization, they are subject to stochastic fluctuations associated with the particular background realization. We do not quantify this realization-to-realization scatter here, since repeating the full analysis over many independent mock datasets and across the full mass range would be computationally prohibitive. The resulting curves should therefore be interpreted as representative expected sensitivities rather than ensemble-averaged limits.

For each coupling scenario, we analyze 81 masses geometrically spaced between $10^{-24}\,\mathrm{eV}$ and $10^{-20}\,\mathrm{eV}$. As discussed in Appendix~\ref{app:mass_resolution}, this spacing is not sufficient to resolve every possible narrow feature in the mass-dependent likelihood. Our goal in this section is instead to characterize the broad expected sensitivity of the analysis across the PTA-accessible ULDM mass range and to compare the finite-correlation treatment with the limiting-case approximations used in previous studies. A higher-resolution mass scan is deferred to the blinded-injection analysis in Sec.~\ref{sec:BlindedTests}, where resolving the signal mass is essential.

For each mass in the analysis grid, we sample the full posterior described in Sec.~\ref{sec:bayesian_procedure}. The sampled parameters include the ULDM signal amplitude, the latent field amplitudes and phases, the SGWB amplitude and spectral index, and the pulsar-intrinsic red-noise parameters. The reported limits are obtained from the marginalized posterior for $A_\phi$ after marginalizing over all other parameters included in the analysis.

We first present our results for a search for linearly coupled ULDM in our null dataset in Fig.~\ref{fig:linear_null}, which should be interpreted as a test of the latent-field prior, rather than as a comparison of conservative and aggressive sensitivity estimates. At low masses, the ULDM coherence length is large compared to the relevant PTA baselines, and the finite-correlation analysis reduces to the fully correlated limit. At high masses, the pulsar terms decohere, and the same analysis reduces to the fully uncorrelated limit for the pulsar contribution. The important regime is the transition between these limits, which lies within the PTA-sensitive mass range. There, the limiting analyses impose the wrong amplitude prior, and the resulting upper limits on \(A_\phi\) can shift by \(\mathcal O(0.1\text{--}0.3)\) dex relative to the finite-correlation result. This shift is therefore a prior-induced bias from applying an asymptotic model outside its regime of validity, not a property of the data themselves. The Earth-term-only case is included as a useful cross-check: because it depends only on the field at Earth, it is insensitive to the pulsar spatial-correlation structure.

Figure~\ref{fig:linear_null} also includes two prior diagnostics: the gray band shows the support of the mass-dependent prior on \(A_\phi\), and the dotted gray curve shows the 95th percentile of that prior at each mass.  These diagnostics help assess whether the reported upper limits are controlled by the likelihood or by the imposed prior range.  At low masses, the prior upper bound scales approximately as \(A_{\rm max}\propto m_\phi^{-3}\), matching the empirical mass dependence of the expected upper limits found in preliminary null runs.  This scaling keeps the prior broad enough to contain the relevant posterior support across the mass range shown, without introducing a much larger unused amplitude range at low masses.  The separation between the posterior upper-limit curves and the prior-only 95th percentile provides a direct visual check that the reported sensitivities are not simply set by the imposed prior cutoff.

\begin{figure*}[!t]
    \centering
    \includegraphics[width=\linewidth]{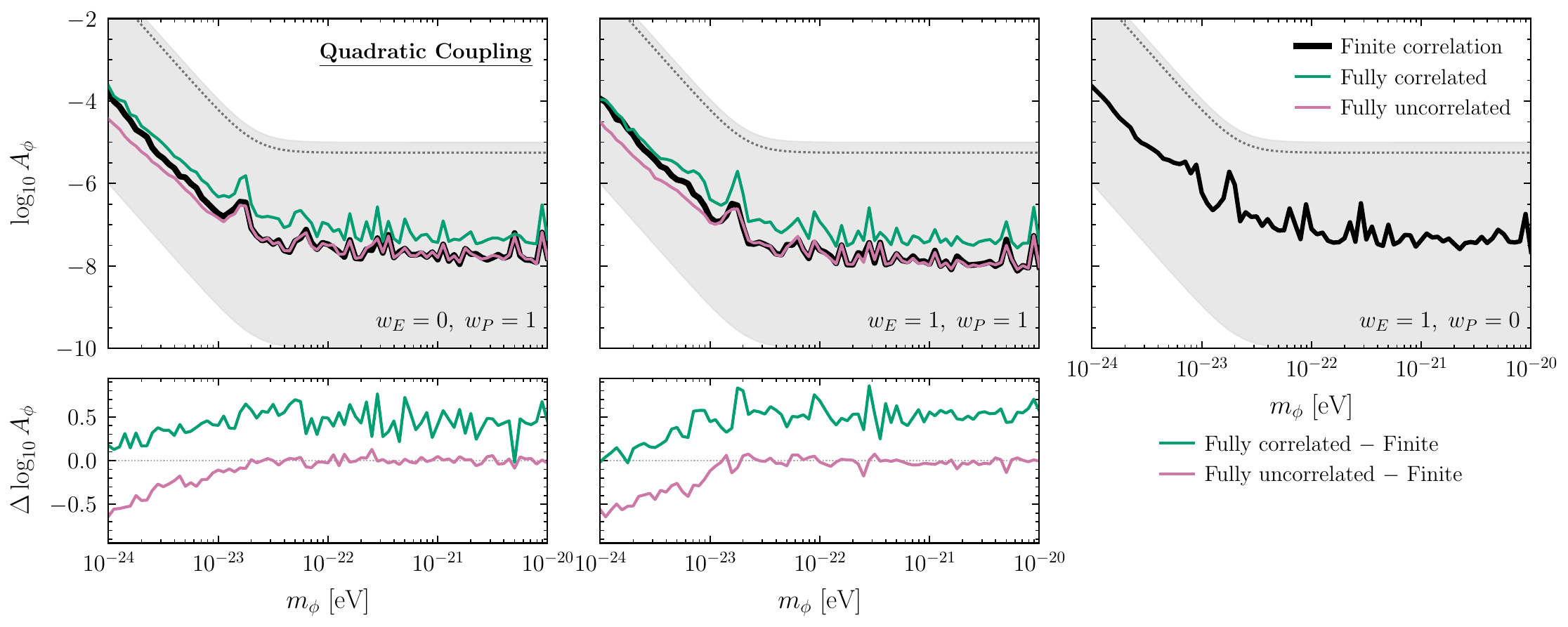}
    \caption{As in Fig.~\ref{fig:linear_null}, but for the quadratically coupled ULDM signal. The quadratic signal analysis exhibits greater dependence on the ULDM correlation, leading to differences between the limiting-case analyses and the new finite-correlation analysis approaching an order of magnitude.}
    \label{fig:quadratic_null}
\end{figure*}

We then repeat the null analysis for the quadratically coupled signals, which contain the gravitational signal as a special case. The pipeline is unchanged from the linear case, using the same mock datasets, sampler, and latent-amplitude prior $\pi_{\rm flow}(\bm A\mid m_\phi)$, with only the linear response map replaced by its quadratic counterpart, see Eq.~\eqref{eq:quadratic_residual_common}; because the squared field oscillates at $2m_\phi$, a trial mass is probed through the Fourier content at twice the linear frequency. Figure~\ref{fig:quadratic_null} shows the projected $95\%$ upper credible limit on the quadratic signal amplitude as a function of mass, comparing the finite-correlation analysis with the fully correlated and fully uncorrelated limits as in Fig.~\ref{fig:linear_null}. Here, we find that mismodeling the correlation structure through the limiting cases has a considerably larger impact on the inferred sensitivities, shifting the resulting upper limits on   \(A_\phi\) by up to nearly $1.0\,\mathrm{dex}$. Notably, even at the lowest mass of $m_\phi = 10^{-24}\,\mathrm{eV}$ considered in this work, where the fully correlated approximation might be expected to perform accurately, we find a $\mathcal{O}(0.2\,\mathrm{dex})$ discrepancy with an analysis that treats the finite correlation structure in detail.  

\begin{figure}[!t]
    \centering
    \includegraphics[width=\linewidth]{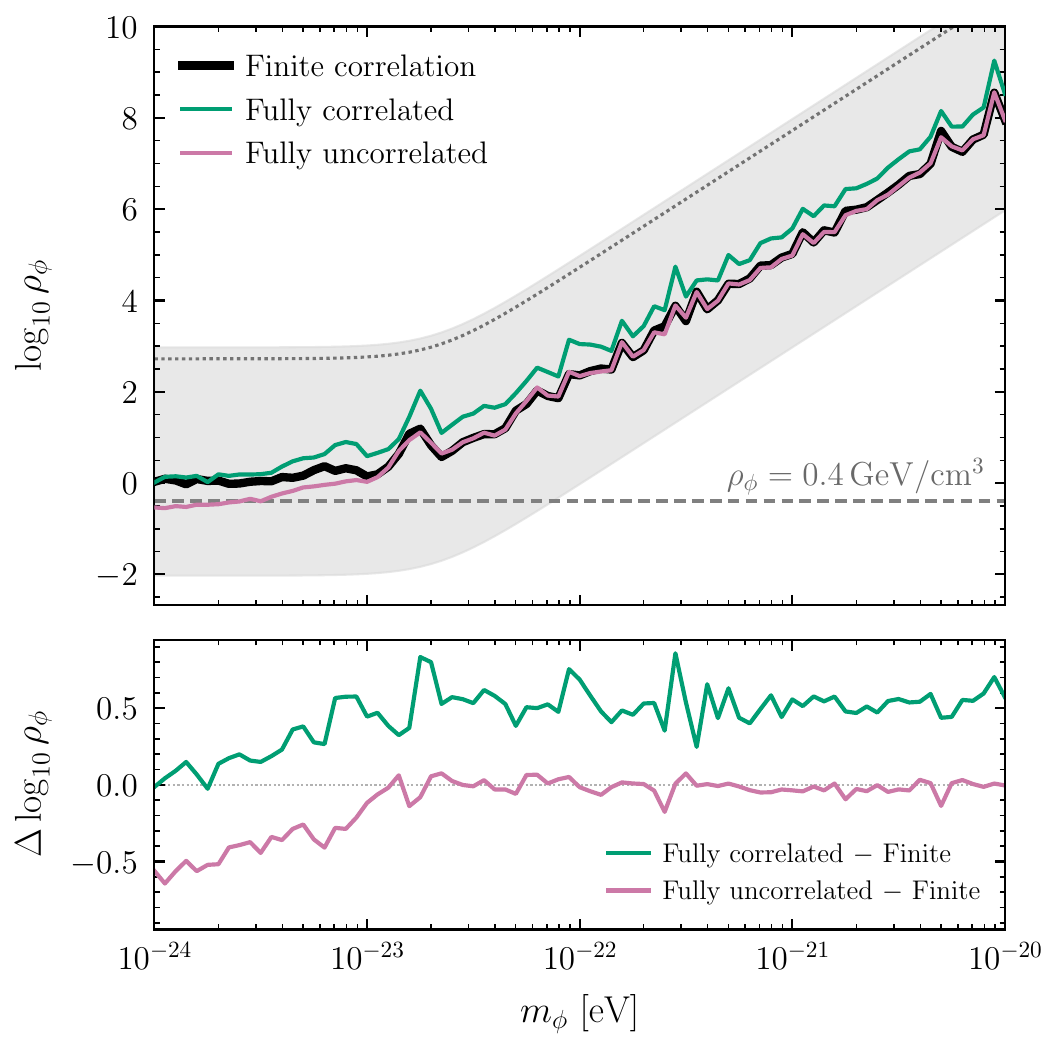}
    \caption{Results of the $w_E = w_P = 1$ null analysis translated into upper limits on the local dark matter density $\rho_\phi$. See text for details.}
    \label{fig:dm_interpretation}
\end{figure}

It is straightforward to relate the projected sensitivities of Figs.~\ref{fig:linear_null} and \ref{fig:quadratic_null} following Sec.~\ref{subsec:signal_parametrization}. As a specific example, Fig.~\ref{fig:dm_interpretation} translates the $w_E = w_P = 1$ null analysis into constraints on the local DM abundance. An inappropriate assumption of the uncorrelated limit at lower masses results in projected sensitivity to sub-unity fractional abundance. By contrast, correctly accounting for the ULDM correlations weakens the projected sensitivities to $\rho_\phi \approx 1\,\mathrm{GeV/cm}^3$, a few times the local density.

\subsection{Unblinded injection tests}
\label{sec:UnblindedTests}

\begin{figure}[!htb]
\includegraphics[width=\linewidth]{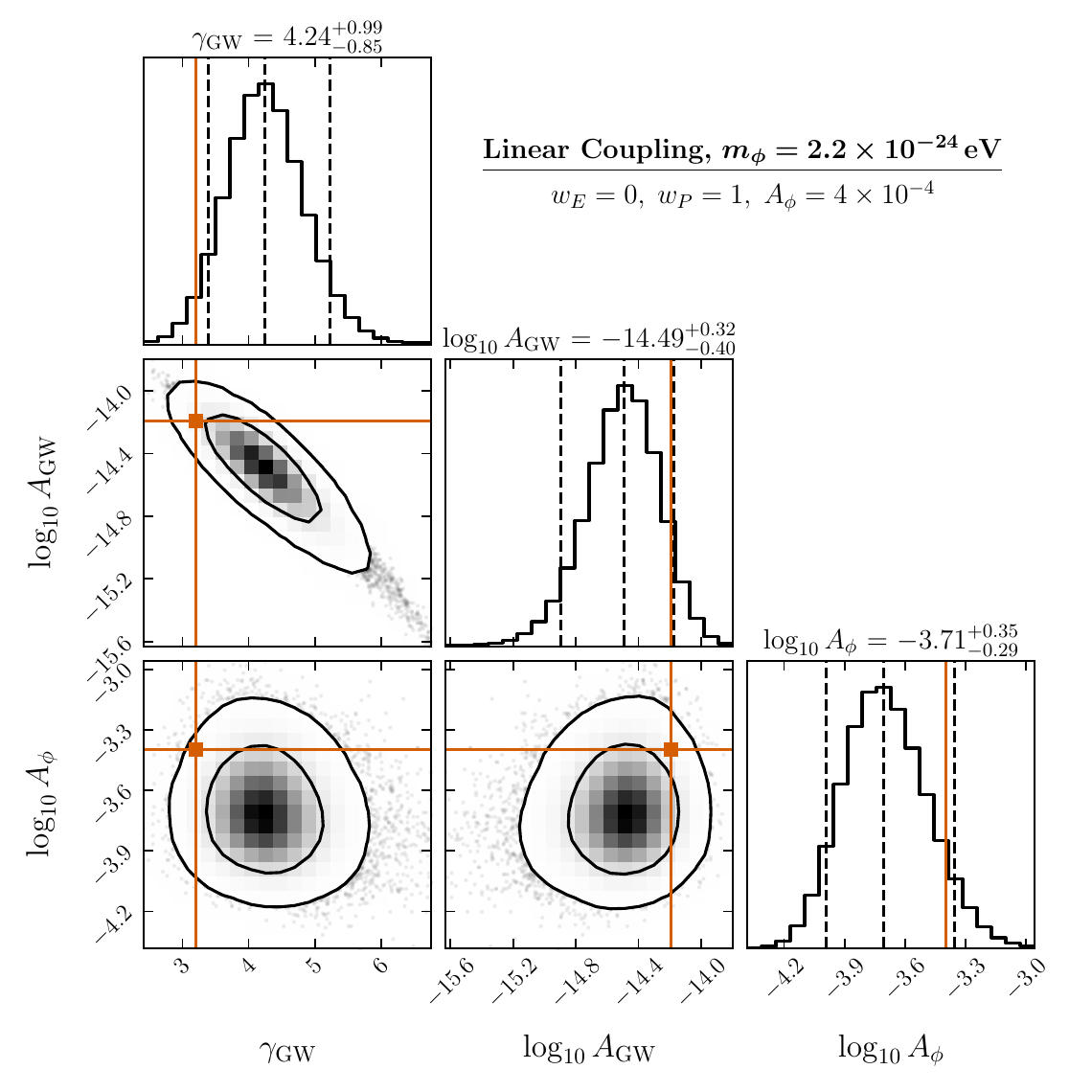}
\caption{Representative posterior distributions for an unblinded ULDM analysis at fixed mass \(m_\phi = 2.2\times 10^{-24}\,{\rm eV}\). The corner plot shows the marginalized posterior over \(\gamma_{\rm GW}\), \(\log_{10} A_{\rm GW}\), and \(\log_{10} A_\phi\), after marginalizing over the full parameter set included in the analysis. The ULDM mass, $m_\phi$, the coupling weights $w_E$ and $w_P$, and the $A_\phi$ used to generate the dataset are indicated in the inset text. Dashed vertical lines mark the 5th, 50th, and 95th percentiles of the one-dimensional marginals, while black contours enclose 68\% and 95\% posterior probability in the two-dimensional marginals. Orange lines indicate the injected parameter values used to generate the data. For this representative case, the true values of the injected ULDM signal amplitude and SGWB amplitude lie within their central 90\% credible intervals, while the SGWB index lies slightly outside its central 90\% credible interval.}
\label{fig:LinearUnblinded_MainExample}
\end{figure}

We next test the analysis on mock datasets containing injected ULDM signals with known parameters.
As in Sec.~\ref{sec:NullData}, we generate synthetic background data following the procedure of Sec.~\ref{sec:mock_data_generation}, but now add a ULDM signal with \textit{a priori} known mass and amplitude $A_\phi$. We consider three injected masses: $m_\phi = 2.2\times 10^{-22}\,\mathrm{eV}$, $2.2\times 10^{-23}\,\mathrm{eV}$, and $2.2\times 10^{-24}\,\mathrm{eV}$. Because the ULDM coherence length scales as $\ell \propto m_\phi^{-1}$, these masses span the nearly uncorrelated, partially correlated, and highly correlated regimes, respectively. For the linearly coupled injections we focus on the pulsar-term scenario, $w_E=0$ and $w_P=1$, because this is the case in which finite pulsar-to-pulsar correlations play the most important role. This provides a stringent stress test of the correlated-signal analysis.

These unblinded tests are designed to verify that the full inference pipeline can recover a signal realization generated from the physical finite-correlation ULDM model, even though the analysis is performed using the augmented latent-field model introduced in Sec.~\ref{subsec:augmented_latent_prior}. They therefore provide a nontrivial validation of the augmentation procedure. We choose injected amplitudes $A_\phi$ that are factors of a few larger than the 95\% upper credible limits obtained from the null analysis in Sec.~\ref{sec:NullData}, ensuring that the signals are detectable while remaining in the weak-signal regime. We therefore perform signal injections for three mass-coupling scenarios for both linear and quadratic signals:
\begin{enumerate}
    \item $m_\phi = 2.2\times 10^{-24}$, $A_\phi = 4\times10^{-4}$
    \item $m_\phi = 2.2\times 10^{-23}$, $A_\phi = 4\times10^{-7}$
    \item $m_\phi = 2.2\times 10^{-22}$, $A_\phi = 7\times10^{-8}$
\end{enumerate}
For linear coupling injections, we use $w_E = 0$ and $w_P = 1$, corresponding to the scenario to which we expect the greatest sensitivity to the ULDM correlation structure. By contrast, for the quadratic coupling injections, we use $w_E = w_P = 1$, which by the weight-sign independence noted in Sec.~\ref{subsec:augmented_latent_prior} is equivalent to the gravitational case $w_E=-w_P=1$, the physically motivated scenario of gravitational fluctuations associated with oscillations in the ULDM field.

\begin{figure}[!t]
    \centering
    \includegraphics[width=\linewidth]{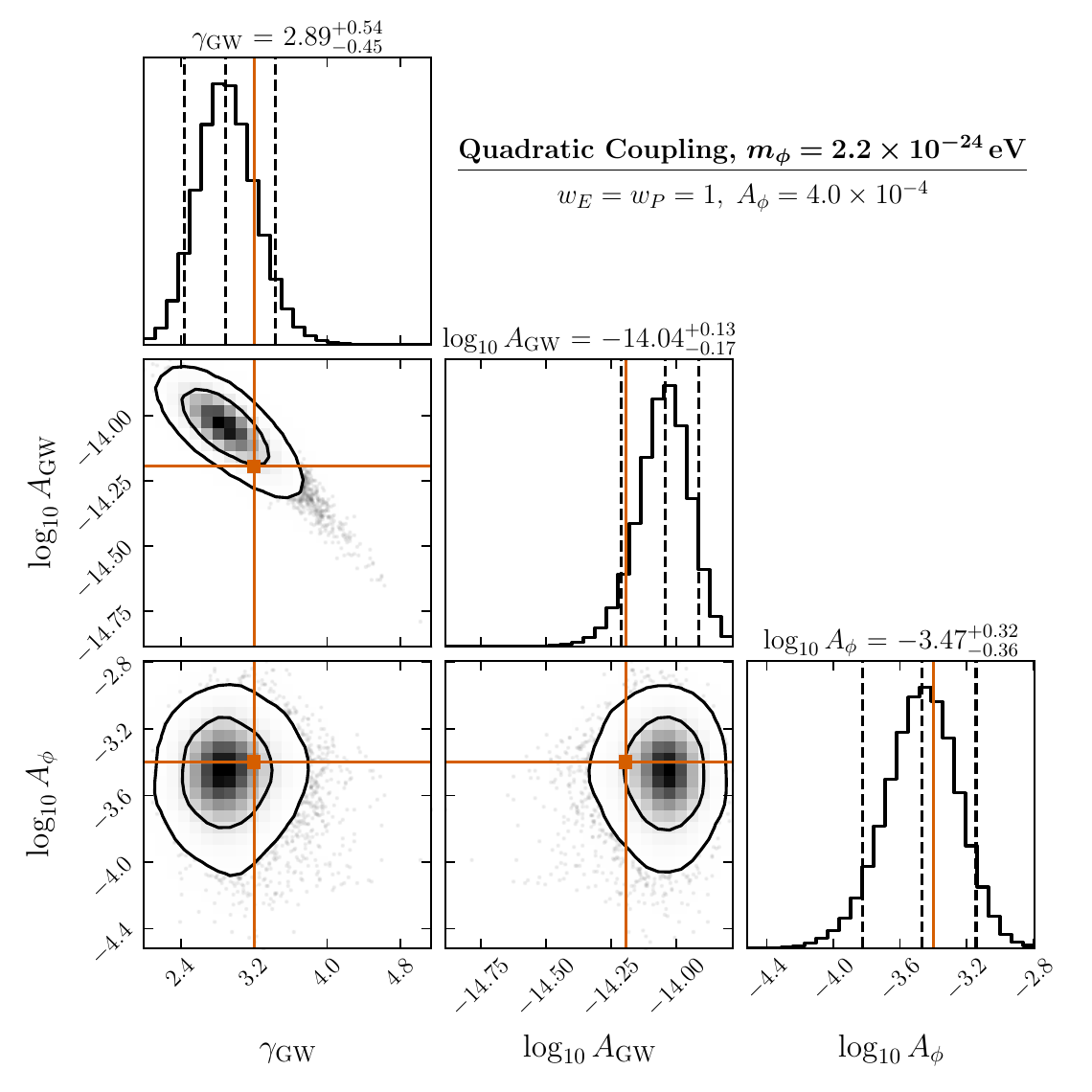}
    \caption{As in Fig.~\ref{fig:LinearUnblinded_MainExample}, but for a quadratic signal injected at an identical mass and $A_\phi$ but with $w_E = w_P = 1$, corresponding to the equal-ratio coupling scenario of the gravitational fluctuations associated with ULDM. All of the ULDM and GW parameters have inferred 90\% credible intervals which contain the true parameter values.}
    \label{fig:QuadraticUnblinded_MainExample}
\end{figure}

For each injected dataset and coupling scenario, we perform a fixed-mass Bayesian analysis as described in Sec.~\ref{sec:bayesian_procedure}, with $m_\phi$ set equal to the injected value. The results of the linear coupling analysis for the $m_\phi = 2.2 \times 10^{-24}\,\mathrm{eV}$ scenario are summarized in Fig.~\ref{fig:LinearUnblinded_MainExample}, which shows the resulting marginalized posteriors for the three parameters of interest, $\gamma_{\rm GW}$, $\log_{10} A_{\rm GW}$, and $\log_{10} A_\phi$. Results for the unblinded linear analyses at $m_\phi = 2.2 \times 10^{-23}\,\mathrm{eV}$ and $m_\phi = 2.2 \times 10^{-22}\,\mathrm{eV}$ are presented in Fig.~\ref{fig:LinearUnblinded_AppExamples} in App.~\ref{app:UnblindedAnalyses}. Likewise, the results of the quadratic coupling analysis for the $m_\phi = 2.2 \times 10^{-24}\,\mathrm{eV}$ scenario are shown in Fig.~\ref{fig:QuadraticUnblinded_MainExample}, with the additional cases presented in Fig.~\ref{fig:QuadraticUnblinded_AppExamples}.

The posterior structure changes qualitatively across the three masses. At high mass, the ULDM oscillation frequency lies above the red-noise-dominated band, so $\log_{10} A_\phi$ is nearly nondegenerate with the SGWB parameters. At lower masses, the ULDM signal overlaps more strongly with the SGWB-supported part of the spectrum, allowing some power to be traded between the monochromatic ULDM signal and the broadband red-noise model.
Additionally, at the lowest masses, timing-model subtraction removes large parts of the signal, hindering the sensitivity even in the white noise limit~\cite{Kaplan:2022lmz}.
Nonetheless, across all three regimes, the injected values lie within the posterior support, demonstrating successful recovery of both the ULDM amplitude and the SGWB parameters in the full analysis.

\subsection{Blinded injection tests}
\label{sec:BlindedTests}

As a final validation of the analysis pipeline, we perform blinded injection tests in which the signal masses are not known to the analyzer.
We consider three simulated datasets, each containing an injected ULDM signal in the range $10^{-24}\,\mathrm{eV}$ to $10^{-22}\,\mathrm{eV}$.  We restrict the blinded scan to $m_\phi \leq 10^{-22}\,\mathrm{eV}$ because larger masses are already deep in the uncorrelated regime for the PTA considered here. Extending the scan to $10^{-20}\,\mathrm{eV}$ would increase the computational cost by approximately an additional factor of 100 without testing a qualitatively new correlation regime. As in the unblinded tests, the injected amplitudes are chosen so that the signals are detectable with high significance while remaining outside the strong-signal regime.

The blinded setup is designed to test the full end-to-end inference procedure.
One author generated the six simulated datasets (three linear coupling scenarios and three quadratic coupling scenarios), while another author performed the analysis without access to the injected masses or amplitudes.
The analyzer was told whether the signal was linear or quadratic, but did not know which dataset corresponded to the low-, intermediate-, or high-mass injection. This procedure tests not only whether the analysis can recover a signal at a known mass, but whether the mass-grid analysis can identify the correct signal frequency without using injection information.

For each dataset, we analyze an adaptive mass grid spanning \(10^{-24}\,\mathrm{eV}\leq m_\phi \leq 10^{-22}\,\mathrm{eV}\).  The grid is constructed recursively from \(m_{\phi,0}=10^{-24}\,\mathrm{eV}\), with each new point chosen as
\begin{equation}
    m_{\phi,i+1}
    =
    \min\!\left[
        m_{\phi,i} + 4\times 10^{-24}\,\mathrm{eV},
        \,
        10^{1/20} m_{\phi,i}
    \right],
\end{equation}
and with the final point clipped to the upper boundary of the search range. Thus neighboring grid points differ by no more than \(4\times 10^{-24}\,\mathrm{eV}\) in absolute mass and have at least 20 grid points per decade in relative spacing.  This adaptive construction satisfies the mass-resolution criterion discussed in Appendix~\ref{app:mass_resolution}, while also maintaining good resolution across the low-mass interval \(10^{-24}\,\mathrm{eV}\lesssim m_\phi \lesssim 10^{-22}\,\mathrm{eV}\), where the spatial correlation structure changes appreciably.

At each mass in the grid, we perform the fixed-mass Bayesian analysis described in Sec.~\ref{sec:bayesian_procedure}, using the free coupling-ratio parametrization. The resulting posterior in the \((A_E,A_P)\) plane indicates whether the data prefer a nonzero Earth-term or pulsar-term ULDM contribution at that trial mass. However, a nonzero amplitude posterior at a given mass is not by itself sufficient to identify the injected signal mass. Because the PTA sampling is irregular and the timing uncertainties are heteroskedastic, power from a monochromatic signal can leak into neighboring or alias-like frequencies. As a result, fixed-mass analyses at masses different from the true injected value can sometimes show an apparent preference for a nonzero ULDM contribution.

For this reason, the blinded analysis uses both the marginalized amplitude posterior at each fixed mass and the Bayesian evidence associated with that mass. Denoting the sampled non-mass parameters at fixed mass by \(\bm\Theta\), the evidence is
\begin{equation}
    Z(m_\phi)
    =
    \int d\bm\Theta\,
    \mathcal L(\mathbf d\mid \bm\Theta,m_\phi)\,
    \pi(\bm\Theta\mid m_\phi).
    \label{eq:mass_grid_evidence}
\end{equation}
The evidence is estimated directly by \texttt{pocomc} during its standard sampling procedure, providing the normalization information needed to compare different fixed-mass analyses without requiring a separate evidence calculation.
For a discrete mass grid with prior weights $\pi(m_i)$, the posterior probability assigned to each trial mass is then
\begin{equation}
    p(m_i\mid \mathbf d)
    =
    \frac{
        Z(m_i)\,\pi(m_i)
    }{
        \sum_j Z(m_j)\,\pi(m_j)
    } .
    \label{eq:discrete_mass_posterior}
\end{equation}
Thus, while the fixed-mass posterior in \((A_E,A_P)\) diagnoses whether a signal-like component is preferred at a particular frequency, the evidence determines which mass provides the best global explanation of the data after accounting for the full parameter volume. This distinction is essential in the blinded tests, where spectral leakage can produce secondary peaks in the fixed-mass amplitude posteriors away from the true injected mass.

\begin{figure}[!t]
    \centering
    \includegraphics[width=\linewidth]{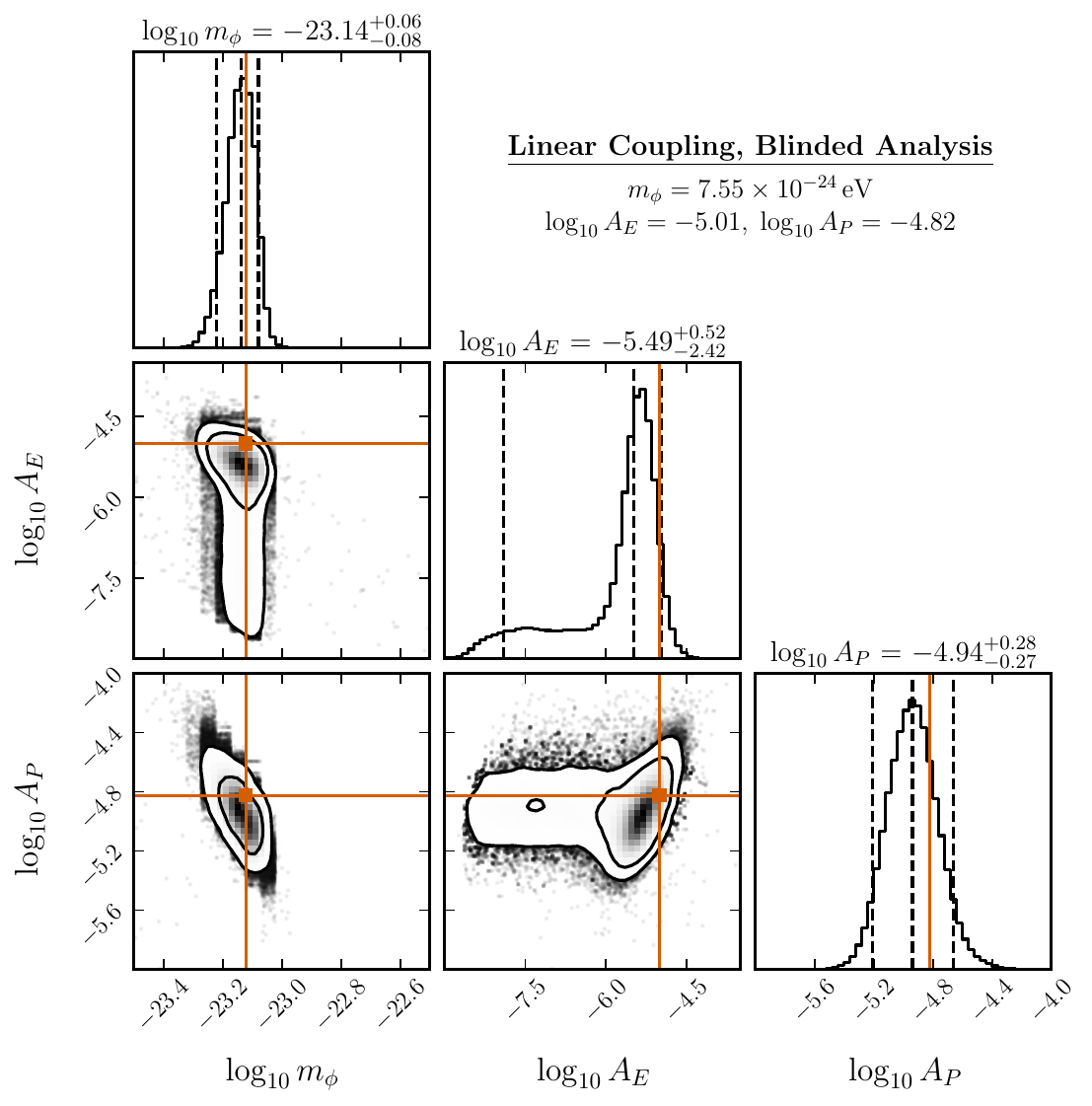}
    \caption{
    Evidence-weighted posterior from the blinded mass scan for one signal-injection test using the free coupling-ratio parametrization.
    The posterior is shown in terms of the scanned ULDM mass \(m_\phi\), the Earth-term amplitude \(A_E\), and the pulsar-term amplitude \(A_P\).
    Fixed-mass posteriors are combined using their Bayesian evidences together with the log-uniform mass prior over the scanned range.
    Black contours enclose the 68\% and 95\% credible regions, while the one-dimensional panels show the marginalized posteriors with dashed lines indicating the 5th, 50th, and 95th percentiles.
    Orange lines and markers indicate the injected parameter values. The true values of the ULDM mass, $m_\phi$, and the signal amplitudes $A_E$ and $A_P$ used to generate the dataset are indicated in the inset text.
    The recovery of the injected mass and amplitudes within the posterior support demonstrates that the mass scan correctly identifies the blinded ULDM signal.
    }
    \label{fig:LinearBlinded_MainExample}
\end{figure}

As a representative example, Fig.~\ref{fig:LinearBlinded_MainExample} shows the evidence-weighted posterior obtained from one of the blinded mass scans.  In these blinded tests, the injected ULDM mass is not supplied to the analysis. We therefore analyze each mass in the grid independently using the free coupling-ratio parametrization and then combine the fixed-mass results using the evidence weights in Eq.~\eqref{eq:discrete_mass_posterior}.  This produces a joint posterior over the scanned mass and the two signal-amplitude parameters, \(A_E\) and \(A_P\).

In the example shown in Fig.~\ref{fig:LinearBlinded_MainExample}, the posterior localizes around the injected mass, and the injected values of \(m_\phi\), \(A_E\), and \(A_P\) all lie within their corresponding one-dimensional 90\% credible intervals.  As seen in the null analyses, the sensitivity to the Earth-term amplitude is generally weaker than the sensitivity to the pulsar-term amplitude.  The broader localization of \(A_E\) in Fig.~\ref{fig:LinearBlinded_MainExample} reflects this reduced sensitivity, together with the fact that the injected value of \(A_E\) is somewhat smaller than that of \(A_P\).  Two additional blinded signal-injection tests are shown in Fig.~\ref{fig:LinearBlinded_AppExamples} of App.~\ref{app:BlindedAnalyses}. Taken as a whole, these blinded tests demonstrate that the analysis methodology provides a robust and accurate pipeline suitable for real-data applications.

\begin{figure}[!t]
    \centering
    \includegraphics[width=\linewidth]{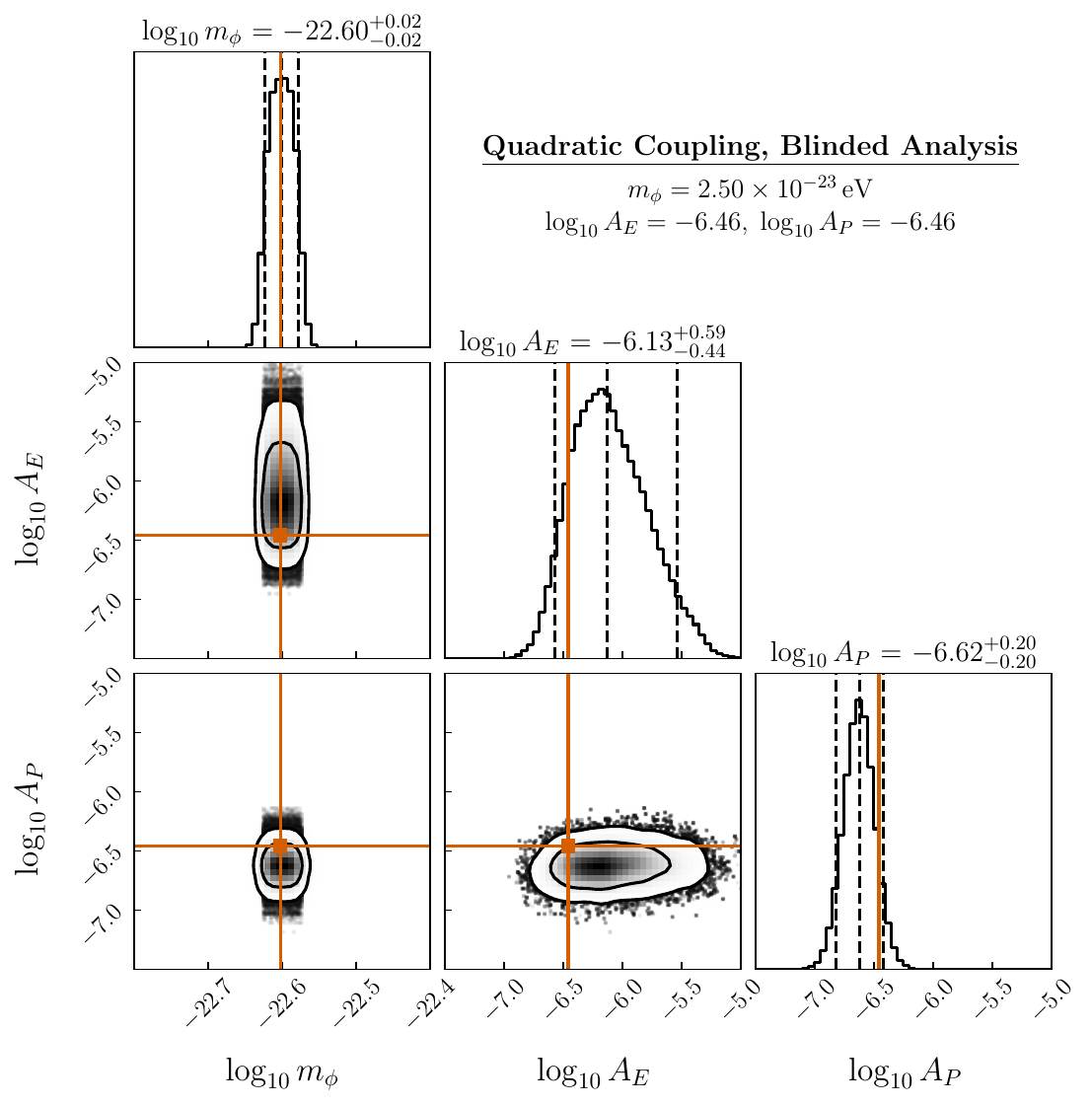}
    \caption{Evidence-weighted posterior from the blinded mass scan for a quadratic signal injection, analogous to Fig.~\ref{fig:LinearBlinded_MainExample}.}
    \label{fig:QuadraticBlinded_MainExample}
\end{figure}

We repeat the blinded mass-scan test for the quadratically coupled signal, combining the fixed-mass evidences over the adaptive grid to localize the injected mass. Figure~\ref{fig:QuadraticBlinded_MainExample} shows a representative example; two further blinded injections appear in Fig.~\ref{fig:QuadraticBlinded_AppExamples} of App.~\ref{app:BlindedAnalyses}. In all three examples, unbeknownst to the analyzer, the blinded data were generated with equal Earth and pulsar couplings so as to realize a gravitational signal. As in the case of the linear signals, the analysis
results in accurate and precise inferred parameters. Taken together, these tests demonstrate that the finite-correlation pipeline applies to the quadratically coupled and gravitational signals with no change beyond the squared response map.

\section{Conclusion}
\label{sec:conclusion}

In this work, we have introduced a self-consistent deterministic framework for analyzing the ULDM fast mode with PTA data while retaining the finite spatial correlations of the underlying ULDM field. We studied ULDM with linear couplings to the SM (Sec.~\ref{sec:linear_signals}), for which the timing residual receives contributions from changes in the pulsar moment of inertia, and hence its rotation rate, as well as from changes in the atomic transition frequencies used as the terrestrial timing reference.  In this case, the map from a latent ULDM field realization to the predicted timing residuals is especially simple: the residual is linear in the field evaluated at the Earth and at the pulsars. We additionally treated quadratically coupled scalar ULDM and the universal gravitational signal (Secs.~\ref{subsec:quadratic_couplings} and~\ref{subsec:gravitational_couplings}), for which the residual is quadratic in the field and oscillates at $2m_\phi$; because it remains a deterministic function of the same latent amplitudes, it is described by the same finite-correlation prior, with only the response map squared.  We therefore use the finite correlation structure of the ULDM field (Sec.~\ref{sec:uldm_signal_correlations}) to define the prior over latent field realizations, and then forward model those realizations into timing residuals as part of the Bayesian inference (Secs.~\ref{subsec:augmented_latent_prior}--\ref{subsec:bayesian_analysis_flow_prior}).

The main advance of this work is to replace the limiting amplitude priors used in previous deterministic searches with a finite-correlation prior derived from the ULDM covariance.  Pulsar-distance uncertainties were incorporated through an augmented latent-field prior (Sec.~\ref{subsec:augmented_latent_prior}), and the resulting distance-marginalized amplitude distribution was represented with a normalizing-flow surrogate (Sec.~\ref{subsec:flow_amplitude_prior}).  This construction preserves the familiar amplitude-phase parametrization used in previous PTA searches while extending it to the intermediate regime where neither limiting description is valid.

We extensively validated the method using mock PTA datasets (Sec.~\ref{sec:mock_data_analysis}).  In the limiting regimes, we reproduced the behavior of the standard deterministic analyses.  We also performed blinded analyses to test signal recovery when the ULDM mass and field realization were not specified in advance.  These tests show that the framework can be used for inference across the fully correlated, intermediate, and fully uncorrelated regimes within a single self-consistent analysis.

This work sets a new standard for PTA searches for linearly and quadratically coupled ULDM: analyses should retain the finite spatial correlations of the ULDM field rather than impose either limiting description across the full mass range. Although we have applied the framework here only to mock datasets, the implementation builds on \texttt{enterprise} and is readily applicable to real PTA data.

While we have focused on scalar ULDM, the same finite-correlation perspective applies to other ULDM signals not treated here. In particular, our methods extend to vector (dark-photon) dark matter, whose direct and gravitational signals require a vector-valued field and a correspondingly generalized latent prior; we leave this to upcoming work. More broadly, the finite spatial correlations of the ULDM field are essential to any PTA search that spans the transition between the fully correlated and fully uncorrelated regimes, and the latent-field framework developed here provides a practical and self-consistent way to incorporate them.

\section*{Acknowledgments}
We thank Hyungjin Kim for collaboration at early stages of the work, and we thank Dan~Hackett and Mike~Wagman for helpful conversations.
KB acknowledges support from the National Science Foundation (NSF) under Grant No.~PHY-2413016.
VL is supported by the Network for Neutrinos, Nuclear Astrophysics and Symmetries (N3AS) through the National Science Foundation Physics Frontier Center, Grant No. PHY-2020275.
AM acknowledges support from a Royal Society University Research Fellowship (URF-R1-251896). KT acknowledges support from the U.S. Department of Defense through the National Defense Science and Engineering Graduate (NDSEG) fellowship. 
TT is supported by the DOE grant DE-SC0015655.
All authors are members of the NANOGrav Collaboration, which is supported by NSF Physics Frontiers Center award \#2020265. 
This research used resources of the National Energy Research Scientific Computing Center (NERSC), a U.S. Department of Energy Office of Science User Facility located at Lawrence Berkeley National Laboratory, operated under Contract No. DE-AC02-05CH11231 using NERSC award HEP-ERCAP0023978.
This work made use of computing resources provided by the Center for High Throughput Computing at the University of Wisconsin, Madison \cite{https://doi.org/10.21231/gnt1-hw21} and by the Texas Advanced Computing Center (TACC) at The University of Texas at Austin.

\appendix

\section{Gaussian statistics for the quadrature parametrization}
\label{app:rayleigh_gaussian_equivalence}

The Rayleigh-amplitude and random-phase parametrization in
Eq.~\eqref{eq:ULDM_field} can be rewritten mode by mode in terms of
Gaussian quadratures.   For an arbitrary velocity mode, define
\begin{equation}
    \Theta_{\mathbf v}
    =
    \omega_{\mathbf v}t
    -
    m_\phi\mathbf v\cdot\mathbf x,
\end{equation}
with the dependence of $ \Theta_{\mathbf v}$ on $t$ and $\mathbf{x}$ left implicit. Then
\begin{equation}
    \alpha_{\mathbf v}
    \cos\!\left(
        \Theta_{\mathbf v}
        +
        \theta_{\mathbf v}
    \right)
    =
    C_{\mathbf v}\cos\Theta_{\mathbf v}
    +
    S_{\mathbf v}\sin\Theta_{\mathbf v},
    \label{eq:single_mode_quadrature_identity}
\end{equation}
with
\begin{equation}
    C_{\mathbf v}
    =
    \alpha_{\mathbf v}\cos\theta_{\mathbf v},
    \qquad
    S_{\mathbf v}
    =
    -\alpha_{\mathbf v}\sin\theta_{\mathbf v}.
    \label{eq:rayleigh_to_gaussian_quadratures}
\end{equation}
The Rayleigh-phase parametrization takes $\theta_{\mathbf v}$ to be
uniformly distributed on $[0,2\pi)$ and $\alpha_{\mathbf v}$ to be
Rayleigh distributed with
\begin{equation}
    \left\langle
        \alpha_{\mathbf v}^2
    \right\rangle
    =
    f(\mathbf v)\Delta\mathbf v .
\end{equation}
The joint density is therefore
\begin{equation}
    p(\alpha_{\mathbf v},\theta_{\mathbf v})
    =
    \frac{
        \alpha_{\mathbf v}
    }{
        \pi f(\mathbf v)\Delta\mathbf v
    }
    \exp\!\left[
        -
        \frac{
            \alpha_{\mathbf v}^2
        }{
            f(\mathbf v)\Delta\mathbf v
        }
    \right].
    \label{eq:rayleigh_phase_density}
\end{equation}
Changing variables from $(\alpha_{\mathbf v},\theta_{\mathbf v})$ to
$(C_{\mathbf v},S_{\mathbf v})$ gives
\begin{equation}
    p(C_{\mathbf v},S_{\mathbf v})
    =
    \frac{
        1
    }{
        \pi f(\mathbf v)\Delta\mathbf v
    }
    \exp\!\left[
        -
        \frac{
            C_{\mathbf v}^2+S_{\mathbf v}^2
        }{
            f(\mathbf v)\Delta\mathbf v
        }
    \right].
\end{equation}
Thus the quadratures are independent Gaussian variables,
\begin{equation}
    C_{\mathbf v}, S_{\mathbf v}
    \sim
    \mathcal N
    \!\left(
        0,
        \frac{1}{2}f(\mathbf v)\Delta\mathbf v
    \right),
\end{equation}
which is the Gaussian-quadrature parametrization used in
Eq.~\eqref{eq:ULDM_field_Gaussian}.

\section{The gravitational signal}
\label{app:gravitational}

Here we derive the gravitational potential quoted in Sec.~\ref{subsec:gravitational_couplings} and isolate its fast and slow parts, the latter feeding the slow-mode analysis of App.~\ref{app:slowmode}. The oscillating ULDM sources the scalar potential, $\Psi$, through the Poisson equation~\cite{Khmelnitsky:2013lxt},
\begin{equation}
    \nabla^2\Psi = 4\pi G\rho,
    \qquad
    \rho = \tfrac12\!\left[\dot\phi^2 + (\nabla\phi)^2 + m_\phi^2\phi^2\right].
    \label{eq:app_poisson}
\end{equation}
We expand the field in the Gaussian quadratures of App.~\ref{app:rayleigh_gaussian_equivalence}, see Eq.~\eqref{eq:ULDM_field_Gaussian}, passing to the continuum limit with unit-normalized quadratures,
\begin{equation}
\begin{gathered}
    \langle C_\mathbf{p}C_{\mathbf{p}'}\rangle=\langle S_\mathbf{p}S_{\mathbf{p}'}\rangle=\delta^{(3)}(\mathbf{p}-\mathbf{p}'),\\ \langle C_\mathbf{p}S_{\mathbf{p}'}\rangle=0.
    \label{eq:unit_quadratures}
\end{gathered}
\end{equation}
Passing to the momentum $\mathbf p=m_\phi\mathbf v$, the discrete per-mode variance $\tfrac12 f(\mathbf v)\Delta\mathbf v$ becomes the explicit measure $\tfrac12 f(\mathbf p)$, with $f(\mathbf p)$ the momentum-space distribution normalized to $\int d^3\mathbf p\, f(\mathbf p)=1$. The field is
\begin{widetext}
\begin{equation}
    \phi(\mathbf{x},t)=\frac{\sqrt{2\rho_\phi}}{m_\phi}\int d^3\mathbf{p}\,\left[\tfrac12 f(\mathbf{p})\right]^{1/2}\big[C_\mathbf{p}\cos(\omega_\mathbf{p}t-\mathbf{p}\!\cdot\!\mathbf{x})+S_\mathbf{p}\sin(\omega_\mathbf{p}t-\mathbf{p}\!\cdot\!\mathbf{x})\big].
    \label{eq:ULDM_field_continuum}
\end{equation}
\end{widetext}
This reproduces $\langle\phi^2\rangle=\rho_\phi/m_\phi^2$. Collecting terms, the energy density separates into parts oscillating at the sum and difference of the mode frequencies,
\begin{widetext}
\begin{equation}
\begin{split}
    \rho = \frac{\rho_\phi}{4m_\phi^2}\int d^3\mathbf p\,d^3\mathbf q\,\left[f(\mathbf p)f(\mathbf q)\right]^{1/2}\Big[
    &W_-(\mathbf p,\mathbf q)\big((C_\mathbf p C_\mathbf q-S_\mathbf p S_\mathbf q)\cos\Theta_+ + (C_\mathbf p S_\mathbf q+S_\mathbf p C_\mathbf q)\sin\Theta_+\big)\\
    +&W_+(\mathbf p,\mathbf q)\big((C_\mathbf p C_\mathbf q+S_\mathbf p S_\mathbf q)\cos\Theta_- + (S_\mathbf p C_\mathbf q-C_\mathbf p S_\mathbf q)\sin\Theta_-\big)\Big],
    \label{eq:app_rho}
\end{split}
\end{equation}
\end{widetext}
where $\Theta_\pm=(\omega_\mathbf p\pm\omega_\mathbf q)t-(\mathbf p\pm\mathbf q)\!\cdot\!\mathbf x$ and
\begin{equation}
    W_\pm(\mathbf p,\mathbf q) = m_\phi^2\pm(\mathbf p\!\cdot\!\mathbf q + \omega_\mathbf p\omega_\mathbf q).
    \label{eq:app_Wpm}
\end{equation}
Each plane wave is an eigenfunction of $\nabla^2$, so Poisson's equation is solved term by term, dividing the sum- and difference-frequency contributions by $|\mathbf p+\mathbf q|^2$ and $|\mathbf p-\mathbf q|^2$ respectively,
\begin{widetext}
\begin{equation}
\begin{split}
    \Psi = -\frac{\pi G\rho_\phi}{m_\phi^2}\int d^3\mathbf p\,d^3\mathbf q\,\left[f(\mathbf p)f(\mathbf q)\right]^{1/2}\Big[
    &\frac{W_-(\mathbf p,\mathbf q)}{|\mathbf p+\mathbf q|^2}\big((C_\mathbf p C_\mathbf q-S_\mathbf p S_\mathbf q)\cos\Theta_+ + (C_\mathbf p S_\mathbf q+S_\mathbf p C_\mathbf q)\sin\Theta_+\big)\\
    +&\frac{W_+(\mathbf p,\mathbf q)}{|\mathbf p-\mathbf q|^2}\big((C_\mathbf p C_\mathbf q+S_\mathbf p S_\mathbf q)\cos\Theta_- + (S_\mathbf p C_\mathbf q-C_\mathbf p S_\mathbf q)\sin\Theta_-\big)\Big].
    \label{eq:app_psi_full}
\end{split}
\end{equation}
\end{widetext}
In the non-relativistic limit the kinematic factors reduce to
\begin{equation}
    W_-(\mathbf p,\mathbf q)\to-\tfrac12|\mathbf p+\mathbf q|^2,\qquad W_+(\mathbf p,\mathbf q)\to2m_\phi^2,
    \label{eq:app_Wpm_NR}
\end{equation}
so that in the sum-frequency term the $|\mathbf p+\mathbf q|^2$ from the Poisson inversion cancels against $W_-$, leaving it local in the field, while the difference-frequency term retains a factor $2m_\phi^2/|\mathbf p-\mathbf q|^2$ and is nonlocal. The potential therefore splits into fast and slow parts, $\Psi=\Psi_{\rm fast}+\Psi_{\rm slow}$, with
\begin{widetext}
\begin{equation}
    \Psi_{\rm fast} = \frac{\pi G\rho_\phi}{2m_\phi^2}\int d^3\mathbf p\,d^3\mathbf q\,\left[f(\mathbf p)f(\mathbf q)\right]^{1/2}\big[(C_\mathbf p C_\mathbf q-S_\mathbf p S_\mathbf q)\cos\Theta_+ + (C_\mathbf p S_\mathbf q+S_\mathbf p C_\mathbf q)\sin\Theta_+\big],
    \label{eq:app_psi_fast}
\end{equation}
\begin{equation}
    \Psi_{\rm slow} = -2\pi G \rho_\phi\int\frac{d^3\mathbf p\,d^3\mathbf q}{|\mathbf p-\mathbf q|^2}\left[f(\mathbf p)f(\mathbf q)\right]^{1/2}\big[(C_\mathbf p C_\mathbf q+S_\mathbf p S_\mathbf q)\cos\Theta_- + (S_\mathbf p C_\mathbf q-C_\mathbf p S_\mathbf q)\sin\Theta_-\big].
    \label{eq:app_psi_slow}
\end{equation}
\end{widetext}
For the fast mode, the subcoherence limit $\omega_\mathbf p t\simeq m_\phi t$ sends $\Theta_+\to2m_\phi t-(\mathbf p+\mathbf q)\!\cdot\!\mathbf x$, so $\Psi_{\rm fast}$ oscillates at $2m_\phi$, while the difference-frequency phase $\Theta_-$ loses its time dependence and $\Psi_{\rm slow}$ is constant over the data span, degenerate with the timing model and unobservable at the masses relevant for the fast mode.

Although $\Psi_{\rm fast}$ is a non-Gaussian quadratic form in the quadratures, it can be sampled directly from a realization of the field. Squaring the field of Eq.~\eqref{eq:ULDM_field_continuum} in the subcoherence limit reproduces the same sum-frequency structure,
\begin{widetext}
\begin{equation}
    \phi^2 = \frac{\rho_\phi}{2m_\phi^2}\int d^3\mathbf p\,d^3\mathbf q\,\left[f(\mathbf p)f(\mathbf q)\right]^{1/2}\big[(C_\mathbf p C_\mathbf q-S_\mathbf p S_\mathbf q)\cos\Theta_+ + (C_\mathbf p S_\mathbf q+S_\mathbf p C_\mathbf q)\sin\Theta_+\big] + [\phi^2]_{\rm slow},
    \label{eq:app_phisq}
\end{equation}
\end{widetext}
whose fast part matches the bracket in Eq.~\eqref{eq:app_psi_fast}. Hence
\begin{equation}
    \Psi_{\rm fast} = \pi G\,[\phi^2]_{\rm fast},
    \label{eq:app_grav_result}
\end{equation}
so a realization of the oscillating potential is obtained by drawing a realization of the field, squaring it, and retaining the $2m_\phi$ component, which is the squared response map used in the main text. The slow potential $\Psi_{\rm slow}$ of Eq.~\eqref{eq:app_psi_slow} and the difference-frequency remainder $[\phi^2]_{\rm slow}$ of Eq.~\eqref{eq:app_phisq} are the slow-mode observables whose exact Gaussianity is established in App.~\ref{app:slowmode}.

\section{Exact Gaussianity of the slow mode}
\label{app:slowmode}

In addition to the fast mode at $2m_\phi$, the quadratic and gravitational signals possess a slow mode: the low-frequency part of the quadratic observable, generated by the beat between field modes of nearly equal frequency, $\omega\sim\omega_\mathbf{p}-\omega_\mathbf{q}\sim m_\phi v_0^2$. Its pulsar-timing phenomenology was studied in Ref.~\cite{Gan:2025icr}, which argued that it is approximately Gaussian. Here we show that in the continuum limit it is exactly Gaussian, so that it may be treated as an ordinary stationary Gaussian process specified entirely by its power spectrum.

We continue to work with the unit-normalized continuum quadratures introduced in App.~\ref{app:gravitational}, see Eq.~\eqref{eq:unit_quadratures}. The slow-mode timing residual is sourced by the difference-frequency part of the quadratic response derived there. For a quadratic SM coupling, it is the time integral of $\delta\nu/\nu\propto\phi^2$, built from $[\phi^2]_{\rm slow}$ of Eq.~\eqref{eq:app_phisq}; for the gravitational signal, it arises from the Doppler shift of the slow potential, $\nabla\Psi_{\rm slow}$ of Eq.~\eqref{eq:app_psi_slow}. At a fixed location, the spatial phase is absorbed into a redefinition of the Gaussian quadratures and does not affect their statistics, so both reduce to a common quadratic form in the quadratures, the slow-mode residual $r_{\rm slow}(t)$,
\begin{widetext}
\begin{equation}
    r_{\rm slow}(t) = \int d^3\mathbf{p}\,d^3\mathbf{q}\,K(\mathbf{p},\mathbf{q})\Big[(C_\mathbf{p} S_\mathbf{q}-S_\mathbf{p} C_\mathbf{q})\cos(\omega_{\mathbf{p}\mathbf{q}}t)+(C_\mathbf{p} C_\mathbf{q}+S_\mathbf{p} S_\mathbf{q})\sin(\omega_{\mathbf{p}\mathbf{q}}t)\Big],
    \qquad \omega_{\mathbf{p}\mathbf{q}}\equiv\omega_\mathbf{p}-\omega_\mathbf{q},
    \label{eq:slow_observable}
\end{equation}
\end{widetext}
with a scalar kernel $K$ fixed by the observable. Time-integrating $[\phi^2]_{\rm slow}$ gives
\begin{equation}
    K(\mathbf{p},\mathbf{q})\propto \frac{\mathbf d_{\rm SM}^{(2)}\cdot\mathbf y^{E,P}}{\Lambda^2}\,\frac{1}{\omega_\mathbf{p}-\omega_\mathbf{q}}\left[f(\mathbf{p})f(\mathbf{q})\right]^{1/2},
\end{equation}
while the line-of-sight Doppler $\hat{\mathbf n}\cdot\nabla\Psi_{\rm slow}$, with $\hat{\mathbf n}$ the Earth-pulsar direction, gives
\begin{equation}
    K(\mathbf{p},\mathbf{q})\propto \frac{\hat{\mathbf n}\cdot(\mathbf{p}-\mathbf{q})}{|\mathbf{p}-\mathbf{q}|^2}\left[f(\mathbf{p})f(\mathbf{q})\right]^{1/2}.
\end{equation}
Both kernels are antisymmetric under $\mathbf{p}\leftrightarrow\mathbf{q}$, a property we use repeatedly below. To expose the frequency content we project $r_{\rm slow}(t)$ onto $\cos\omega t$ and $\sin\omega t$ at fixed $\omega>0$. Because $\cos(\omega_{\mathbf{p}\mathbf{q}}t)$ is even and $\sin(\omega_{\mathbf{p}\mathbf{q}}t)$ is odd in $\omega_{\mathbf{p}\mathbf{q}}$, the cosine projection adds the $\omega_{\mathbf{p}\mathbf{q}}=\pm\omega$ contributions while the sine projection subtracts them, so the latter carries a factor $\mathrm{sgn}(\omega_{\mathbf{p}\mathbf{q}})$. The frequency content is then carried by the integrands
\begin{widetext}
\begin{equation}
    I_c(\mathbf{p},\mathbf{q}) = K(\mathbf{p},\mathbf{q})\,(C_\mathbf{p} S_\mathbf{q}-S_\mathbf{p} C_\mathbf{q})\,\delta(\omega-|\omega_{\mathbf{p}\mathbf{q}}|), \qquad
    I_s(\mathbf{p},\mathbf{q}) = K(\mathbf{p},\mathbf{q})\,(C_\mathbf{p} C_\mathbf{q}+S_\mathbf{p} S_\mathbf{q})\,\mathrm{sgn}(\omega_{\mathbf{p}\mathbf{q}})\delta(\omega-|\omega_{\mathbf{p}\mathbf{q}}|),
\label{eq:slow_integrands}
\end{equation}
\end{widetext}
in terms of which
\begin{equation}
\begin{aligned}
    C(\omega) &= \int d^3\mathbf{p}\,d^3\mathbf{q}\,I_c(\mathbf{p},\mathbf{q}),\\
    S(\omega) &= \int d^3\mathbf{p}\,d^3\mathbf{q}\,I_s(\mathbf{p},\mathbf{q}),
\end{aligned}
    \label{eq:slow_CS}
\end{equation}
and $r_{\rm slow}(t)=\int d\omega\,[C(\omega)\cos\omega t+S(\omega)\sin\omega t]$. For $\omega\neq0$ the support requires $\mathbf{p}\neq\mathbf{q}$, where $K$ is smooth, and this support is noncompact.

Both integrands have zero mean: $\langle I_c\rangle=0$ because $\langle C_\mathbf{p}S_\mathbf{q}-S_\mathbf{p}C_\mathbf{q}\rangle=0$, and $\langle I_s\rangle=0$ because $\langle C_\mathbf{p}C_\mathbf{q}+S_\mathbf{p}S_\mathbf{q}\rangle=2\delta^{(3)}(\mathbf{p}-\mathbf{q})$ vanishes on the support $\mathbf{p}\neq\mathbf{q}$. We now evaluate their covariances, from which both the Gaussianity and the stationarity of $r_{\rm slow}(t)$ follow.

Expanding the cosine-integrand covariance by Wick contraction, and keeping only the pairings that join each $C$ to a $C$ and each $S$ to an $S$,
\begin{widetext}
\begin{equation}
\begin{aligned}
    \langle I_c(\mathbf{p},\mathbf{q})\,I_c(\mathbf{p}',\mathbf{q}')\rangle
    &= K(\mathbf{p},\mathbf{q})\,K(\mathbf{p}',\mathbf{q}')\,
       \big\langle(C_\mathbf{p}S_\mathbf{q}-S_\mathbf{p}C_\mathbf{q})(C_{\mathbf{p}'}S_{\mathbf{q}'}-S_{\mathbf{p}'}C_{\mathbf{q}'})\big\rangle\,
       \delta(\omega-|\omega_{\mathbf{p}\mathbf{q}}|)\,\delta(\omega'-|\omega_{\mathbf{p}'\mathbf{q}'}|)\\
    &= 2\,K(\mathbf{p},\mathbf{q})\,K(\mathbf{p}',\mathbf{q}')
       \big[\delta^{(3)}(\mathbf{p}-\mathbf{p}')\delta^{(3)}(\mathbf{q}-\mathbf{q}')
          - \delta^{(3)}(\mathbf{p}-\mathbf{q}')\delta^{(3)}(\mathbf{q}-\mathbf{p}')\big]\\
    &\qquad\times\,\delta(\omega-|\omega_{\mathbf{p}\mathbf{q}}|)\,\delta(\omega'-|\omega_{\mathbf{p}'\mathbf{q}'}|).
\end{aligned}
\label{eq:slow_ic_cov}
\end{equation}
\end{widetext}
Integrating over all four momenta to form $\langle C(\omega)C(\omega')\rangle$, the first delta pair sets $(\mathbf{p}',\mathbf{q}')=(\mathbf{p},\mathbf{q})$ and the second sets $(\mathbf{p}',\mathbf{q}')=(\mathbf{q},\mathbf{p})$, where the kernel changes sign, $K(\mathbf{q},\mathbf{p})=-K(\mathbf{p},\mathbf{q})$:
\begin{widetext}
\begin{equation}
\begin{aligned}
    \langle C(\omega)\,C(\omega')\rangle
    &= \int d^3\mathbf{p}\,d^3\mathbf{q}\,d^3\mathbf{p}'\,d^3\mathbf{q}'\;
       \langle I_c(\mathbf{p},\mathbf{q})\,I_c(\mathbf{p}',\mathbf{q}')\rangle\\
    &= 2\int d^3\mathbf{p}\,d^3\mathbf{q}\;
       \big[K(\mathbf{p},\mathbf{q})^2
          - K(\mathbf{p},\mathbf{q})K(\mathbf{q},\mathbf{p})\big]\,
       \delta(\omega-|\omega_{\mathbf{p}\mathbf{q}}|)\,\delta(\omega'-|\omega_{\mathbf{p}\mathbf{q}}|)\\
    &= 4\int d^3\mathbf{p}\,d^3\mathbf{q}\;
       K(\mathbf{p},\mathbf{q})^2\,
       \delta(\omega-|\omega_{\mathbf{p}\mathbf{q}}|)\;\delta(\omega-\omega')
    \;\equiv\; P(\omega)\,\delta(\omega-\omega').
\end{aligned}
\label{eq:slow_CC}
\end{equation}
\end{widetext}
The antisymmetry of $K$ turns the relative minus sign into the sum of two equal terms, so the autocovariance is nonvanishing and defines the power spectrum $P(\omega)=4\int d^3\mathbf{p}\,d^3\mathbf{q}\,K(\mathbf{p},\mathbf{q})^2\,\delta(\omega-|\omega_{\mathbf{p}\mathbf{q}}|)$.

The sine integrand is treated identically. On the support $\mathbf{p}\neq\mathbf{q}$, the same Wick contraction, now with the symmetric quadrature combination and the signum factor of each $I_s$, gives
\begin{widetext}
\begin{equation}
\begin{aligned}
    \langle I_s(\mathbf{p},\mathbf{q})\,I_s(\mathbf{p}',\mathbf{q}')\rangle
    &= K(\mathbf{p},\mathbf{q})\,K(\mathbf{p}',\mathbf{q}')\,\mathrm{sgn}(\omega_{\mathbf{p}\mathbf{q}})\,\mathrm{sgn}(\omega_{\mathbf{p}'\mathbf{q}'})\,
       \big\langle(C_\mathbf{p}C_\mathbf{q}+S_\mathbf{p}S_\mathbf{q})(C_{\mathbf{p}'}C_{\mathbf{q}'}+S_{\mathbf{p}'}S_{\mathbf{q}'})\big\rangle\\
    &\qquad\times\,\delta(\omega-|\omega_{\mathbf{p}\mathbf{q}}|)\,\delta(\omega'-|\omega_{\mathbf{p}'\mathbf{q}'}|)\\
    &= 2\,K(\mathbf{p},\mathbf{q})\,K(\mathbf{p}',\mathbf{q}')\,\mathrm{sgn}(\omega_{\mathbf{p}\mathbf{q}})\,\mathrm{sgn}(\omega_{\mathbf{p}'\mathbf{q}'})
       \big[\delta^{(3)}(\mathbf{p}-\mathbf{p}')\delta^{(3)}(\mathbf{q}-\mathbf{q}')
          + \delta^{(3)}(\mathbf{p}-\mathbf{q}')\delta^{(3)}(\mathbf{q}-\mathbf{p}')\big]\\
    &\qquad\times\,\delta(\omega-|\omega_{\mathbf{p}\mathbf{q}}|)\,\delta(\omega'-|\omega_{\mathbf{p}'\mathbf{q}'}|).
\end{aligned}
\label{eq:slow_is_cov}
\end{equation}
\end{widetext}
On the swapped pair $(\mathbf{p}',\mathbf{q}')=(\mathbf{q},\mathbf{p})$ the kernel and the signum each change sign, $K(\mathbf{q},\mathbf{p})=-K(\mathbf{p},\mathbf{q})$ and $\mathrm{sgn}(\omega_{\mathbf{q}\mathbf{p}})=-\mathrm{sgn}(\omega_{\mathbf{p}\mathbf{q}})$. The two flips cancel, converting the plus sign into the same positive contribution found above, so
\begin{equation}
    \langle S(\omega)\,S(\omega')\rangle
    = P(\omega)\,\delta(\omega-\omega')
    = \langle C(\omega)\,C(\omega')\rangle.
    \label{eq:slow_SS}
\end{equation}
The cross-covariance vanishes identically, $\langle C(\omega)\,S(\omega')\rangle=0$, since $\langle(C_\mathbf{p}S_\mathbf{q}-S_\mathbf{p}C_\mathbf{q})(C_{\mathbf{p}'}C_{\mathbf{q}'}+S_{\mathbf{p}'}S_{\mathbf{q}'})\rangle=0$: every Wick pairing leaves three quadratures of one type and one of the other.

These covariances give the two properties we need. First, $r_{\rm slow}(t)$ is stationary: the relations in Eqs.~\eqref{eq:slow_CC} and~\eqref{eq:slow_SS}, together with the vanishing cross-covariance, are exactly the conditions for a stationary process with power spectrum $P(\omega)$. Second, $C(\omega)$ and $S(\omega)$ are Gaussian: each is an integral over the noncompact set of mode pairs of contributions that are uncorrelated between distinct pairs, by the delta functions in Eqs.~\eqref{eq:slow_ic_cov} and~\eqref{eq:slow_is_cov}, and individually infinitesimal; with no single pair dominating, the central limit theorem makes the integral exactly Gaussian in the continuum limit. This is what separates the slow mode from the fast mode, where in the coherence limit the mode phases collapse to $2m_\phi t$, the response factorizes into the square of a single complex field amplitude and is strongly non-Gaussian. Being zero-mean, Gaussian, and stationary, $r_{\rm slow}(t)$ is completely specified by $P(\omega)$, strengthening the approximate-Gaussianity argument of Ref.~\cite{Gan:2025icr}: it does not rely on a large but finite number of contributing modes, but holds exactly in the continuum limit.

\section{Details of the Fourier basis covariance}
\label{app:FourierCovariances}

Here we derive the coefficient covariances quoted in Eq.~\eqref{eq:FourierCoefficientCovariances} using the projectors defined in Eq.~\eqref{eq:projectors} and the frequency-domain covariance in Eq.~\eqref{eq:FreqDomainCovariance}.

As a representative example, consider the cosine-cosine correlator:
\begin{widetext} 
\begin{equation}
\begin{split} 
\langle \hat C_I \hat C_J \rangle &= \lim_{T\rightarrow \infty}\frac{1}{T^2}\int_{-T}^T dt \int_{-T}^T dt' \langle \hat\phi_I(t) \hat\phi_J(t') \rangle \cos(m_\phi t) \cos(m_\phi t') \\ 
&= \lim_{T\rightarrow \infty}\frac{1}{T^2}\int_{-T}^T dt \int_{-T}^T dt' \int df df' \, e^{2 \pi i (f t + f' t')} \langle \hat{\phi}_I(f) \hat{\phi}_J(f') \rangle \cos(m_\phi t) \cos(m_\phi t') \\ 
&= \lim_{T\rightarrow \infty} \frac{1}{4 T^2}\int_{-T}^T dt \int_{-T}^T dt' \int df df' \, e^{2 \pi i (f t + f' t')} \delta(f + f') \\
&\qquad\times \bigg[\delta \left(f - \frac{m_\phi}{2\pi} \right) \hat\Gamma_{IJ} + \delta \left(f + \frac{m_\phi}{2\pi} \right) \hat\Gamma_{IJ}^* \bigg] \cos(m_\phi t) \cos(m_\phi t') \\
&= \lim_{T\rightarrow \infty} \frac{1}{4 T^2} \int_{-T}^T dt \int_{-T}^T dt'\bigg[ e^{ i m_\phi (t- t')} \hat\Gamma_{IJ} + e^{-i m_\phi (t- t')} \hat\Gamma_{IJ}^* \bigg] \cos(m_\phi t) \cos(m_\phi t') \\
&= \frac{1}{2}\,\mathrm{Re}(\hat\Gamma_{IJ}). \end{split}
\end{equation}
\end{widetext}
Proceeding identically for the remaining projections yields
\begin{equation}
\begin{gathered}
\langle \hat S_I \hat S_J \rangle
= \frac12\,\mathrm{Re}(\hat\Gamma_{IJ}), \\
\langle \hat C_I \hat S_J \rangle
= \frac12\,\mathrm{Im}(\hat\Gamma_{IJ}), \\
\langle \hat S_I \hat C_J \rangle
= \frac12\,\mathrm{Im}(\hat\Gamma_{JI})
= -\frac12\,\mathrm{Im}(\hat\Gamma_{IJ}) ,
\end{gathered}
\end{equation}
where in the last step we used \(\hat\Gamma_{JI}=\hat\Gamma_{IJ}^*\). These expressions reproduce Eq.~\eqref{eq:FourierCoefficientCovariances}.

\section{Details of pulsar distance uncertainties}
\label{app:distance_uncertainties}

In this appendix, we specify the pulsar distance priors used in our analyses. Table~\ref{tab:distances} lists the 30 pulsars used in our analysis, selected from the NANOGrav 15-year dataset, together with the distance values and uncertainties adopted here. Pulsar distances are inferred either from parallax measurements, indicated by ``PX,'' or from dispersion measures using the NE2001 free-electron density model, indicated by ``DM.'' For additional discussion of the distance determinations used in the NANOGrav dataset, see Refs.~\cite{NANOGrav:2023bts,NANOGrav:2023pdq}. Additional distance measurements not yet incorporated in NANOGrav analyses include Refs.~\cite{Mingarelli:2018myc,Antoniadis:2020gos,Egleston:2026mcg}.

For pulsars with parallax-based distance measurements, we adopt a Gaussian prior,
\begin{equation}
    \pi(x) = \frac{1}{\sqrt{2\pi}\sigma_x}
    \exp\!\left[-\frac{(x-\mu_x)^2}{2\sigma_x^2}\right],
\end{equation}
where \(\mu_x\) is the tabulated distance and \(\sigma_x\) is the quoted uncertainty.

For pulsars with distances inferred from dispersion measures, we follow Refs.~\cite{NANOGrav:2023bts,NANOGrav:2023pdq} and adopt a prior centered on the quoted DM distance \(x_{\rm DM}\) that is uniform on the interval \(0.8\,x_{\rm DM} \le x \le 1.2\,x_{\rm DM}\), with half-Gaussian tails outside this range. With \(\Delta_\pm \equiv x-(1\pm0.2)\,x_{\rm DM}\),
\begin{equation}
\pi(x) \propto
\begin{cases}
\exp\!\left[-\frac{\Delta_-^2}{2\sigma_{\rm DM}^2}\right], & x < 0.8\,x_{\rm DM}, \\[0.5ex]
1, & 0.8\,x_{\rm DM} \le x \le 1.2\,x_{\rm DM}, \\[0.5ex]
\exp\!\left[-\frac{\Delta_+^2}{2\sigma_{\rm DM}^2}\right], & x > 1.2\,x_{\rm DM},
\end{cases}
\end{equation}
where \(\sigma_{\rm DM}\) is taken to be one quarter of the quoted DM-distance uncertainty.

\begin{table*}[!ht]{
\renewcommand{\arraystretch}{1.2}
    \ra{1.3}
    \begin{center}
    \tabcolsep=0.25cm
    \begin{tabular}{| c| l l l || c| l l l |}
    \hlinewd{1pt}
    \textbf{Pulsar}& \textbf{Type} & $x_I$ [kpc] & $\delta x_I$ [kpc] & 
    \textbf{Pulsar}& \textbf{Type} & $x_I$ [kpc] & $\delta x_I$ [kpc] \\
    \hlinewd{1pt}
    B1855+09 & PX & 1.18 & 0.12 \cite{2009MNRAS.400..951V, 1994ApJ...428..713K, NANOGrav:2023hde}  
    & J0023+0923 & PX & 1.02 & 0.11  \cite{NANOGrav:2023hde} \\

    J0030+0451 & PX & 0.3296 & 0.0036  \cite{Ding:2022luk, Lommen:2006jq, NANOGrav:2023hde}
    & J0340+4130 & DM & 1.71 & 0.34 \\

    J0613$-$0200 & PX & 1.07 & 0.10  \cite{2006MNRAS.369.1502H, NANOGrav:2023hde}
    & J0645+5158 & PX & 1.37 & 0.19  \cite{Stovall:2014gua, NANOGrav:2023hde} \\

    J0931$-$1902 & DM & 1.88 & 0.38
    & J1024$-$0719 & PX & 1.080 & 0.042  \cite{Ding:2022luk,Guillemot:2016skn, NANOGrav:2023hde} \\

    J1455$-$3330 & PX & 1.01 & 0.22  \cite{Guillemot:2016skn}
    & J1600$-$3053 & PX & 1.84 & 0.26  \cite{NANOGrav:2023hde} \\

    J1614$-$2230 & PX & 0.699 & 0.026  \cite{Guillemot:2016skn, Fermi-LAT:2013svs, NANOGrav:2023hde}
    & J1640+2224 & PX & 1.404 & 0.095  \cite{Ding:2022luk} \\

    J1713+0747 & PX & 1.138 & 0.019  \cite{2009MNRAS.400..951V, Chatterjee_2009, 2006MNRAS.369.1502H, Splaver:2004du, NANOGrav:2023hde}
    & J1741+1351 & PX & 2.36 & 0.56  \cite{NANOGrav:2023hde} \\

    J1832$-$0836 & PX & 2.00 & 0.47  \cite{NANOGrav:2023hde}
    & J1853+1303 & PX & 1.91 & 0.17  \cite{Ding:2022luk, NANOGrav:2023hde} \\

    J1909$-$3744 & PX & 1.159 & 0.013 \cite{2009MNRAS.400..951V, 2006MNRAS.369.1502H, Jacoby:2005qg, NANOGrav:2023hde}
    & J1910+1256 & PX & 3.52 & 0.41  \cite{Ding:2022luk} \\

    J1911+1347 & DM & 2.08 & 0.42
    & J1918$-$0642 & PX & 1.44 & 0.11  \cite{Ding:2022luk, NANOGrav:2023hde} \\

    J1923+2515 & PX & 0.94 & 0.21  \cite{NANOGrav:2023hde}
    & J1944+0907 & PX & 1.38 & 0.36  \cite{NANOGrav:2023hde} \\

    J2010$-$1323 & PX & 1.94 & 0.41  \cite{NANOGrav:2023hde}
    & J2017+0603 & DM & 1.57 & 0.31 \\

    J2033+1734 & DM & 1.99 & 0.40
    & J2043+1711 & PX & 1.58 & 0.11  \cite{NANOGrav:2023hde} \\

    J2214+3000 & DM & 1.54 & 0.31
    & J2229+2643 & DM & 1.43 & 0.29 \\

    J2234+0611 & PX & 1.23 & 0.17  \cite{NANOGrav:2023hde}
    & J2317+1439 & PX & 1.57 & 0.29  \cite{NANOGrav:2023hde} \\

    \hlinewd{1pt}
    \end{tabular}
    \end{center}}
\caption{The table of 30 pulsars included in our analysis, selected from the 15-year dataset generated by the NANOGrav collaboration. For each pulsar, we provide the distance of the pulsar from Earth, the reported error on that pulsar distance, and the method of distance measurement, with PX indicating a parallax measurement and DM indicating a dispersion-measure-based distance estimate. This table has been adapted from \cite{NANOGrav:2023pdq}.}
\label{tab:distances}
\end{table*}

\section{Conditions on mass resolution}
\label{app:mass_resolution}

In this appendix, we give a simple scaling argument for why it is computationally advantageous to analyze ULDM signals on a mass-by-mass grid, rather than treating \(m_\phi\) as a continuously sampled inference parameter.

For a PTA dataset of duration \(T_{\rm obs}\), the Fourier frequency resolution is of order \(1/T_{\rm obs}\). Since a ULDM field of mass \(m_\phi\) oscillates at frequency \(f_\phi = m_\phi/(2\pi)\), this corresponds to a mass resolution
\begin{equation}
    \Delta m_\phi \approx \frac{2\pi}{T_{\rm obs}}
    \simeq 8.7\times 10^{-24}\,\mathrm{eV}
    \left(\frac{15\,\mathrm{yr}}{T_{\rm obs}}\right).
\end{equation}
Physically, shifting the ULDM mass by more than \(\Delta m_\phi\) moves the signal by more than one resolved Fourier bin, so one expects the likelihood as a function of \(m_\phi\) to vary on this scale.

Across a search range \([m_\phi^{\rm min},\,m_\phi^{\rm max}]\), the number of independently resolvable ULDM masses is therefore approximately
\begin{equation}
    N_m \approx \frac{m_\phi^{\rm max}-m_\phi^{\rm min}}{\Delta m_\phi}
    \approx 1.1\times 10^{3}
    \left(\frac{m_\phi^{\rm max}}{10^{-20}\,\mathrm{eV}}\right)
    \left(\frac{T_{\rm obs}}{15\,\mathrm{yr}}\right),
\end{equation}
where in the second estimate we assumed \(m_\phi^{\rm min}\ll m_\phi^{\rm max}\). Thus even before considering any additional nuisance or latent parameters, the mass direction alone contains \(\mathcal{O}(10^3)\) independently resolved locations in parameter space.

Now, consider an analysis in which \(m_\phi\) is sampled directly with a local random-walk proposal. In order to resolve likelihood structure on the scale of distinct Fourier bins, the characteristic proposal size must satisfy \(\delta m_\phi \lesssim \Delta m_\phi\). After \(N\) accepted random-walk updates, the typical excursion in mass space is then
\begin{equation}
    \Delta M \sim \sqrt{N}\,\delta m_\phi \lesssim \sqrt{N}\,\Delta m_\phi.
\end{equation}
To explore the full prior range, this excursion must be at least as large as \(m_\phi^{\rm max}-m_\phi^{\rm min}\), which implies
\begin{equation}
    N \gtrsim
    \left(\frac{m_\phi^{\rm max}-m_\phi^{\rm min}}{\Delta m_\phi}\right)^2
    \sim N_m^2.
\end{equation}
For the mass range relevant to PTA ULDM searches, this already requires
\begin{equation}
    N \gtrsim 10^6.
\end{equation}
This estimate is optimistic, since it assumes that all other parameters are fixed and ignores correlations between \(m_\phi\) and the remaining nuisance, background, and latent variables. For additional context, in our implementation even a baseline analysis with no ULDM signal already requires \(\gtrsim 5\times 10^6\) Monte Carlo samples to achieve satisfactory convergence. Treating \(m_\phi\) as a continuously sampled parameter would therefore add a sampling burden comparable to that of the entire baseline inference problem.

There is also a separate issue for the final scientific product of the search. ULDM results are usually reported as constraints on the signal coupling at fixed \(m_\phi\). If \(m_\phi\) were sampled continuously, these fixed-mass constraints would have to be reconstructed by conditioning the joint posterior samples on a narrow mass interval of width \(\sim \Delta m_\phi\). For a search range containing \(N_m\) independently resolved masses, the effective sample size available in any one such interval is at best of order
\begin{equation}
    {\rm ESS}_{\rm bin}
    \sim
    \frac{{\rm ESS}_{\rm tot}}{N_m},
\end{equation}
up to variations in the posterior mass density and further losses from autocorrelation. Thus even a globally converged chain can have too few effectively independent samples in any particular mass bin to give reliable conditional marginals for the coupling. This problem is especially acute for exclusion limits, where one requires trustworthy coupling constraints across the full mass range rather than only near the masses favored by the data.

Together, these considerations motivate our choice to perform ULDM analyses on a discrete mass grid, with independent inference at each mass, rather than sampling \(m_\phi\) continuously.

\section{Statistics of the latent variables}
\label{app:amplitude_phase_statistics}

This appendix derives the properties of the latent amplitude-phase variables used in our deterministic analysis. The main results are as follows. First, at fixed pulsar distances, each local amplitude is Rayleigh distributed and each local phase is uniform. Second, in the exact distance-uncertain model, the one-point marginals remain unchanged but the full joint prior generally does not factorize. Third, in the augmented model introduced in Sec.~\ref{subsec:augmented_latent_prior}, the phases are i.i.d. uniform and independent of the amplitudes, so all nontrivial spatial information is carried by the joint amplitude prior.

\subsection{Latent amplitude-phase statistics}

For a specified ULDM mass \(m_\phi\), the monochromatic retarded-time field at location \(I\) may be written as
\begin{equation}
    \hat\phi_I(t)=\hat C_I\cos(m_\phi t)+\hat S_I\sin(m_\phi t),
\end{equation}
with \(\{(\hat C_I,\hat S_I)\}\) jointly Gaussian distributed according to Eq.~\eqref{eq:FourierCoefficientCovariances}. It is convenient to combine these variables into the complex amplitudes
\begin{equation}
    \hat z_I \equiv \hat C_I+i\hat S_I = A_I e^{i\alpha_I}.
\end{equation}
In terms of Eq.~\eqref{eq:FourierCoefficientCovariances}, one finds
\begin{equation}
\begin{gathered}
    \langle \hat z_I \hat z_J \rangle = 0,\\
    \langle \hat z_I \hat z_J^\ast \rangle = \hat\Gamma_{IJ}^\ast.
\end{gathered}
\label{eq:complex_latent_covariances_compact}
\end{equation}
Thus the latent vector \(\hat{\bm z}\) is jointly proper complex Gaussian.

For fixed pulsar distances, the one-point marginal of each \(\hat z_I\) is completely determined by \(\hat\Gamma_{II}=1\), which implies
\begin{equation}
    p(\hat z_I)=\frac{1}{\pi}e^{-|\hat z_I|^2}.
\end{equation}
Transforming to polar variables \(\hat z_I=A_I e^{i\alpha_I}\) gives
\begin{equation}
    p(A_I,\alpha_I)
    =
    \frac{A_I}{\pi}e^{-A_I^2}
    =
    \left(2A_I e^{-A_I^2}\right)\left(\frac{1}{2\pi}\right).
\label{eq:single_polar_density_compact}
\end{equation}
Hence, for each location \(I\), the amplitude and phase are independent, with \(A_I\) Rayleigh distributed and \(\alpha_I\) uniform on \([0,2\pi)\). In particular, \(\langle A_I\rangle=\sqrt{\pi}/2\).

In the exact distance-uncertain model, both the correlation magnitudes \(R_{IJ}\) and the effective distance phases \(\psi_I\), see Eqs.~\eqref{eq:corr-amplitude} and~\eqref{eq:corr-phase}, depend on the pulsar distances. The full latent prior at fixed \(m_\phi\) is therefore the mixture
\begin{equation}
    \pi(\bm A,\bm\alpha \mid m_\phi)
    =
    \int \prod_{I=1}^{N_p} dx_I\,
    \pi_I(x_I)\,
    \pi(\bm A,\bm\alpha \mid \{x_I\},m_\phi).
\label{eq:distance_marginalized_amp_phase_prior_compact}
\end{equation}
Because varying the distances changes both the correlation envelope and the pulsar-term phase delays, the resulting joint distribution generally does not admit a simple factorization in \((\bm A,\bm\alpha)\). This is the sense in which the exact distance-marginalized latent prior is complicated.

Nevertheless, the one-point marginals remain unchanged. For any fixed choice of distances, the conditional marginal \(p(\hat z_I\mid \{x_I\},m_\phi)\) depends only on \(\hat\Gamma_{II}=1\), and is therefore independent of the distances. It follows that the distance-marginalized one-point density is still
\begin{equation}
    p(\hat z_I\mid m_\phi)=\frac{1}{\pi}e^{-|\hat z_I|^2}.
\end{equation}
Equivalently, each \(A_I\) remains Rayleigh distributed and each \(\alpha_I\) remains uniform at the one-point level. The complication lies only in the full joint prior across the PTA.

A substantial simplification occurs in the augmented model of Sec.~\ref{subsec:augmented_latent_prior}, where the correlation magnitudes \(R_{IJ}\) are determined by the pulsar distances but the effective distance phases \(\psi_I\) for \(I\ge 1\) are treated as independent nuisance parameters distributed uniformly on \([0,2\pi)\). In that case,
\begin{equation}
    \langle \hat z_I \hat z_J^\ast \rangle
    =
    e^{i(\psi_I-\psi_J)}R_{IJ}.
\end{equation}
Defining rotated latent variables
\begin{equation}
    \hat w_I \equiv e^{-i\psi_I}\hat z_I = A_I e^{i\beta_I},
\end{equation}
with \(\psi_0=0\) at Earth, we obtain a proper complex Gaussian vector \(\hat{\bm w}\) with real covariance matrix \(\mathbf R\). Since this rotation changes only the phases, the amplitudes are unchanged.

The joint density of \((\bm A,\bm\beta)\) at fixed distances depends on the phases only through differences \(\beta_I-\beta_J\). Therefore it is invariant under a common phase shift of all \(\beta_I\), which implies that the overall phase is uniform and independent of the amplitudes. Writing \(\alpha_I=\beta_I+\psi_I\) mod \(2\pi\), and using the fact that the nuisance phases \(\psi_I\) for \(I\ge 1\) are i.i.d. uniform, it follows that all physical phases \(\alpha_I\) are themselves i.i.d. uniform and independent of \(\bm A\). Hence the conditional prior factorizes as
\begin{equation}
    \pi(\bm A,\bm\alpha \mid \{x_I\},m_\phi)
    =
    \pi(\bm A \mid \{x_I\},m_\phi)
    \prod_{I=0}^{N_p}\frac{1}{2\pi}.
\label{eq:conditional_factorized_prior_compact}
\end{equation}
Marginalizing over the pulsar distance priors preserves this structure, so the augmented-model latent prior takes the form
\begin{equation}
    \pi(\bm A,\bm\alpha \mid m_\phi)
    =
    \pi(\bm A \mid m_\phi)
    \prod_{I=0}^{N_p}\frac{1}{2\pi}.
\label{eq:augmented_factorized_prior_compact}
\end{equation}
Thus, in the augmented model, the phases are i.i.d. uniform and independent of the amplitudes, while all nontrivial spatial information is encoded in the correlated amplitude prior \(\pi(\bm A\mid m_\phi)\).

\subsection{Relation to previous analyses}
\label{app:prior_analyses}

The augmented formulation also makes it straightforward to recover the limiting signal models used in previous PTA analyses of ULDM.

In the fully correlated limit, \(\ell\to\infty\), the correlation-envelope matrix approaches
\begin{equation}
    R_{IJ}\to 1,
    \qquad
    \mathbf R \to \mathbf 1 \mathbf 1^T.
\end{equation}
The rotated latent vector \(\hat{\bm w}\) therefore becomes rank one, so there exists a single proper complex Gaussian random variable \(\xi\sim\mathcal{CN}(0,1)\) such that \(\hat w_I=\xi\) for all \(I\). Writing \(\xi=A_\star e^{i\beta_\star}\), one finds that all local amplitudes are equal,
\begin{equation}
    A_I=A_\star
    \qquad
    \forall I,
\end{equation}
with \(A_\star\) Rayleigh distributed; we denote this common amplitude by its Earth-term value, \(A_0\equiv A_\star\). After marginalizing over the nuisance phases \(\psi_I\), the physical phases remain independent and uniform, and the latent prior reduces to
\begin{equation}
    \pi_{\rm corr}(\bm A,\bm\alpha)
    =\frac{ 2A_0 e^{-A_0^2}}{2 \pi}
    \prod_{I=1}^{N_p}\frac{\delta(A_I-A_0)}{2 \pi}.
\label{eq:fully_corr_latent_prior_compact}
\end{equation}
This is precisely the prior structure assumed in the fully correlated limit, where all locations share a common local ULDM amplitude.

In the opposite limit, \(\ell\to 0\), distinct locations fall into independent coherence patches, so
\begin{equation}
    R_{IJ}\to \delta_{IJ},
    \qquad
    \mathbf R \to \mathbf I.
\end{equation}
The rotated latent variables then become independent proper complex Gaussians,
\begin{equation}
    \hat w_I \sim \mathcal{CN}(0,1),
\end{equation}
and the latent prior factorizes across locations:
\begin{equation}
    \pi_{\rm uncorr}(\bm A,\bm\alpha)
    =
    \prod_{I=0}^{N_p}
    \left[
        2A_I e^{-A_I^2}\,
        \frac{1}{2\pi}
    \right].
\label{eq:fully_uncorr_latent_prior_compact}
\end{equation}
This reproduces the fully uncorrelated limit used in earlier analyses, in which each Earth or pulsar term carries an independent local amplitude and phase.

Equations~\eqref{eq:fully_corr_latent_prior_compact} and \eqref{eq:fully_uncorr_latent_prior_compact} show that the finite-correlation latent-field prior continuously interpolates between the two limiting models adopted previously, while also providing a principled treatment of the intermediate partially correlated regime.

\section{Neural spline flow implementation}
\label{app:nsf}

The training samples for the normalizing-flow surrogate are generated according to the procedure described in Sec.~\ref{sec:deterministic_analysis}. At fixed ULDM mass \(m_\phi\), this yields samples from the distance-marginalized amplitude prior \(\pi(\bm A\mid m_\phi)\). This appendix records the additional implementation details used to represent that distribution efficiently with a neural spline flow.

A direct flow model for the amplitudes \(\bm A\) is possible, but is inefficient because each component has a Rayleigh one-point marginal, while the flow is built on a Gaussian base distribution. To improve training efficiency, we therefore apply an analytic change of variables that maps each amplitude's Rayleigh marginal onto a standard normal, and train the flow only on the remaining joint dependence structure.

Since each amplitude \(A_I\) has marginal distribution
\begin{equation}
    p(A_I)=2A_I e^{-A_I^2},
\end{equation}
we define transformed variables
\begin{equation}
    y_I=\Phi^{-1}\!\left(1-e^{-A_I^2}\right),
\end{equation}
where \(\Phi\) is the standard normal cumulative distribution function. By construction, each \(y_I\) has an exactly standard normal marginal distribution. The transformed vector \(\bm y\) therefore has Gaussian one-point marginals, with all remaining nontrivial structure encoded in its correlations and higher-dimensional dependence.

For each analysis mass, we model the transformed distribution with a separate neural spline flow,
\begin{equation}
    q_{\theta}^{(m_\phi)}(\bm y).
\end{equation}
As discussed in Sec.~\ref{sec:deterministic_analysis}, we found a single conditional flow in \(m_\phi\) to be significantly less stable than training separate mass-specific models. Our implementation uses the \texttt{zuko} package and a neural spline flow architecture with \(8\) spline transforms, hidden-layer widths \((256,256)\), and \(16\) spline bins in each transform. These choices were found to provide adequate accuracy across the mass range considered in this work.

The training objective is the negative log-likelihood of the transformed samples,
\begin{equation}
    \mathcal L(\theta)
    =
    -\frac{1}{N_{\rm batch}}
    \sum_{n=1}^{N_{\rm batch}}
    \log q_{\theta}^{(m_\phi)}(\bm y^{(n)}).
\end{equation}
Optimization is performed with Adam, using learning rate \(10^{-3}\), batch size \(2048\), and a cosine-annealing learning-rate schedule over \(1000\) epochs. These settings were used uniformly across the mass-specific flow models.

At inference time, the learned prior in amplitude space is recovered by a change of variables,
\begin{equation}
    \log \pi_{\rm flow}(\bm A\mid m_\phi)
    =
    \log q_{\theta}^{(m_\phi)}(\bm y(\bm A))
    +
    \sum_{I=0}^{N_p}\log\left|\frac{\partial y_I}{\partial A_I}\right|.
\end{equation}
Writing \(\varphi\) for the standard normal probability density function, the Jacobian factor is
\begin{equation}
    \frac{\partial y_I}{\partial A_I}
    =
    \frac{2A_I e^{-A_I^2}}{\varphi(y_I)}.
\end{equation}
Thus the learned amplitude prior combines an analytic Gaussianization of the Rayleigh marginals with a flow-based representation of the remaining correlated structure. This is the finite-correlation amplitude prior used in the deterministic analyses in Sec.~\ref{sec:deterministic_analysis}.

\section{Additional unblinded analyses}
\label{app:UnblindedAnalyses}

\begin{figure*}[!t]
    \centering
    \includegraphics[width=.49\textwidth]{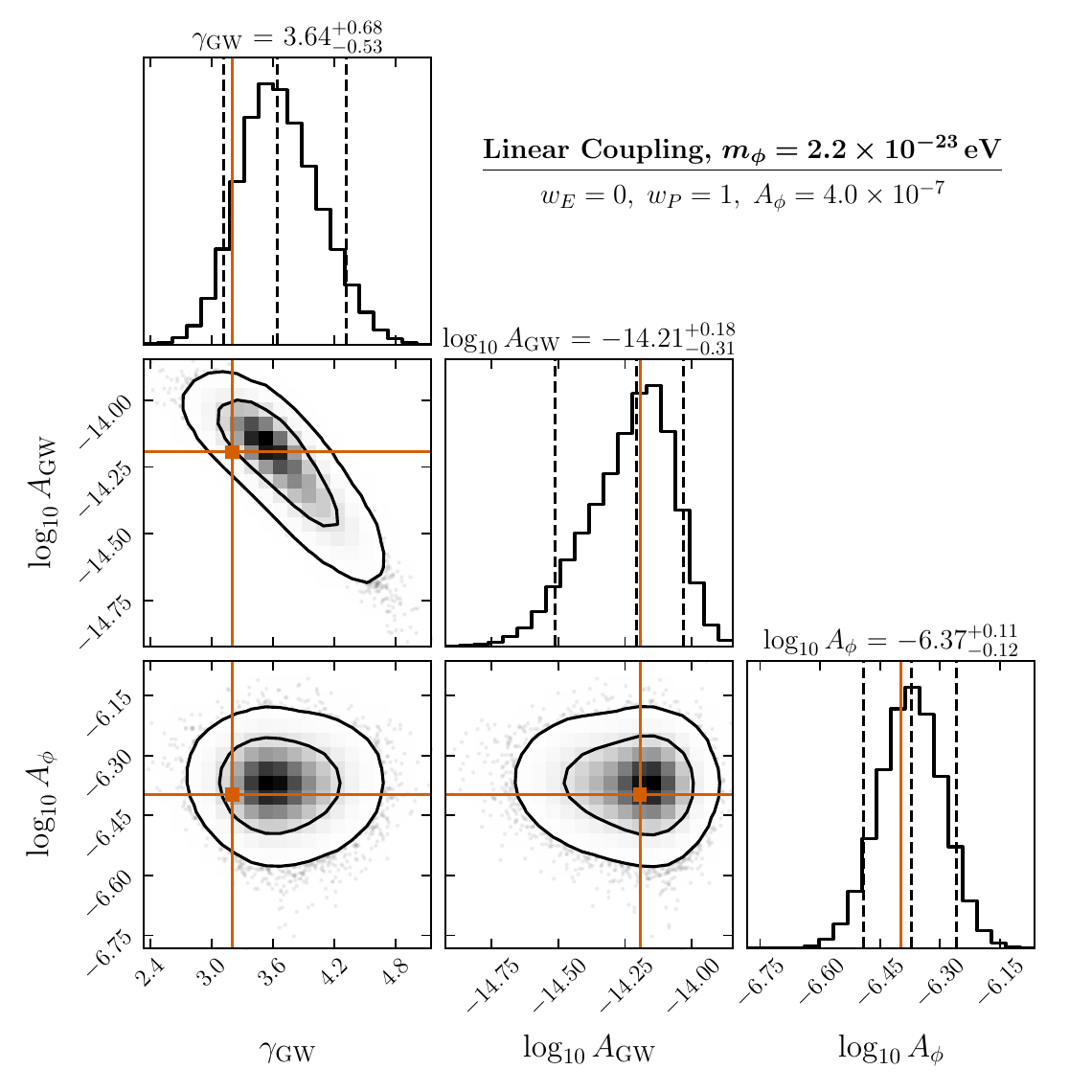}
    \includegraphics[width=.49\textwidth]{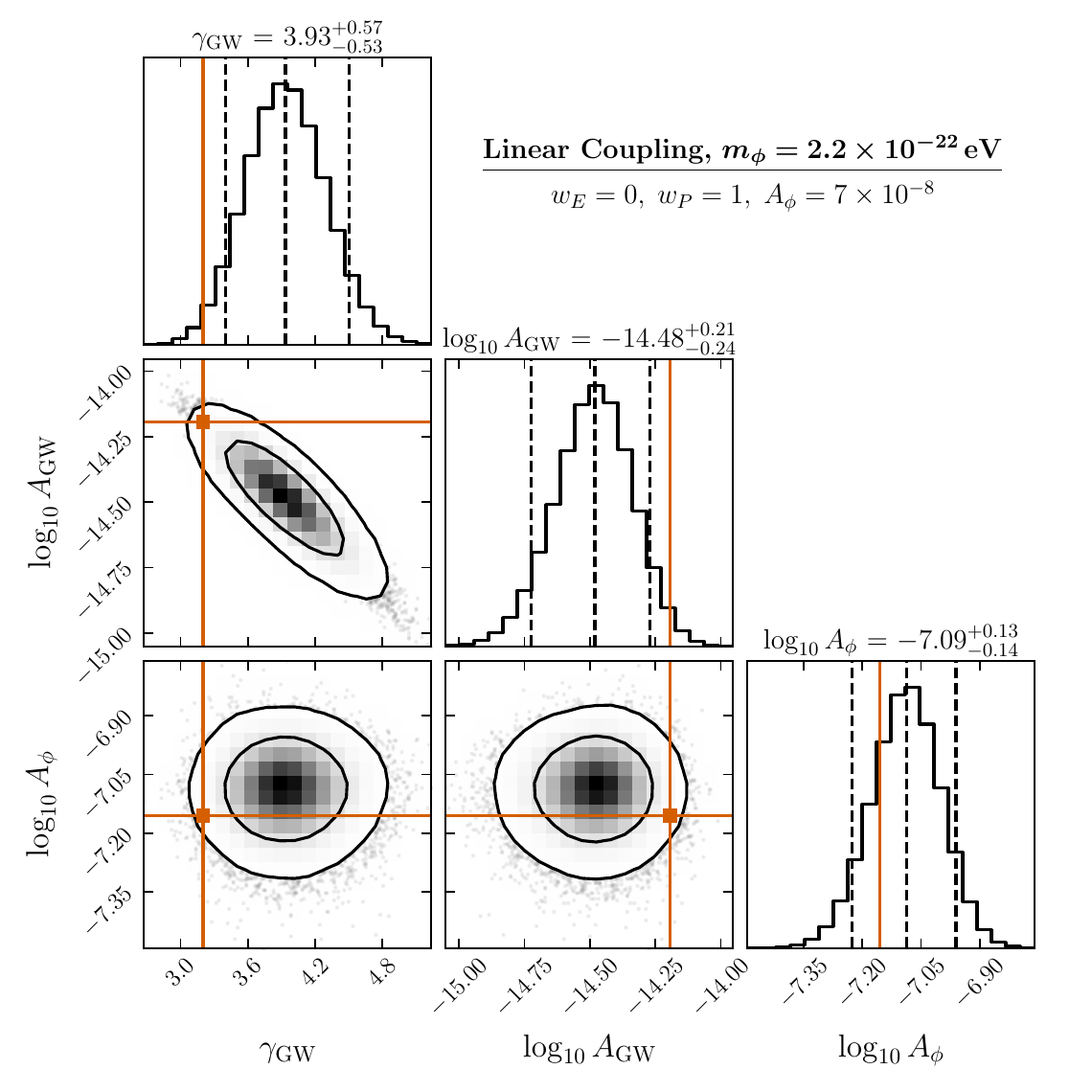}
    \caption{As in Fig.~\ref{fig:LinearUnblinded_MainExample}, but for $m_\phi = 2.2 \times 10^{-23}\,\mathrm{eV}$ in the left corner plot and for $m_\phi = 2.2 \times 10^{-22}\,\mathrm{eV}$ in the right corner plot. At these masses and associated signal frequencies, there is less overlap between SGWB power and signal power, leading to more precise SGWB parameter reconstruction in these examples than realized in the analysis at $m_\phi = 2.2 \times 10^{-24}\,\mathrm{eV}$ presented in the main text.}
    \label{fig:LinearUnblinded_AppExamples}
\end{figure*}

\begin{figure*}[!t]
    \centering
    \includegraphics[width=.49\textwidth]{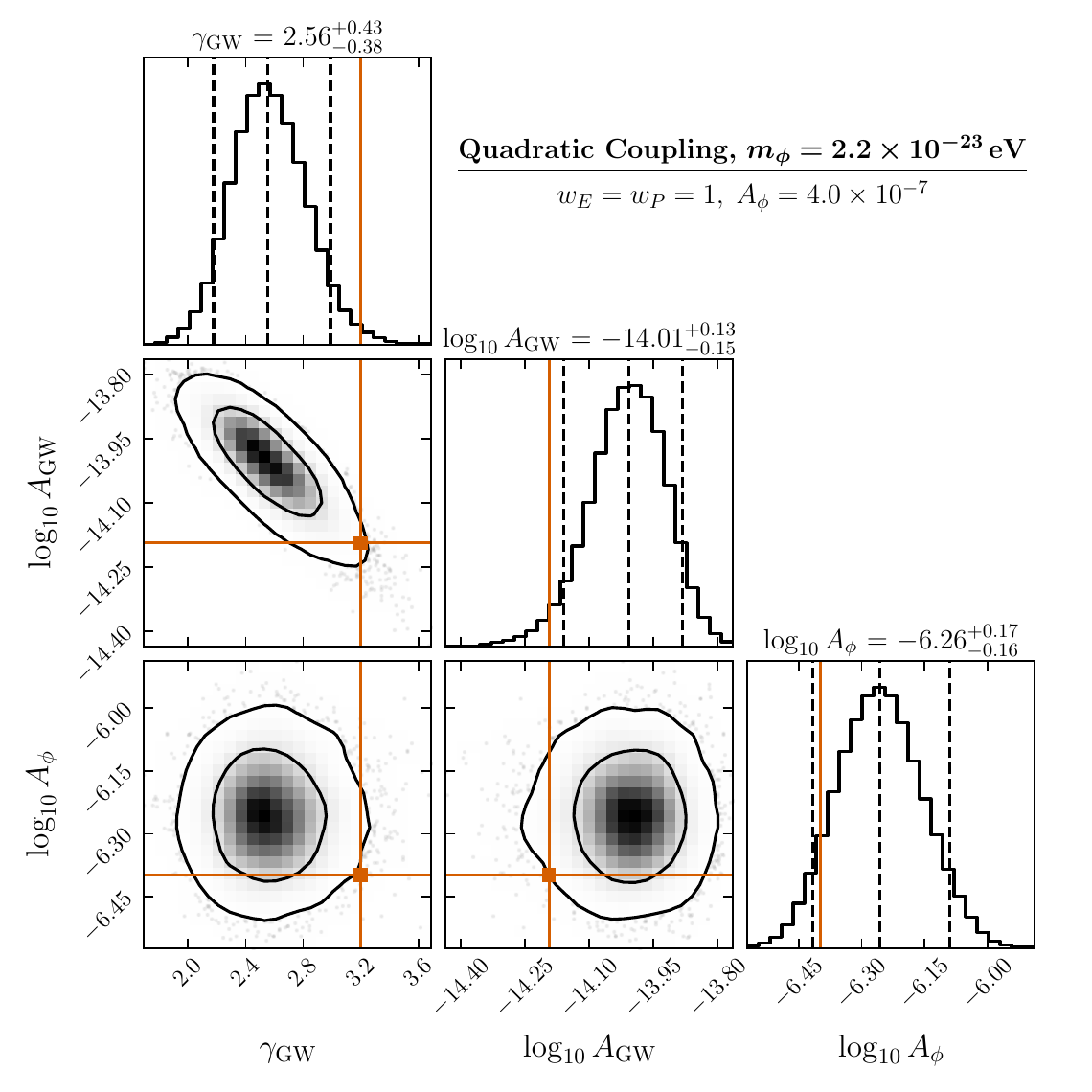}
    \includegraphics[width=.49\textwidth]{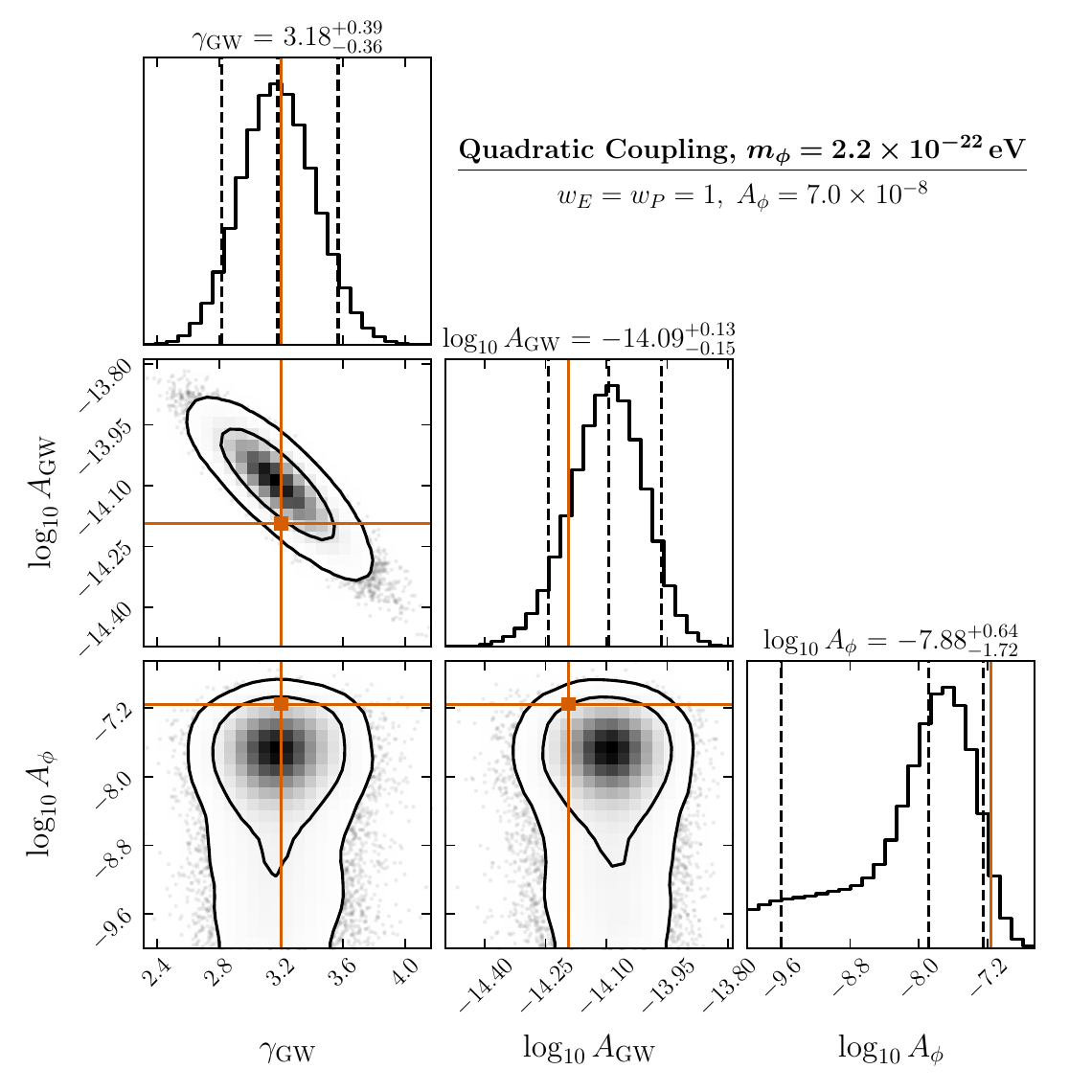}
    \caption{As in Fig.~\ref{fig:QuadraticUnblinded_MainExample}, but for two additional unblinded injections of the quadratically coupled signal at the intermediate and uncorrelated masses. Parameter estimation continues to perform at a high level, though for the large mass example, we find the true value of $A_\phi$ for which the data were generated to be slightly outside the 90\% credible interval.}
    \label{fig:QuadraticUnblinded_AppExamples}
\end{figure*}

In this appendix, we present additional results from our unblinded signal injection tests. Complementing our accurate signal reconstruction in the main text for a signal in the highly correlated limit at $m_\phi = 2.2 \times 10^{-24}\,\mathrm{eV}$, we demonstrate similarly successful signal reconstructions in the moderately correlated regime at $m_\phi = 2.2 \times 10^{-23}\,\mathrm{eV}$ and in the uncorrelated regime at $m_\phi = 2.2 \times 10^{-22}\,\mathrm{eV}$ for linearly coupled ULDM signals in Fig.~\ref{fig:LinearUnblinded_AppExamples} and for quadratically coupled ULDM signals in Fig.~\ref{fig:QuadraticUnblinded_AppExamples}. See main text for details.

\section{Additional blinded analyses}
\label{app:BlindedAnalyses}

\begin{figure*}[!t]
    \centering
    \includegraphics[width=.48\textwidth]{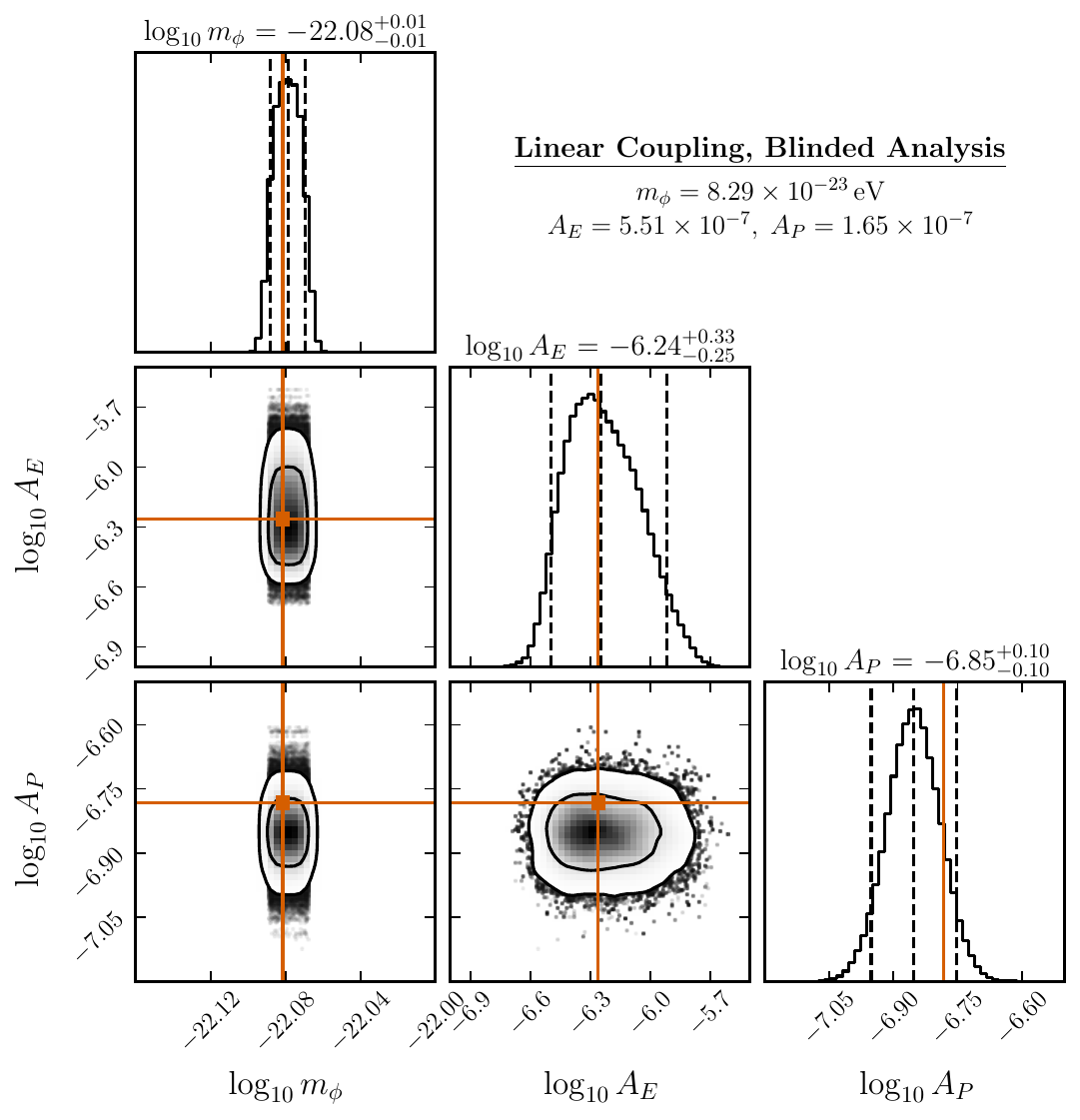}
    \includegraphics[width=.48\textwidth]{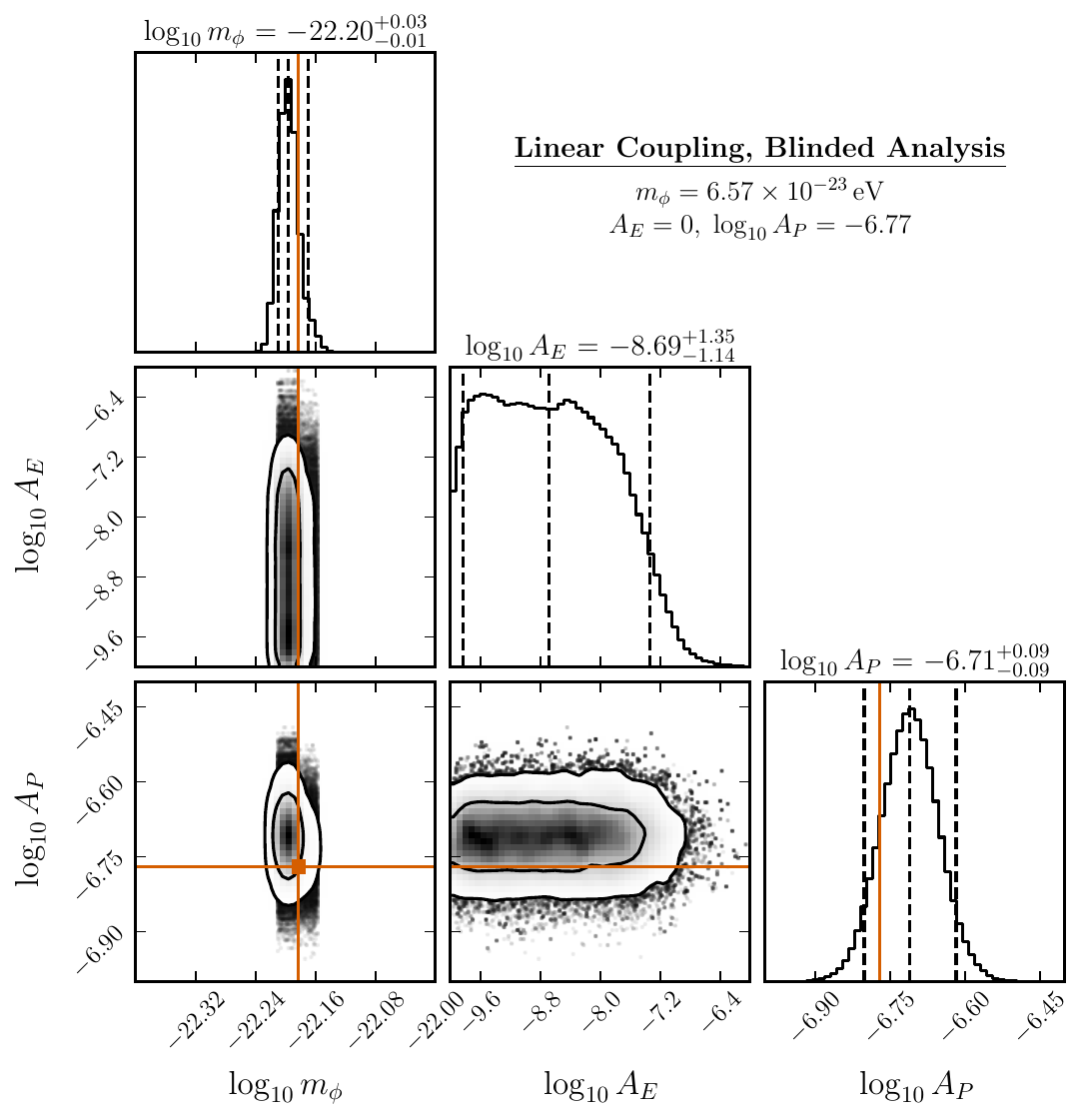}
    \caption{As in Fig.~\ref{fig:LinearBlinded_MainExample}, but for two additional, independently generated blinded signal analysis tests. Parameter reconstruction continues to perform at the expected level.
    }
    \label{fig:LinearBlinded_AppExamples}
\end{figure*}

\begin{figure*}[!t]
    \centering
    \includegraphics[width=.48\textwidth]{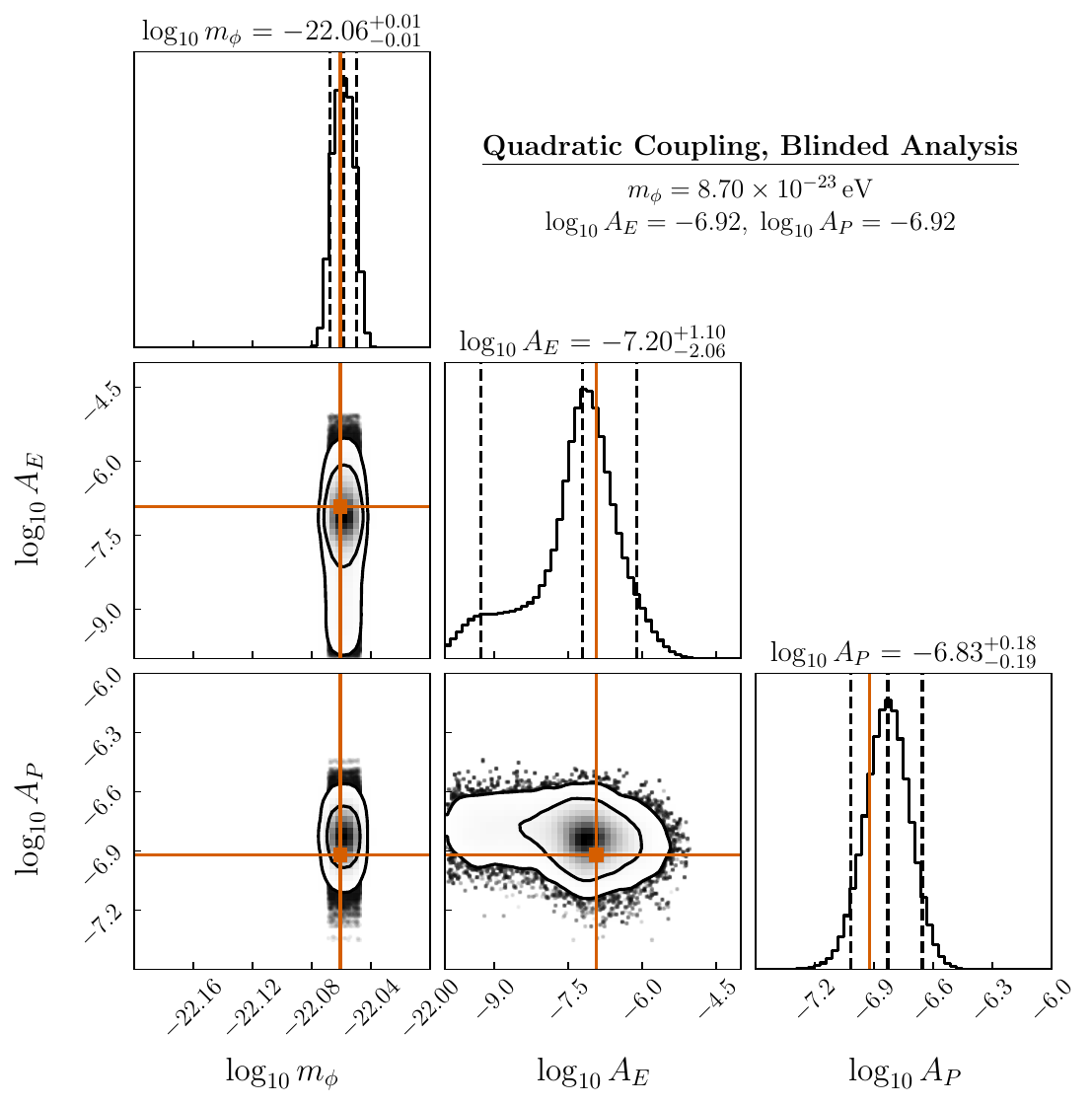}
    \includegraphics[width=.48\textwidth]{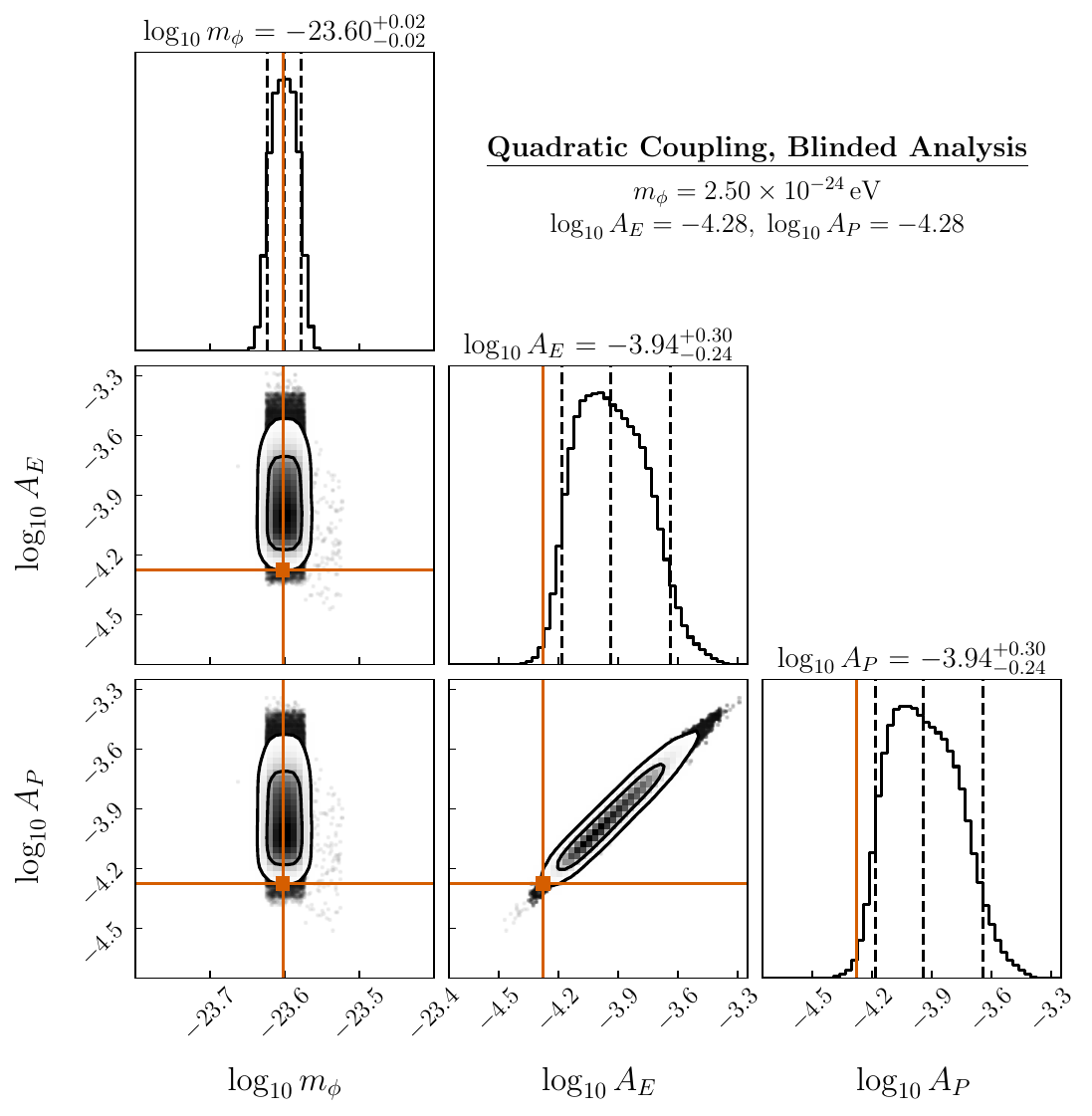}
    \caption{As in Fig.~\ref{fig:QuadraticBlinded_MainExample}, but for two additional, independently generated blinded injections of the quadratically coupled signal. As before, signals were generated at identical $w_E = w_P$ to realize the coupling-independent gravitational signal.}\label{fig:QuadraticBlinded_AppExamples}
\end{figure*}

For the linearly coupled signal, Fig.~\ref{fig:LinearBlinded_AppExamples} presents two further blinded injections, complementing the representative example in the main text. Likewise, Fig.~\ref{fig:QuadraticBlinded_AppExamples} presents two further blinded injections for the quadratically coupled signal.

\section{Gaussian-process formulation of the linear signal}
\label{app:stochastic_analysis}

For completeness, we describe an alternative formulation of the linearly coupled ULDM signal as a correlated Gaussian process in the timing residuals.
This construction follows directly from the Gaussian statistics of the ULDM field.
It is mathematically interesting, but it is not the primary analysis used in this work because the distance-induced nuisance structure remains, and the resulting covariance likelihood can be numerically less stable than the deterministic formulation.

\subsection{Signal construction}
For the fixed coupling-ratio parametrization, starting from Eq.~\eqref{eq:useful_residual}, we write the induced timing residual for pulsar \(I\) as
\begin{equation}
    r_I(t)
    =
    A_\phi
    \left[
        C_I \cos(m_\phi t)
        +
        S_I \sin(m_\phi t)
    \right],
    \label{eq:stochastic_residual_cos_sin}
\end{equation}
where the residual-level cosine-sine coefficients are
\begin{align}
    C_I
    &\equiv
    w_E \hat C_0
    +
    w_P \hat C_I ,
    &
    S_I
    &\equiv
    w_E \hat S_0
    +
    w_P \hat S_I .
    \label{eq:residual_level_coefficients}
\end{align}
Here \(I,J\geq 1\) label pulsars, while the index \(0\) denotes Earth.
Using Eq.~\eqref{eq:FourierCoefficientCovariances}, these coefficients have covariance
\begin{align}
    \langle C_I C_J\rangle
    &=
    \langle S_I S_J\rangle
    =
    \mathrm{Re}\,\Gamma_{IJ},
    \\
    \langle C_I S_J\rangle
    &=
    -
    \langle S_I C_J\rangle
    =
    \mathrm{Im}\,\Gamma_{IJ},
    \label{eq:residual_level_coefficient_covariances}
\end{align}
with the residual-level complex overlap matrix
\begin{equation}
    \Gamma_{IJ}
    =
    \frac12
    \left[
        w_E^2
        +
        w_Ew_P
        \left(
            \hat\Gamma_{I0}
            +
            \hat\Gamma_{0J}
        \right)
        +
        w_P^2
        \hat\Gamma_{IJ}
    \right].
    \label{eq:residual_orf}
\end{equation}
This overlap matrix is defined for the timing residuals of the \(N_p\) pulsars and should be distinguished from the field-level overlap matrix \(\hat\Gamma_{IJ}\), which also includes the Earth index as a field location.

We now collect the timing residuals into a single data vector.
Let pulsar \(I\) have observing times \(\mathbf t_I\), with length \(N_I\), and define
\begin{equation}
    \mathbf c_I
    =
    \cos(m_\phi \mathbf t_I),
    \qquad
    \mathbf s_I
    =
    \sin(m_\phi \mathbf t_I).
\end{equation}
The single-pulsar monochromatic design matrix is
\begin{equation}
    \mathbf F_I
    =
    \begin{bmatrix}
        \mathbf c_I & \mathbf s_I
    \end{bmatrix},
\end{equation}
and the full PTA design matrix is the block-diagonal matrix
\begin{equation}
    \mathbf F_\phi
    =
    \mathrm{blockdiag}
    \left(
        \mathbf F_1,
        \mathbf F_2,
        \ldots,
        \mathbf F_{N_p}
    \right).
    \label{eq:uldm_gp_design_matrix}
\end{equation}
We also define the coefficient vector
\begin{equation}
    \mathbf a
    =
    \begin{bmatrix}
        C_1 & S_1 & C_2 & S_2 & \cdots & C_{N_p} & S_{N_p}
    \end{bmatrix}^{T}.
    \label{eq:residual_coefficient_vector}
\end{equation}
The ULDM timing residual vector is then
\begin{equation}
    \mathbf r_\phi
    =
    A_\phi
    \mathbf F_\phi
    \mathbf a .
    \label{eq:gp_residual_vector}
\end{equation}
Since \(\mathbf a\) is Gaussian, \(\mathbf r_\phi\) is also Gaussian.

The covariance of \(\mathbf a\) is a \(2N_p\times 2N_p\) block matrix,
\begin{equation}
    \bm\Phi_\phi
    \equiv
    \langle
        \mathbf a \mathbf a^T
    \rangle,
    \qquad
    \bm\Phi_{\phi,IJ}
    =
    \begin{bmatrix}
        \mathrm{Re}\,\Gamma_{IJ}
        &
        \mathrm{Im}\,\Gamma_{IJ}
        \\
        -\mathrm{Im}\,\Gamma_{IJ}
        &
        \mathrm{Re}\,\Gamma_{IJ}
    \end{bmatrix}.
    \label{eq:residual_coefficient_covariance}
\end{equation}
The induced ULDM covariance in the timing residuals is therefore
\begin{equation}
    \bm\Sigma_\phi
    =
    A_\phi^2
    \mathbf F_\phi
    \bm\Phi_\phi
    \mathbf F_\phi^T .
    \label{eq:signal_covariance_low_rank}
\end{equation}
This gives a low-rank representation of the time-domain covariance of the linearly coupled ULDM signal.

\subsection{Gaussian-process likelihood}

At fixed pulsar distances and effective phases, the Gaussian-process likelihood is obtained by adding the ULDM signal covariance in Eq.~\eqref{eq:signal_covariance_low_rank} to the background covariance,
\begin{equation}
    \bm\Sigma_{\rm tot}
    =
    \bm\Sigma_{\rm bkg}
    +
    \bm\Sigma_\phi .
    \label{eq:stochastic_total_covariance}
\end{equation}
For a PTA residual vector $\mathbf d_{\rm PTA}$, the corresponding likelihood is
\begin{equation}
\begin{split}
    \mathcal L_{\rm GP}
    (&\mathbf d_{\rm PTA}
    \mid
    \bm\theta_{\rm bkg},
    A_\phi,
    \{x_I\},
    \{\psi_I\})
    =
    \frac{1}
    {|2\pi\bm\Sigma_{\rm tot}|^{1/2}}
    \\
    &\times
    \exp\!\left[
        -\frac12
        \mathbf d_{\rm PTA}^{T}
        \bm\Sigma_{\rm tot}^{-1}
        \mathbf d_{\rm PTA}
    \right].
\end{split}
\label{eq:gp_uldm_likelihood}
\end{equation}
Here $\bm\Sigma_{\rm bkg}$ contains the non-ULDM timing-residual covariance, while $\bm\Sigma_\phi$ is the low-rank ULDM covariance constructed from the residual-level coefficient covariance $\bm\Phi_\phi$.
This formulation analytically marginalizes over the Gaussian coefficients $\mathbf a$ at fixed $\{x_I,\psi_I\}$.

In the exact distance-dependent model, the same pulsar distances determine both the correlation magnitudes $R_{IJ}$ and the effective phases $\psi_I$ appearing in the overlap matrix $\Gamma_{IJ}$.
The augmented model modifies this prescription in the same way as in Sec.~\ref{subsec:augmented_latent_prior}: the distances $\{x_I\}$ are used to compute the correlation magnitudes $R_{IJ}$, while the phases $\{\psi_I\}$ are treated as independent nuisance parameters with uniform priors on $[0,2\pi)$, with $\psi_0=0$ at Earth.
Thus, the residual-level overlap matrix in Eq.~\eqref{eq:residual_orf} is evaluated with
\begin{equation}
    \hat\Gamma_{IJ}
    =
    R_{IJ} e^{-i(\psi_I-\psi_J)},
    \label{eq:augmented_field_orf_gp}
\end{equation}
where the amplitudes $R_{IJ}$ and phases $\psi_I$ are no longer tied to the same distance variables.
This prescription preserves the distance dependence of the spatial correlation envelope while treating the effectively unconstrained retarded-time phases as independent nuisance parameters.

We implemented this covariance model in \texttt{enterprise} by treating the ULDM signal as a low-rank correlated Gaussian process.
In standard PTA applications, the Fourier-basis coefficient covariance is usually diagonal in the cosine-sine basis.
The ULDM covariance in Eq.~\eqref{eq:residual_coefficient_covariance} is more general: for each pulsar pair, the imaginary part of $\Gamma_{IJ}$ mixes sine and cosine coefficients across pulsars.
Implementing the Gaussian-process formulation, therefore, required an overlap-reduction function that returns the full $2\times2$ block
\begin{equation}
    \bm\Phi_{\phi,IJ}
    =
    \begin{bmatrix}
        \mathrm{Re}\,\Gamma_{IJ}
        &
        \mathrm{Im}\,\Gamma_{IJ}
        \\
        -\mathrm{Im}\,\Gamma_{IJ}
        &
        \mathrm{Re}\,\Gamma_{IJ}
    \end{bmatrix},
\end{equation}
rather than a diagonal covariance for independent sine and cosine modes.

In total, the Gaussian-process formulation provides a compact representation of the linearly coupled signal after analytic marginalization over the Gaussian coefficients $\mathbf a$. However, the covariance still depends on the distance-dependent correlation envelope through $R_{IJ}$ and on the augmented phase variables through $\psi_I$. Thus, the main inference challenge is not removed; instead, it appears as repeated numerical linear algebra involving a signal covariance matrix whose entries vary nontrivially with the same nuisance parameters. More seriously, for the ULDM masses and pulsar distance priors considered here, we found that the Gaussian-process covariance can become numerically ill-conditioned, leading to inference failures. By contrast, the latent-field formulation proved more convenient and numerically robust, especially because the distance and phase marginalization can be incorporated into the flow prior described in Sec.~\ref{subsec:flow_amplitude_prior}. We, therefore, use the deterministic latent-field formulation as our primary analysis framework.

\bibliographystyle{utphys}
\bibliography{main}

\end{document}